\documentclass[5p,english,authoryear]{elsarticle}
\usepackage{url}
\usepackage{amsmath}
\usepackage[english]{babel}
\usepackage{mathtools}
\mathtoolsset{showonlyrefs}

\journal{Astronomy \& Computing}

\newcommand{\figureref}[1]{Figure \ref{#1}}
\newcommand{\multifigref}[2]{Figures \ref{#1} - \ref{#2}}
\newcommand{\tableref}[1]{Table \ref{#1}}
\newcommand{\multitableref}[2]{Tables \ref{#1} - \ref{#2}}
\newcommand{\code}[1]{{\normalfont\fontfamily{cmvtt}\selectfont #1}}

\newcommand{\cmacionize}{\textsc{CMacIonize}}
\newcommand{\shadowfax}{\textsc{Shadowfax}}

\begin{document}

\begin{frontmatter}

\title{The Monte Carlo photoionization and moving-mesh radiation hydrodynamics
code \cmacionize{}}
\author[adres]{B.~Vandenbroucke\corref{cor1}}
\ead{bv7@st-andrews.ac.uk}
\author[adres]{K.~Wood}
\ead{kw25@st-andrews.ac.uk}

\cortext[cor1]{Corresponding author}
\address[adres]{SUPA, School of Physics \& Astronomy, University of St Andrews,
North Haugh, St Andrews, KY16 9SS, United Kingdom}

\begin{abstract}
We present the public Monte Carlo photoionization and moving-mesh radiation
hydrodynamics code \cmacionize{}, which can be used to simulate the
self-consistent evolution of HII regions surrounding young O and B stars, or
other sources of ionizing radiation. The code combines a Monte Carlo
photoionization algorithm that uses a complex mix of hydrogen, helium and
several coolants in order to self-consistently solve for the ionization and
temperature balance at any given type, with a standard first order hydrodynamics
scheme. The code can be run as a post-processing tool to get the line emission
from an existing simulation snapshot, but can also be used to run full radiation
hydrodynamical simulations. Both the radiation transfer and the hydrodynamics
are implemented in a general way that is independent of the grid structure that
is used to discretize the system, allowing it to be run both as a standard fixed
grid code, but also as a moving-mesh code.
\end{abstract}


\end{frontmatter}
%%%%%%%%%%%%%%%%%%%%%%%%%%%%%%%%%%%%%%%%%%%%%%%%%%%%%%%%%%%%%%%%%%%%%%%%%%%%%%%%
\section{Introduction}
%%%%%%%%%%%%%%%%%%%%%%%%%%%%%%%%%%%%%%%%%%%%%%%%%%%%%%%%%%%%%%%%%%%%%%%%%%%%%%%%

Photoionization of hydrogen and helium in the interstellar medium (ISM) by
luminous UV sources has an important effect on the evolution and properties of
the ISM. Absorption of ionizing radiation through ionization is an important
source of energy that feeds the expansion of bubbles surrounding young O and B
stars, and hence shapes the structure of the ISM on small scales
\citep{2017Harries}. In cases where the dynamical effect of ionizing radiation
is less pronounced, the presence of ionizing radiation will still alter the
overall ionization balance, not only of hydrogen and helium, but also of other
elements. This, together with an increase in temperature in ionized regions,
will have a visible impact on their emission spectrum, making them stand out as
HII regions. Detailed observations of HII emission spectra contain a wealth of
information about the local ISM and the incident radiation field, and modeling
them is important in understanding observational signatures of star formation
\citep{2012Klassen, 2016Mackey}, and diffuse emission from galactic discs
\citep{2014Barnes, 2018Vandenbroucke}.

On larger scales, photoionization also has important dynamical effects. The
combined UV emission of quasars and young stars in the early Universe generates
a UV background radiation field that is responsible for the reionization of the
Universe by redshift 6 \citep{2001Becker, 2009Alvarez}. This UV background field
affects the formation of galaxies by altering the abundances of ISM coolants
\citep{2013DeRijcke}, and is responsible for suppressing galaxy formation in low
mass haloes \citep{2015BenitezLlambay, 2016Vandenbrouckeb}. Futhermore,
radiative feedback might be an important mechanism to regulate star formation in
galactic discs \citep{2017Peters}.

Modeling photoionization and in particular HII regions requires solving a very
complex system of ionization balance equations for the various elements present
in these regions, which is only possible if strong assumptions are made. The
widely used code \textsc{Cloudy} ascl:9910.001 \citep{2017Ferland} for example
assumes a simple 1D geometry, but keeps track of a large number of elements and
ionization stages. When a real 3D geometry is necessary, it is no longer
feasible to keep track of so many elements, and a selection has to be made,
depending on the problem at hand.

When the effect of ionizing radiation on the dynamics of the ISM is studied,
further assumptions need to be made about how to deal with the coupling between
radiation transfer and hydrodynamics, to get a radiation hydrodynamics (RHD)
scheme. Some methods treat the radiation field as a fluid governed by diffusion
equations \citep{2013Kolb, 2013Rosdahl}. These methods have the advantage that
they do not require extensive algorithmic changes and are relatively efficient.
However, they have undesired side effects, like e.g. the absence of shadows in
optically thin regions, and the fact that extra assumptions need to be made
about the propagation speed of the radiation field to prevent the integration
time step from getting very small. Alternative methods use an approximate ray
tracing scheme \citep{2008Pawlik, 2009Bisbas, 2015Baczynski} which is more
complex to implement and preserves some directional information. However, these
schemes require careful fine tuning to make sure the radiation field accurately
covers the density structure, especially if the density structure is asymmetric
or clumpy.

A more accurate, albeit less efficient way to treat the radiation field is
provided by using Monte Carlo based RHD codes like \textsc{torus} ascl:1404.006
\citep{2000Harries} and \textsc{mocassin} ascl:1110.010 \citep{2005Ercolano}.
These codes have the advantage that they are also much more flexible and easier
to extend with extra physics, e.g. extra chemistry \citep{2015Bisbasb}.
Furthermore, Monte Carlo techniques are also widely used to model dust
scattering and absorption \citep{2013Steinacker}, making it straightforward to
include dust scattering in Monte Carlo based RHD modeling.

In this work, we present our own Monte Carlo RHD code called \cmacionize{}, that
couples a basic finite volume hydrodynamics scheme with a Monte Carlo
photoionization code. Our code can use a variety of different grid types to
discretize the ISM, and can run with both a fixed grid and a fully adaptive
moving mesh. Apart from running as a RHD code, \cmacionize{} can also be run as
a pure Monte Carlo photoionization code, and can be used to post-process density
fields from other simulations.

Our code is written in modular C++ and is meant to be both user-friendly and
efficient by combining a well-structured and documented design with an
implementation that makes use of new features of modern C++11. The code has a
limited number of dependencies and can be run in parallel using a hybrid OpenMP
and MPI parallelization strategy. Some parts of the code are wrapped into a
Python library using Boost
Python\footnote{\url{http://www.boost.org/doc/libs/release/libs/python/doc/html/index.html}}.
The photoionization part of the code can also be used as an external C or
Fortran library, facilitating coupling our code to other simulation codes.

This paper is structured as follows: in section \ref{section:physics} we discuss
the physics that has been implemented in \cmacionize{}, and give a short
overview of the Monte Carlo photoionization technique and the finite volume
hydrodynamics scheme. In section \ref{section:code} we describe the design
considerations that were used during the development of the code, and detail
their implementation. We conclude in section \ref{section:benchmarks} with the
results of a number of benchmark tests that are part of the public code
repository and that show its accuracy and performance.

%%%%%%%%%%%%%%%%%%%%%%%%%%%%%%%%%%%%%%%%%%%%%%%%%%%%%%%%%%%%%%%%%%%%%%%%%%%%%%%%
\section{Physics}
\label{section:physics}
%%%%%%%%%%%%%%%%%%%%%%%%%%%%%%%%%%%%%%%%%%%%%%%%%%%%%%%%%%%%%%%%%%%%%%%%%%%%%%%%

The emission line spectrum of a star forming nebula is determined by its thermal
equilibrium, which is a steady-state equilibrium between heating through
photoionization by UV sources, and cooling by various atomic processes in the
nebula. \citet{2006Osterbrock} identify four important sources of cooling:
\begin{enumerate}
  \item{} Energy loss by recombination of hydrogen and helium, i.e. the reverse
  of the photoionization process,
  \item{} energy loss by bremsstrahlung emitted by free electrons,
  \item{} energy loss by collisionally excited line radiation from some abundant
  metals, and
  \item{} energy loss by collisionally excited line radiation from hydrogen.
\end{enumerate}
In order to compute photoionization and recombination rates, we need to know the
ionization structure of the gas in the nebula, and the temperature of the
nebula. Since the temperature itself is the solution of the thermal equilibrium,
this can only be solved for iteratively.

The thermal and ionization equilibrium is also important for the dynamics of the
nebula: ionized regions have more free particles and hence a higher specific
energy than neutral regions, so that photoionization effectively acts as a
heating term in the hydrodynamics of the gas. In order to properly model this
effect, combined radiation hydrodynamics (RHD) simulations are necessary.

\cmacionize{} can be run in two different modes: either as a pure Monte-Carlo
photoionization code that ray traces the radiation of an ionizing UV radiation
field through a density field and self-consistently solves for the ionization
and temperature structure, or as a radiation hydrodynamics code that uses the
output of the photoionization code as a heating source in a hydrodynamical
simulation. The former is essentially a completely rewritten version of the
photoionization code of \citet{2004Wood}, while the latter combines this code
with a standard finite volume method which is a simplified version of the
algorithm implemented in \shadowfax{} ascl:1605.003 \citep{2016Vandenbroucke}.
We will summarize the most important physical ingredients of both methods below.

Note that in the current version of the code, we do not include a treatment of
non-ionizing radiation, nor do we take into account dust scattering and the
dynamical effect of radiation pressure on dust. The treatment of these processes
uses algorithms that are very similar to the ones used for photoionisation, and
it is straightforward to extend the code with these processes in the future.

%%%%%%%%%%%%%%%%%%%%%%%%%%%%%%%%%%%%%%%%%%%%%%%%%%%%%%%%%%%%%%%%%%%%%%%%%%%%%%%%
\subsection{Photoionization}

As our initial research focusses on diffuse ionized gas in star forming nebulae,
we only model photoionization of hydrogen and helium self-consistently, for UV
radiation in the energy range $[13.6, 54.4]$~eV, corresponding to the ionization
threshold for hydrogen and the second ionization threshold for helium. As in
\citet{2004Wood}, we do not trace double ionized helium, and we only care about
photons that are energetic enough to ionize hydrogen.

To model the various cooling mechanisms correctly, we also need to know the
ionization structure of a number of coolants, i.e. C$^+$, C$^{++}$, N$^0$,
N$^+$, N$^{++}$, O$^0$, O$^+$, O$^{++}$, Ne$^+$, Ne$^{++}$, S$^+$, S$^{++}$, and
S$^{+++}$ (see \ref{subsection:line_cooling}). These are treated approximately,
where we make the assumption that the number of free electrons released by
photoionization of these elements is neglible compared to the total number of
free electrons, which allows us to use a simplified ionization balance equation.

This approximation only holds in regions that are sufficiently ionized, as the
total number of free electrons is mainly determined by the ionization of
hydrogen and helium for realistic elemental abundances.

Note that \citet{2004Wood} did not include cooling due to S$^{+++}$, and does
not mention the use of carbon cooling rates (although they were used). However,
we found that not including these coolants leads to excessively high
temperatures in the Lexington benchmark tests (see
\ref{subsection:lexington_benchmarks}).

\subsubsection{Monte Carlo technique}

The local photoionization rate depends on various factors, the most important of
which are the position, direction and energy of the incoming ionizing radiation,
and the local ionization state. Due to the strong non-linearity of the
photoionization process, it is impossible to exactly solve for the ionization
balance except for a very limited number of cases, so that approximate
techniques are required.

As a first step, we discretize the density field of interest on a geometrical
grid structure consisting of a (large) number of small cells. Each cell contains
a compact subset of the total physical region of interest and is bounded by a
discrete number of planar faces, which separate it from neighbouring cells. Our
grid can be a regular Cartesian grid consisting of cubical cells, but can also
be a hierarchical adaptive mesh refinement (AMR) grid \citep{2014Saftly}, or an
unstructured grid \citep{2013Camps}.

The radiation field is also split up into a (very large) number of photon
packets with a certain weight, which are sampled using a Monte Carlo technique.
Each packet represents a fraction of the total radiation field, and is emitted
by randomly sampling its properties (origin, travel direction and
energy/wavelength) from underlying distribution functions. The photon packet is
then ray-traced through the density grid by computing geometric path lengths
until a randomly sampled optical depth is reached, or the photon packet leaves
the simulation box. For each discrete cell that lies in the path of the photon
packet, we keep track of the total path length covered by the photon packet
within that cell, to get an approximation to the photoionization integral
$I_{i,X^0 \rightarrow{} X^+}$ for photoionization from ion $X^0$ to level $X^+$
in that cell \citep{2004Wood,2006Osterbrock}:
\begin{eqnarray}
I_{i,X^0 \rightarrow{} X^+} &=& \int_{\nu{}_{X^0 \rightarrow{} X^+}}^\infty{}
\frac{4\pi{}J_i(\nu{}')}{h\nu{}'} \sigma{}_{X^0 \rightarrow{} X^+}(\nu{}')
\text{d} \nu{}' \\
&\approx{}& \frac{Q}{W_tV_i} \sum_j w_j l_{i,j}
\sigma{}_{X^0 \rightarrow{} X^+}(\nu{}_j),
\end{eqnarray}
where $J_i(\nu{})$ is the mean intensity of radiation in the cell as a function
of frequency, $\nu{}_{X^0 \rightarrow{} X^+}$ is the threshold ionization
frequency for ionization of ion $X^0$, $\sigma{}_{X^0 \rightarrow{} X^+}(\nu{})$
is the ionization cross section as a function of photon frequency, $Q$ is the
total luminosity of all UV sources, $V_i$ the volume of the cell, $w_j$ is the
weight of the individual photon packets $j$ that pass through the cell (with
$W_t = \sum_j w_j$ the total weight of all packets), each of which covers a path
length $l_{i,j}$ through the cell, and
$h=6.626\times{}10^{-34}~\text{J}~\text{s}$ is Planck's constant.

Note that the original code of \citet{2004Wood} assumed equal weights for all
photon packets. We generalized this to improve the sampling of low luminosity
external radiation fields.

To account for the diffuse radiation field caused by recombination of ionized
hydrogen and helium, we perform an extra sampling step when a photon packet has
reached the desired optical depth and is still within the simulation box. We
first randomly decide if the photon is absorbed by hydrogen or helium, with the
probability for absorption by hydrogen given by \citep{2004Wood}
\begin{equation}
P_{i,j}(\text{H}^0) = \frac{n_{i,\text{H}^0}\sigma{}_{\text{H}^0 \rightarrow{}
\text{H}^+}(\nu{}_j)}{n_{i,\text{H}^0}\sigma{}_{\text{H}^0 \rightarrow{}
\text{H}^+}(\nu{}_j) + n_{i,\text{He}^0}\sigma{}_{\text{He}^0 \rightarrow{}
\text{He}^+}(\nu{}_j)},
\end{equation}
where $n_{i,\text{H}^0}$ and $n_{i,\text{He}^0}$ are the number densities of
neutral hydrogen and neutral helium in the cell respectively.

Depending on the element that absorbed the photon, there are various reemission
channels, some of which give rise to ionizing UV radiation:
\begin{itemize}
  \item{} hydrogen Lyman continuum radiation,
  \item{} helium Lyman continuum radiation,
  \item{} 19.8~eV radiation from the resonant $2^3S \rightarrow{} 1^1S$
  transition in neutral helium,
  \item{} ionizing radiation for one of the two photons in the helium two photon
  continuum,
  \item{} helium Lyman $\alpha{}$ radiation.
\end{itemize}
Each reemission channel has a specific probability associated with it, which
will depend on the local temperature in the grid cell \citep{2004Wood}. We use
these probabilities to randomly pick a channel. For each channel there is an
associated probability of actually producing an ionizing photon, and an
associated spectrum for the resulting reemitted radiation. We randomly decide if
the photon is reemitted in the ionizing part of the spectrum. If it is, we
sample a new random direction and optical depth for the photon and repeat the
ray-tracing step. If it is not, the photon is assumed to be reemitted as
non-ionizing line continuum which escapes from the system, and it is terminated.

As in \citet{2004Wood}, we assume a medium that is optically thick to Lyman
alpha radiation, so that we do not explicitly ray-trace helium Lyman $\alpha{}$
photons, but assume that they are absorbed on the spot. We need to explicitly
take this into account as an extra term when solving the ionization balance
within the cell:
\begin{multline}
n_{i,\text{H}^0} I_{i,\text{H}^0 \rightarrow{} \text{H}^+} +
P_i(\text{H}_\text{OTS}) n_{i,e} n_{i,\text{He}^+} \alpha{}_{\text{He}^0,2^1P}
\\ = n_{i,e} n_{i,\text{H}^+} \alpha{}_{\text{H}^+ \rightarrow{}
\text{H}^0}(T_i),
\label{eq:ionization_balance_hydrogen}
\end{multline}
where $\alpha{}_{\text{He}^0,2^1P}$ is the recombination rate of the $2^1P$
level of neutral helium, $\alpha{}_{\text{H}^+ \rightarrow{} \text{H}^0}(T)$
is the recombination rate from ionized to neutral hydrogen as a function of
temperature, and $T_i$ is the temperature in the cell.

$n_{i,e}$ is the electron density in the cell, which is approximately given by
\begin{equation}
n_{i,e} = n_{i,\text{H}^+} + n_{i,\text{He}^+}
\label{eq:charge_conservation}
\end{equation}
if we neglect the free electrons released by ionized metals.

$P_i(\text{H}_\text{OTS})$ is the probability of on the spot absorption of
helium Lyman $\alpha{}$ radiation, which is approximately given by
\citep{2004Wood}
\begin{equation}
P_i(\text{H}_\text{OTS}) = \left( 1 +
\frac{0.77}{\sqrt{\frac{T_i}{10^4~\text{K}}}}
\frac{f_{i,\text{He}^0}}{f_{i,\text{H}^0}} \right)^{-1},
\end{equation}
with $f_{i,\text{H}^0}$ and $f_{i,\text{He}^0}$ the neutral fractions of
hydrogen and helium in the cell, defined as
\begin{equation}
f_{i,X^0} = \frac{n_{i,X^0}}{n_{i,X^0} + n_{i,X^+}}.
\end{equation}

For helium the ionization balance is simply given by
\begin{equation}
n_{i,\text{He}^0} I_{i,\text{He}^0 \rightarrow{} \text{He}^+} = n_{i,e}
n_{i,\text{He}^+} \alpha{}_{\text{He}^+ \rightarrow{} \text{He}^0}(T_i).
\label{eq:ionization_balance_helium}
\end{equation}

Equations (\ref{eq:ionization_balance_hydrogen}) and
(\ref{eq:ionization_balance_helium}) are solved simultaneously for a given input
temperature $T_i$. Once the ionization state of hydrogen and helium is known, we
can compute the density of free electrons using equation
(\ref{eq:charge_conservation}). With these densities, we can solve for the
ionization state of the metals, for which the ionization balance is generally
given by
\begin{equation}
n_{i,X^0} I_{\text{tot},i,X^0 \rightarrow{} X^+}(T_i) = n_{i,X^+}
\alpha{}_{\text{tot},i,X^+ \rightarrow{} X^0}(T_i),
\end{equation}
with $I_{\text{tot},i,X^0 \rightarrow{} X^+}(T_i)$ and
$\alpha{}_{\text{tot},i,X^+ \rightarrow{} X^0}(T_i)$ the total ionization and
recombination rate in the cell.

The ionization rate is generally given by
\begin{align}
I_{\text{tot},i,X^0 \rightarrow{} X^+}(T_i) &= I_{i,X^0 \rightarrow{} X^+} \\
& + n_{i,\text{H}^+} I_{C,\text{H},X^0 \rightarrow{} X^+}(T_i) \\
& + n_{i,\text{He}^+} I_{C,\text{He},X^0 \rightarrow{} X^+} (T_i),
\end{align}
where $I_{C,\text{H},X^0 \rightarrow{} X^+}(T)$ and $I_{C,\text{He},X^0
\rightarrow{} X^+}(T)$ are the ionization rates for ion $X^0$ through a charge
transfer reaction with ionized hydrogen or helium respectively, as a function of
temperature.

Likewise, the total recombination rate is given by
\begin{align}
\alpha{}_{\text{tot},i,X^+ \rightarrow{} X^0}(T_i) &= n_{i,e} \alpha{}_{X^+
\rightarrow{} X}(T_) \\
& + n_{i,\text{H}^0} \alpha{}_{C,\text{H},X^+ \rightarrow{} X}(T_i) \\
& + n_{i,\text{H}^0} \alpha{}_{C,\text{He},X^+ \rightarrow{} X}(T_i),
\end{align}
with $\alpha{}_{C,\text{H},X^+ \rightarrow{} X}(T)$ and
$\alpha{}_{C,\text{He},X^+ \rightarrow{} X}(T)$ charge transfer recombination
rates.

We will generally not include charge transfer ionization rates for reactions
involving ionized helium, and only include charge transfer ionization rates for
hydrogen and charge transfer recombination rates for some of the metals.

\subsubsection{Data}

\paragraph{Spectra}

We support a number of input spectra for the ionizing radiation, ranging from
single frequency spectra that are used for benchmark tests, over black body
spectra, to realistic stellar atmosphere spectra from the models of
\citet{2003Hoffmann}. For complex spectra, we precompute the cumulative number
distribution function of ionizing photons for a discrete number of frequencies,
and then use linear interpolation on this table to sample random frequencies at
runtime.

We also support external radiation fields, like a redshift-dependent cosmic UV
background field which can be used to model the ISM of high redshift galaxies.
For this, we use the spectra of \citet{2009FaucherGiguere}, as downloaded from
their website.

\paragraph{Ionization cross sections}

We use fits to the photoionization cross sections of hydrogen, helium, and the
various coolants from \citet{1996Verner_cross}. These fits smooth out over
resonances, but have the advantage that they are relatively cheap to compute at
runtime.

\paragraph{Recombination rates}

We use radiative recombination rate fits of \citet{1996Verner_rec}. For the
coolants, these are supplemented with dielectronic recombination rate fits of
\citet{1983Nussbaumer}, \citet{1987Nussbaumer}, \citet{1998Mazzotta}, and
\citet{2012AbdelNaby}.

\paragraph{Charge transfer rates}

We use the charge transfer ionization and recombination rates for hydrogen from
\citet{1996Kingdon}, and the helium charge transfer recombination rates of
\citet{1985Arnaud}.

\paragraph{Alternative data}

For some of the benchmark tests described in section \ref{section:benchmarks} we
need simplified values for the photoionization cross sections and radiative
recombination rates. To this end, we made sure that the rates can be easily
changed in our implementation, as part of the modular design of our code (see
\ref{subsubsection:object_oriented_design}).

%%%%%%%%%%%%%%%%%%%%%%%%%%%%%%%%%%%%%%%%%%%%%%%%%%%%%%%%%%%%%%%%%%%%%%%%%%%%%%%%
\subsection{Heating and cooling}

When a photon packet is absorbed by hydrogen or helium, an amount of energy
equal to the excess w.r.t. the ionization energy of that element is converted
into heating of the local gas. The total integrated heating $H_{i,X^0
\rightarrow{} X^+}$ for a grid cell is given by \citep{2004Wood}
\begin{multline}
H_{i,X^0 \rightarrow{} X^+} = \int_{\nu{}_{X^0 \rightarrow{} X^+}}^\infty{}
\frac{4\pi{}J_i(\nu')}{h\nu'} \sigma{}_{X^0 \rightarrow{} X^+}(\nu{}') \\
h \left( \nu{}' - \nu{}_{X^0 \rightarrow{} X^+} \right) \text{d}\nu{}'.
\end{multline}
and is hence very similar to the ionization rate estimator, and can be treated
in the same way during our Monte Carlo photon propagation scheme. As for the
ionizing luminosity, the optical depth for a cell depends on the temperature and
ionization state of the gas in that cell, so that a self-consistent solution can
only be obtained with an iterative scheme.

For cooling by recombination of hydrogen and helium, we use the recombination
cooling rates of \citet{1981Black}. The cooling rate due to bremsstrahlung from
free electrons is given by \citet{2006Osterbrock}, where we use the fits to the
mean Gaunt factor given by \citet{1996Katz}.

%%%%%%%%%%%%%%%%%%%%%%%%%%%%%%%%%%%%%%%%%%%%%%%%%%%%%%%%%%%%%%%%%%%%%%%%%%%%%%%%
\subsection{Line cooling}
\label{subsection:line_cooling}

Despite the low abundances of metals such as C, N, O, Ne, and S in star forming
nebulae, line emission from these elements contributes signicantly to the
radiative cooling, as their low-lying energy levels can be easily excited
through collisions with free electrons \citep{2006Osterbrock}. To model this
process, we need to keep track of the ionization state of these coolants, and
model their line emission. The details of this treatment are explained below.

\subsubsection{Mechanism and data}

In general, the level population $x_{X,i} = \frac{n_{X,i}}{n_{X}}$ of the $i$th
energy level of an ion $X$ with density $n_{X}$ is the solution of
\citep{2006Osterbrock}
\begin{multline}
\sum_{j > i}x_{X,j}n_e q_{X,ji} + \sum_{j > i} x_{X,j} A_{X,ji} \\
= \sum_{j < i} x_{X,i} n_e q_{X,ij} + \sum_{j < i} x_{X,i} A_{X,ij},
\label{equation:level_population}
\end{multline}
where $q_{X,ij}$ is the collisional excitation or deexcitation rate from level
$j$ to level $i$, $A_{X,ij}$ is the radiative deexcitation rate from level $i$
to level $j$, and $n_e$ is the electron density.

The collisional deexcitation rate $q_{X,ji}$ is given by
\begin{equation}
q_{X,ji} = \frac{h^2}{\sqrt{k}\left(2\pi{}m_e\right)^{\frac{3}{2}}}
\frac{\Upsilon{}_X(i, j)}{\omega{}_{X,j}},
\end{equation}
with $k=1.38 \times{} 10^{-23}~\text{J}~\text{K}^{-1}$ Boltzmann's constant,
$m_e = 9.109 \times{} 10^{-31}~\text{kg}$ the mass of an electron,
$\Upsilon{}_X(i,j)$ the velocity-averaged collision strength, and
$\omega{}_{X,j}$ the statistical weight of the $j$th level of ion $X$. It is
linked to the collisional excitation rate $q_{X,ij}$ through the relation
\begin{equation}
q_{X,ij} = \frac{\omega{}_{X,j}}{\omega{}_{X,i}} q_{X,ji}
\exp\left(\frac{-\chi{}_{X,ij}}{kT}\right),
\end{equation}
with $\chi{}_{X,ij}$ the energy difference between level $i$ and level $j$.

Velocity-averaged collision strengths, radiative recombination rates, energy
differences, and statistical weights can be measured experimentally or modelled
quantum mechanically. We use data from a large number of sources, as detailed in
\tableref{table:line_cooling_data}.

\begin{table}
\centering{}
\caption{Data values used for the line cooling computations of the various
coolants. The symbols in the table are explained below the table.
\label{table:line_cooling_data}}

\vspace{12pt}
\begin{tabular}{p{1.5cm} p{1.5cm} p{1.5cm} p{1.5cm}}
\hline
ion & A, B & C & D \\
\hline
C$^+$ & FF04 & FF04 & T08 \\
C$^{++}$ & FF04 & FF04 & B85 \\
\hline
N & FF04 & FF04 & T00 \\
N$^+$ & G97 & G97 & L94 \\
N$^{++}$ & B92 & G98 & B92 \\
\hline
O & G97 & G97 & B88, Z03 \\
O$^+$ & FF04 & FF04 & K09 \\
O$^{++}$ & G97 & G97 & L94 \\
\hline
Ne$^+$ & S94 & K86 & G01 \\
Ne$^{++}$ & G97 & G97 & B94 \\
\hline
S$^+$ & T10 & T10 & T10 \\
S$^{++}$ & M82 & M82 & H12 \\
S$^{+++}$ & M90 & P95 & S99 \\
\hline
\end{tabular}

\vspace{12pt}
\begin{tabular}{p{1.0cm} p{5.5cm}}
A & energy levels \\
B & statistical weights \\
C & radiative recombination rates \\
D & velocity-averaged collision strengths \\
\end{tabular}

\vspace{12pt}
\begin{tabular}{p{1.0cm} p{5.5cm}}
B85 & \citet{1985Berrington} \\
B88 & \citet{1988Berrington} \\
B92 & \citet{1992Blum} \\
B94 & \citet{1994Butler} \\
FF04 & \citet{2004FroeseFischer} \\
G97 & \citet{1997Galavis} \\
G98 & \citet{1998Galavis} \\
G01 & \citet{2001Griffin} \\
H12 & \citet{2012Hudson} \\
K86 & \citet{1986Kaufman} \\
K09 & \citet{2009Kisielius} \\
L94 & \citet{1994Lennon} \\
M82 & \citet{1982Mendoza} \\
M90 & \citet{1990Martin} \\
P95 & \citet{1995Pradhan} \\
S94 & \citet{1994Saraph} \\
S99 & \citet{1999Saraph} \\
T00 & \citet{2000Tayal} \\
T08 & \citet{2008Tayal} \\
T10 & \citet{2010Tayal} \\
Z03 & \citet{2003Zatsarinny} \\
\end{tabular}
\end{table}

The specific ions we use can be classified into two categories: ions with two
low lying energy levels (N$^{++}$ and Ne$^+$), and ions with five low lying
levels (C$^+$, C$^{++}$, N, N$^+$, O, O$^+$, O$^{++}$, Ne$^{++}$, S$^+$ and
S$^{++}$). For the former, we solve the system of two equations
(\eqref{equation:level_population} for $i = 2$, and $x_{X,1} + x_{X,2} = 1$)
analytically, while for the latter we need to solve the full set of five coupled
linear equations.

\subsubsection{New data fits}

The velocity-averaged collision strengths used above vary with temperature. To
account for this fact, we fitted a general curve of the form
\begin{multline}
\Upsilon{}_X(i,j,T) = T^{a_{ij} + 1} \left( b_{ij} + \frac{c_{ij}}{T} + d_{ij}
\log(T) + \right. \\
\left. e_{ij} T \left( 1 + (f_{ij} - 1) T^{g_{ij}} \right) \right)
\end{multline}
to all data, where $T$ is the temperature (in K) and $a_{ij}$, $b_{ij}$,
$c_{ij}$, $d_{ij}$, $e_{ij}$, $f_{ij}$ and $g_{ij}$ are fitting parameters. The
general form of this curve was inspired by \citep{1992Burgess}. It is worth
noticing that there are no reliable data for some of the fine structure levels
of O$^+$ above 10,000~K, so that we needed to extrapolate the low temperature
data. The same is true for N$^{++}$ above 40,000~K. For C$^{++}$ and Ne$^{+}$
the data points are sparse, so the fits are less reliable. For S$^+$ and
S$^{++}$ the values of $e$, $f$ and $g$ were kept zero, as this provided a
better fit than when they were allowed to vary.

The values of the fitting parameters are listed in
\multitableref{table:collision_strength_fits_carbon}
{table:collision_strength_fits_sulphur}, the corresponding fits and relative
differences between fit and data are shown in
\multifigref{figure:CII_fit}{figure:SIV_fit}.

%%%%%%%%%%%%%%%%%%%%%%%%%%%%%%%%%%%%%%%%%%%%%%%%%%%%%%%%%%%%%%%%%%%%%%%%%%%%%%%%
\subsection{Hydrodynamics}

Hydrodynamical integration is performed using a finite volume method
\citep{2009Toro} on a generally unstructured, (co-)moving mesh
\citep{2010Springel}. The Euler equations of hydrodynamics in conservative form
\begin{equation}
\frac{\partial{}U}{\partial{}t} + \vec{\nabla{}}.\vec{F}\left( U \right) = 0,
\label{equation:conservation_law}
\end{equation}
with
\begin{equation}
U = \begin{pmatrix}
\rho{} \\
\vec{v} \\
P
\end{pmatrix},
\end{equation}
\begin{equation}
\vec{F}(U) = \begin{pmatrix}
\rho{} \vec{v} \\
\rho{} \vec{v} \vec{v} + P \vec{\vec{1}} \\
\rho{} \left( u + \frac{1}{2}\left|\vec{v}\right|^2 \right) + P\vec{v}
\end{pmatrix},\label{equation:fluxes}
\end{equation}
and an adiabatic equation of state $P = \left(\gamma{}-1\right)\rho{}u$, where
$\rho{}$ is the mass density, $\vec{v}$ the flow velocity, $P$ the pressure, and
$u$ the thermal energy per unit mass, are discretized on an unstructured Voronoi
mesh. Integration of the primitive variables over the volume $V_i$ of each cell
leads to a set of conserved variables (mass $m_i$, momentum $\vec{p}_i$, and
total energy $E_i$) for each cell. Integrating the conservation law
\eqref{equation:conservation_law} over the volume of the cell allows us to
reduce the time integration of these conserved quantities as a flux exchange
between the cell and its neighbouring cells:
\begin{equation}
\frac{\text{d}Q_i}{\text{d}t} = -\sum_j \vec{F}_{ij} . \vec{A}_{ij},
\label{equation:finite_volume}
\end{equation}
where $\vec{A}_{ij}$ represents the oriented surface area of the geometrical
face between cell $i$ and cell $j$, $\vec{F}_{ij}$ is an appropriate estimate of
the fluxes \eqref{equation:fluxes} at a representative location on the face, and
$Q_i$ is the vector of conserved variables for that cell.

To obtain appropriate fluxes, we use the values on both sides of the face as
input for an exact Riemann solver, which gives the exact physical solution for
the two state problem defined by the variables at both sides up to a desired
precision. By sampling this analytic solution we obtain new values for the
primitive variables that can be used to compute fluxes to be used in
\eqref{equation:finite_volume}.

Note that our current implementation is only first order in space and time. It
is however very straightforward to extend this to higher order by the
introduction of appropriate spatial gradients, see e.g.
\citet{2016Vandenbroucke}; this will be implemented in future versions of the
code. Our current implementation also does not yet include external forces or
self-gravity.

\subsubsection{Moving mesh scheme}

For the specific case of an unstructured Voronoi grid, we can make the method
Lagrangian by allowing the generators of the Voronoi grid to move in between
time steps of the integration scheme. To account for this movement, we add
correction terms to the flux expressions \eqref{equation:fluxes}, as detailed in
\citep{2010Springel, 2015Hopkins, 2016Vandenbroucke}. Note that the
hydrodynamical integration is completely independent of the movement of the
generators. If the generators do not move, the method is Eulerian. If on the
other hand the generator movement is set to the local fluid velocity in the
corresponding cell, the method is fully Lagrangian. In the latter case, we can
solve the Riemann problem across the cell faces in the rest frame of the faces,
which leads to better accuracy in the presence of large bulk velocities, and
allows us to use a larger integration time step than would be used in an
equivalent Eulerian scheme, as the time step only depends on the relative
velocity of the fluid w.r.t. the grid.

\subsubsection{Radiation hydrodynamics}

To couple the radiation to the hydrodynamics, an operator splitting method is
used, whereby the photoionization heating term is added after each step of the
hydrodynamics scheme, assuming photoionization equilibrium. The current values
of the density in each cell are converted into number densities that are then
used as input for the photoionization code. The photoionization code computes a
self-consistent ionization structure for each cell, which is then used to decide
what the temperature of the cell should be. The number of iterations and number
of photon packets used in this step is a simulation parameter. The resulting
temperature is compared with the actual hydrodynamical temperature of the cell,
and the energy difference is added to the cell as an energy source term.

In our current implementation, we do not subtract energy for cells that were
ionized and become neutral again, since this involves a careful treatment of the
time step to prevent negative energies. We use an almost isothermal equation of
state with an adiabatic index $\gamma{}=1.0001$, and assume a two-temperature
medium, whereby the temperature in the neutral phase is assumed to be $T_n =
100~{\rm{}K}$, while the temperature in the ionized medium is assumed to be $T_i
= 10,000~{\rm{}K}$. For a cell with hydrogen neutral fraction $x_{\rm{}H}$, the
assumed average temperature is then
\begin{equation}
T = x_{\rm{}H} T_n + \left( 1 - x_{\rm{}H} \right) T_i.
\end{equation}
This approach is sufficient to reproduce a basic benchmark expansion test (see
\ref{subsection:STARBENCH}). We could of course also get the photoionization
heating directly from the photoionization step; this more general approach will
be subject of future work.

Note that it is possible to couple \cmacionize{} as an external library to other
hydrodynamics codes using a very similar approach, see
\ref{subsection:library_exposure}.

%%%%%%%%%%%%%%%%%%%%%%%%%%%%%%%%%%%%%%%%%%%%%%%%%%%%%%%%%%%%%%%%%%%%%%%%%%%%%%%%
\section{Code}
\label{section:code}
%%%%%%%%%%%%%%%%%%%%%%%%%%%%%%%%%%%%%%%%%%%%%%%%%%%%%%%%%%%%%%%%%%%%%%%%%%%%%%%%

The most important difference between \cmacionize{} and the original
photoionization code of \citet{2004Wood} is a complete redesign of the structure
of the code (accompanied by a migration from Fortran to C++11), including a full
in-line documentation of the code using
Doxygen\footnote{\url{http://www.doxygen.org}} (the documentation for the latest
stable version of the code can be found on a dedicated
webpage\footnote{\url{http://www-star.st-and.ac.uk/~bv7/CMacIonize_documentation/}}).
Below we outline the main design considerations and detail how they are
implemented in the code.

%%%%%%%%%%%%%%%%%%%%%%%%%%%%%%%%%%%%%%%%%%%%%%%%%%%%%%%%%%%%%%%%%%%%%%%%%%%%%%%%
\subsection{Design considerations}

\subsubsection{User friendliness}

The photoionization part of \cmacionize{} is primarily focused on
post-processing output from other simulation codes as part of a simulation
analysis work flow. This means that the code will be used by researchers that
are not necessarily very familiar with the details of the code, but that still
want to produce reproducible science products. To accommodate this, we aim to
minimize the learning curve for using \cmacionize{}. In cases where writing
additional code is inevitable (like for example when reading a new type of input
file), we want to limit the effort necessary to accomplish this: the new code
should be limited to a single function or C++ class, and the user should be able
to write this code without worrying about the details of the photoionization
algorithm.

\subsubsection{Reproducibility}

A properly designed computer algorithm should be deterministic, so that running
the same simulation twice with the same input and the same version of a code
should produce exactly the same result, independent of the hardware
architecture. Parallelization and system specific optimizations might cause tiny
changes in round off that could cause minuscule changes in result between
different runs (especially in Monte Carlo algorithms), but even then a
simulation code should be very close to a unique mapping from input data to an
output solution. Reproducibility is hence an inherent feature of computer
simulations.

However, keeping track of input parameters and code versions can be tedious,
especially when simulations are combined with the design of improved algorithms
and code changes are made. For this reason, we aim to provide a robust system to
log parameters and code versions, so that all published \cmacionize{} results
should be perfectly reproducible.

\subsubsection{Modularity}

Complex algorithms combine a large number of components that each have their
specific complexities. However, most of these components are predominantly
independent of each other, and only interact with the other components through
narrowly defined interfaces. Isolating complex components into separate entities
or modules increases the readability of a code, and makes the code more robust
if combined with an appropriate unit testing strategy. We will therefore aim to
produce a modular code, whereby separate components are identified and isolated
into separate code entities.

\subsubsection{Scalability}

Modern computing architecture is highly parallel, with the computing power of a
typical high performance computer spread out over a large number of separate
computing units or \emph{nodes} that are interconnected through a high speed
network, and with each of these nodes in turn consisting of up to 128 separate
CPU \emph{cores} that share a single memory space. In order to efficiently use
these machines, it is crucial that an algorithm is designed with a parallel
mindset. We cannot think of the algorithm as a serial list of instructions that
are executed one by one, but instead need to think of the \emph{tasks} that need
to be performed by the algorithm, the \emph{data} that is needed to perform
these tasks (and that might be shared with other tasks), and the
\emph{dependencies} that govern which tasks can be executed in parallel and
which tasks are mutually exclusive due to conflicts.

In the current version of \cmacionize{}, we aim to provide a reasonable scaling
by distributing computations across multiple cores on a single node, and across
multiple nodes. Our current parallelization strategy does not address the need
to distribute data across multiple nodes in order to efficiently use the
available memory. This will be addressed in future versions of the code.

%%%%%%%%%%%%%%%%%%%%%%%%%%%%%%%%%%%%%%%%%%%%%%%%%%%%%%%%%%%%%%%%%%%%%%%%%%%%%%%%
\subsection{Design implementation}

\subsubsection{User interface}
\label{subsubsection:user_interface}

For standard users that do not plan to add additional code, the interaction with
\cmacionize{} is limited to
\begin{itemize}
  \item{} calling the command line program \code{CMacIonize}, and
  \item{} writing a parameter file that contains the parameters for the
  simulation.
\end{itemize}
The command line program has a very limited set of options that control the
number of shared memory parallel threads used to run, and the mode in which to
run (photoionization only or RHD). The only other parameter to the program is
the name of the parameter file. An example command line call to \cmacionize{}
could like this:
\begin{verbatim}
./CMacIonize --params parameterfile.param \
             --threads 8
\end{verbatim}
This will run \cmacionize{} in the standard photoionization mode using 8 shared
memory parallel threads, and using the parameters provided in the file
\code{parameterfile.param}.

The parameter file contains all information needed to set up and run the
simulation, and maps to the underlying modular structure of the code (all
parameters are linked to a specific C++ class). It is a simple text file in YAML
format\footnote{\url{http://yaml.org/}}, and a very basic example could look
like this:
\begin{verbatim}
SimulationBox:
  anchor: [-5. pc, -5. pc, -5. pc]
  sides: [10. pc, 10. pc, 10. pc]
  periodicity: [false, false, false]

DensityGrid:
  type: Cartesian
  number of cells: [64, 64, 64]

DensityFunction:
  type: Homogeneous
  density: 100. cm^-3
  temperature: 8000. K

PhotonSourceDistribution:
  type: SingleStar
  position: [0. pc, 0. pc, 0. pc]
  luminosity: 4.26e49 s^-1

PhotonSourceSpectrum:
  type: Monochromatic
  frequency: 13.6 eV

IonizationSimulation:
  number of photons: 1e6
  number of iterations: 20
\end{verbatim}
This parameter file sets up a Str\"{o}mgren test in a box of $10~
{\rm{}pc}\times{}10~{\rm{} pc}\times{}10~{\rm{} pc}$ containing gas with a
density of $100~{\rm{}cm}^{-3}$ at a temperature of 8,000~K, with a star at the
centre with a total luminosity of $4.26\times{}10^{49}~{\rm{}s}^{-1}$ and a
monochromatic spectrum with a frequency equivalent to a photon of
$13.6~{\rm{}eV}$. The simulation uses a Cartesian grid of
$64\times{}64\times{}64$ cells, and uses 20 iterations with $10^6$ photon
packets for each iteration (see \ref{subsection:stromgren_sphere} for details of
this test).

The example above illustrates how easy it is to read and understand a parameter
file. It also illustrates another key feature of the code: the use of units.
Internally, we consistently use SI units throughout the code to avoid any
confusion about units. However, we also require the user to specify units for
all physical quantities that are used as an input, so that the user does not
need to worry about unit conversions at all. We support a variety of different
units, including complex unit conversions (e.g. photon energy in eV to photon
frequency in Hz), and adding new units is very straightforward.

Apart from supporting units, the parameter file also supports various number
formats and 3D vectors.

When the program is started, the parameter file is parsed and translated into a
corresponding simulation structure. Before the actual simulation starts, the
actually used parameters are written to a reference file. Most parameters have
default values and need not be specified in the parameter file; when written to
the reference file, the default values will be displayed, and the file will
clearly state that the default value was used. If a parameter is not used, the
reference file will mention this as well. For parameters that have units, the
reference file will contain the value in SI units, as well as the original
value.

If no parameter file is given, default values will be used for all parameters
(that correspond to a low resolution version of the Str\"{o}mgren benchmark
test, see \ref{subsection:stromgren_sphere}). In this case it is possible to use
the reference file as a first guess for sensible parameter values, and iterate
on it to construct an actual useful parameter file. All available parameters are
also extensively documented in the Doxygen documentation of the corresponding
classes.

\subsubsection{Reproducibility}
\label{subsubsection:reproducibility}

In order to guarantee reproducibility, we use a strategy that consists of three
pillars:
\begin{enumerate}
  \item{} version control as a way to uniquely identify a specific code version,
  \item{} parameter logging in output files as a way to keep track of used
  parameters, and
  \item{} unit testing to guarantee the same results across code versions.
\end{enumerate}

\paragraph{Version control}

The code is stored in a public online Git
repository\footnote{\url{https://github.com/bwvdnbro/CMacIonize}} to make it
easier to keep track of the code history, and to facilitate collaboration on the
code. Git keeps track of the changes that are made to the code in between so
called \emph{commits}, i.e. logged checkpoints of the code status. Each commit
has an associated key that uniquely identifies it, and it is always possible to
return to a specific version of the code using the appropriate commit key.

Moreover, Git also provides a command line tool called \code{git describe} that
can be used to check the current version of the local copy of the repository
that a user is using, and that checks if the repository is \emph{dirty}, i.e.
contains uncommitted code changes.

We have incorporated \code{git describe} into our code configuration chain, so
that the compiled code knows (a) what the unique commit key of the current
version is, and (b) if the current code version is exactly equal to that code
version, or contains uncommitted changes. If the code contains uncommitted
changes, it will refuse to run any simulation, so that the user is forced to
commit changes and make sure the code is identifiable before running scientific
simulations.

\paragraph{Logging}

The output of any simulation consists of snapshots, i.e. dumps of specific
quantities at some time during the simulation. \cmacionize{} supports various
types of snapshot files, the default being the same HDF5 format that is also
used by \textsc{Gadget2} ascl:0003.001,
\textsc{SWIFT}\footnote{\url{https://gitlab.cosma.dur.ac.uk/swift/swiftsim}},
and \shadowfax{}.

In order to exactly reproduce a snapshot file, we need to know
\begin{itemize}
  \item{} what version of the code was used to generate it, and
  \item{} which parameters were used to run the code.
\end{itemize}
The former can be easily realised by storing the unique commit key for the
current code version in the snapshot files. To guarantee the latter, we also
store all parameters in the snapshot files, as they were used. This corresponds
to the values that are part of the reference parameter file (see
\ref{subsubsection:user_interface}). By also storing the parameters for which
default values were used, we guarantee reproducibility across different code
versions, if at some point the default value for a parameter were to change.
The public version of the code contains a Python script that can extract the
parameter file that was used from a snapshot file in the default HDF5 format.

For simulations that use input from external files, as e.g. simulations that
post-process the density field from another simulation, we also need these
external files in order to reproduce the results. Since these files can be quite
large, storing them as part of the snapshot files is not an option. In this
case, we rely on the user using a convenient method of keeping track of those
files to guarantee reproducibility.

Apart from the code version and the parameter values we also log configuration
flags and system specific information. This is not strictly necessary in order
to reproduce simulation results, but might nonetheless be helpful e.g. during
debugging.

\paragraph{Unit testing}

One of the key issues when developing a simulation code is making sure that the
results are scientifically accurate, and that they stay accurate throughout the
further development of the code. Unit testing is a very powerful tool to achieve
this, especially when combined with a good modular design (see
\ref{subsubsection:object_oriented_design}).

A unit test is a small independent program that calls a small part of the code
with known input values and checks its output against a known solution. If the
output matches the expected result, the test passes and we know that part of the
code behaves as expected.

When properly designed, a unit test covers all possible paths through the code
that is being tested, e.g. if the code contains conditions that check for
strange input values, the test will call the code with strange input values and
check that these are handled correctly.

Unit tests were an integral part of the early development process of
\cmacionize{}, as all new code was tested against reference values of the old
code of \citet{2004Wood}. During the addition of new features that were not part
of the old code, we still tried to start from the unit test as much as possible,
which meant first thinking about what the expected behaviour of a code component
(function or class) should be, before actually implementing it. The overhead
this implies is quickly recovered by the ease with which we can locate bugs in
new code.

The current version of the code contains almost 70 separate unit tests, which
are managed as part of our code configuration and run using CTest, the
CMake\footnote{\url{https://cmake.org/}} unit testing framework. Depending on
the hardware and system configuration, the tests take less than a minute to a
few minutes to complete, and can be run as part of the standard compilation
process.

When new code is added to the stable version on the Git repository, the new code
is automatically compiled with a number of different compilers on different
systems using the continuous integration environment Travis
CI\footnote{\url{https://travis-ci.org}}, and the unit tests are run. Code is
only allowed to be merged into the stable repository if it passes all the tests.
This way we ensure that new code never breaks or alters old functionality,
unless this is done on purpose (in which case the corresponding unit test needs
to be modified).

\subsubsection{Object oriented design}
\label{subsubsection:object_oriented_design}

To implement modularity in our code design, we use C++ objects as the building
blocks of the code. An object has a limited number of responsibilities, and is
as unaware as possible of the rest of the code (unless the interaction of
various objects is the responsibility of the class). Most objects are covered by
a corresponding unit test (see \ref{subsubsection:reproducibility}), although
the unit tests for some classes are grouped together if this makes more sense.

We use a number of different design patterns \citep{1995Gamma}. Basic simulation
components like the density grid and the source distribution use
\emph{inheritance} combined with a \emph{factory class} to provide different
interchangeable implementations (e.g. the density grid can be a regular
Cartesian grid, an AMR mesh, or an unstructured Voronoi grid). The density grid
itself makes extensive use of \emph{iterators} to provide grid type unaware
access to cells, while most grid computations are performed using
\emph{visitors} that perform a single task for each cell of the grid.

The object oriented design is tightly interwoven with the parameter file used as
the user interface (see \ref{subsubsection:user_interface}), with objects
mapping to specific blocks in the parameter file (and the \verb+type+ keyword
always referring to a factory class that provides multiple implementations of a
general interface). Most objects have a so-called parameter file constructor,
which can create an object instance based on the parameter values given in the
parameter file, with parameters mapping directly to object properties.

\subsubsection{Task based design}
\label{subsubsection:task_based_design}

\begin{figure*}
\centering{}
\includegraphics{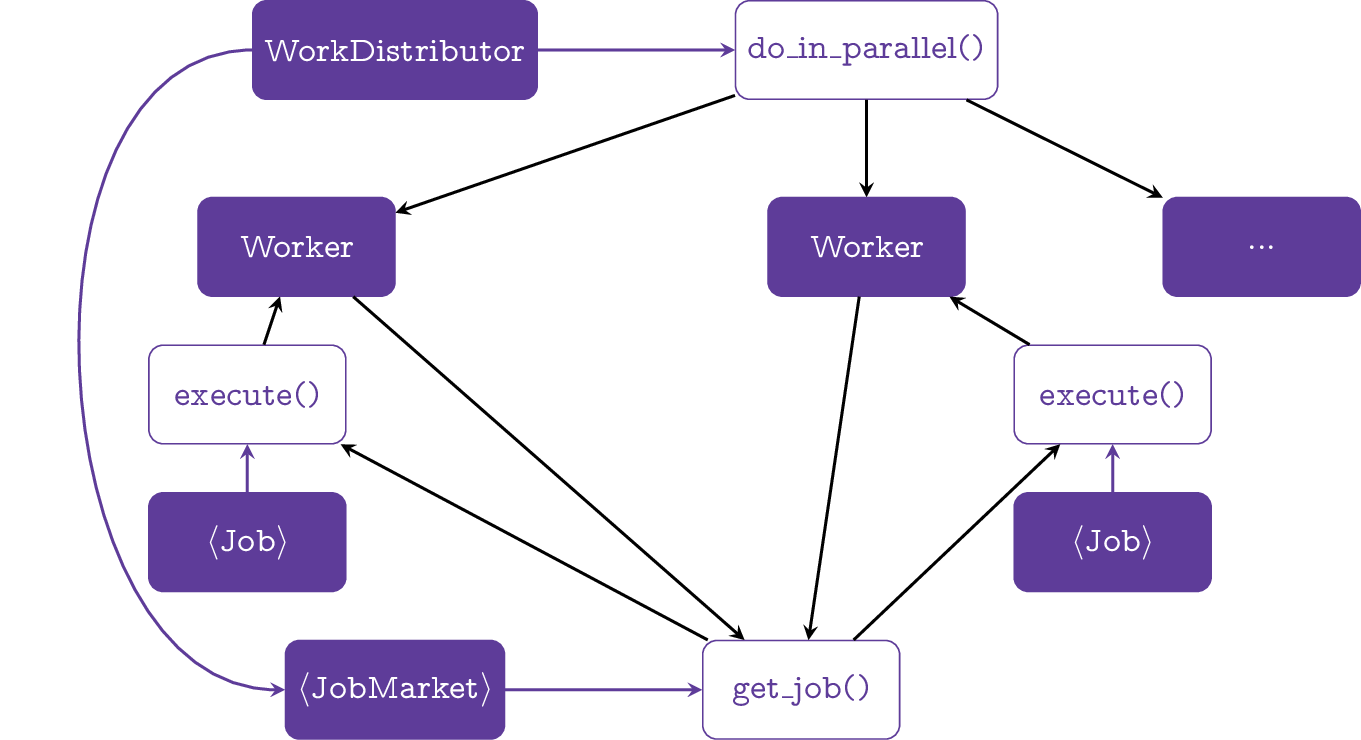}
\caption{Schematic representation of a typical shared memory parallel run. A
\code{WorkDistributor} spawns a number of parallel \code{Worker} objects that
get tasks from a template \code{JobMarket} object that acts as a taskpool. The
different threads keep drawing tasks from the pool until all tasks have been
executed.
\label{figure:workdistributor}}
\end{figure*}

To provide inherent parallelism in our code, we think of the actions that need
to be performed by the algorithm in terms of small \emph{tasks}, that perform a
limited set of actions on a small part of the computational domain. By limiting
the amount of work done by a single task, we can improve the load balancing
between different parallel threads significantly, while limiting the fraction of
the computational domain that is affected minimizes conflicts.

The photon traversal algorithm e.g. can be done independently for small batches
of photons. Each part of the path of a photon is only a single cell, so that we
only need to worry about two threads accessing the same cell at the same time to
prevent conflicts. Similarly, the ionization balance computation for different
cells can be done completely independently, and so can the temperature balance
computation or even the cell initialization.

In practice, our task based design is implemented using a number of interacting
classes called \code{Worker}, \code{Job}, and \code{JobMarket}. \code{Job} and
\code{JobMarket} make use of compile time polymorphism and are template
interfaces, meaning that these classes do not actually exist, but are abstract
interfaces that define common functionality for classes that can be used as C++
template arguments for other classes. This offers the same flexibility as run
time polymorphism, but without the computational overhead. A \code{Worker} is
our abstract representation of a thread, while a \code{Job} is the abstract
representation of a task that needs to be performed. The \code{JobMarket} is
responsible for spawning \code{Job} instances. The \code{Worker} instances in
turn are spawned by a \code{WorkDistributor}, which is the only class that needs
to know about the underlying parallel environment that is used. Our current
implementation only supports OpenMP, but it would be straightforward to replace
this by e.g. a POSIX threads or Intel Threading Building Blocks implementation.

The general workflow for a shared memory parallel run is illustrated in
\figureref{figure:workdistributor}. When a parallel part of the execution is
started, a corresponding \code{JobMarket} implementation is created and passed
on as a template argument to the \code{WorkDistributor}. The
\code{WorkDistributor} then generates a number of \code{Worker} instances that
are run in parallel, and that call the \code{get\_job()} function of the
\code{JobMarket} instance to get actual \code{Job} instances that need to be
executed. The worker then calls the \code{execute()} method of the \code{Job}
instance to perform the task at hand. When the task is finished, the
\code{Worker} goes back to the \code{JobMarket} to get the next \code{Job},
until no more tasks are available.

\begin{figure*}
\centering{}
\includegraphics{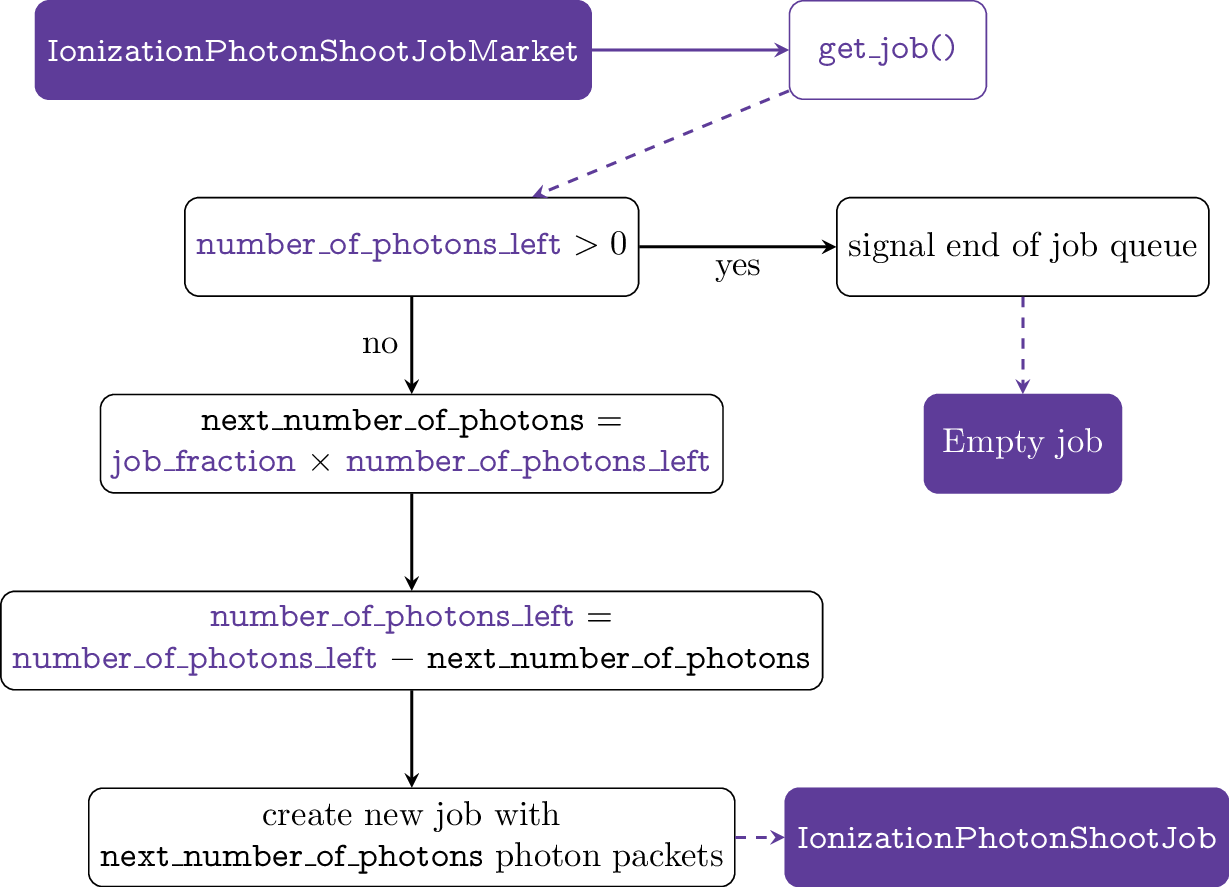}
\caption{\code{JobMarket} implementation that controls the load balancing of a
parallel photoionization run. We create a new \code{Job} instance that will
propagate a fraction of the total number of photons left to propagate. This
ensures that tasks gradually get smaller, which in turn guarantees a minimal
load imbalance, as the maximal load imbalance is the time it takes to execute a
single task. Note that this function is not thread safe and hence requires a
locking mechanism to ensure safe access.
\label{figure:photon_shoot_jobmarket}}
\end{figure*}

\begin{figure*}
\centering{}
\includegraphics[width=0.98\textwidth]{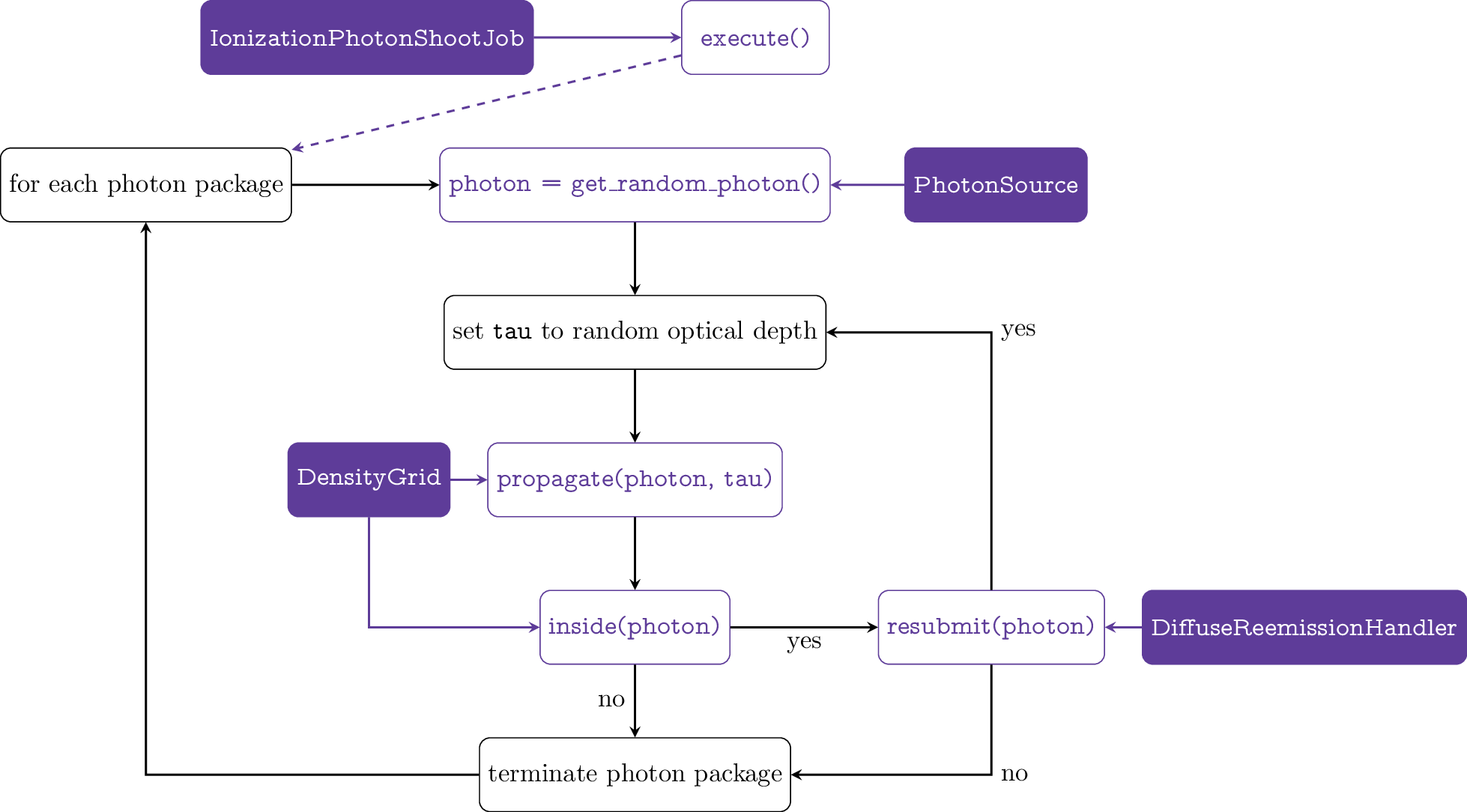}
\caption{\code{Job} implementation that propagates a number of photons through
the grid. Note that the call to \code{DensityGrid::propagate} requires a locking
mechanism for thread safe cell access (the rest of the function is thread safe).
\label{figure:ionization_photon_shoot_job}}
\end{figure*}

This paradigm nicely divides the responsibilities of the parallelization
process: the \code{Job} provides the actual task at hand, the \code{JobMarket}
regulates how tasks are divided and hence controls the load balancing, while the
\code{WorkDistributor} is responsible for handling the underlying parallel
environment. \figureref{figure:photon_shoot_jobmarket} and
\figureref{figure:ionization_photon_shoot_job} show how this works for a
photoionization simulation.

We need a locking mechanism to protect common variables in the shared memory
domain, like e.g. cells in the photon traversal algorithm or counter values
within the \code{JobMarket}. To this end, we provide our own \code{Lock} class
that is unaware of the underlying parallel environment. We also experimented
with using atomic data operations, but found this to be slower than a locking
mechanism in most cases, predominantly because of the lack of hardware support
for floating point atomic operations, and because the large number of variables
updated per cell access in the photon traversal algorithm reduces the impact of
using an expensive locking mechanism.

%%%%%%%%%%%%%%%%%%%%%%%%%%%%%%%%%%%%%%%%%%%%%%%%%%%%%%%%%%%%%%%%%%%%%%%%%%%%%%%%
\subsection{Library exposure}
\label{subsection:library_exposure}

To improve the usability of \cmacionize{}, we also provide a library interface
to the code, which can be used in both C, C++ and Fortran2003. This library
interface is in all ways equivalent to the standard \cmacionize{} program (the
same \code{IonizationSimulation} class is used to run an actual photoionization
simulation), but uses special input and output classes to directly obtain the
density field from another code and return the resulting ionization structure to
that code without the need to output anything to disk.

Our current library implementation has already successfully been used to couple
the code to the SPH code \textsc{PHANTOM} ascl:1709.002.

The library needs to be initialized using a parameter file that is identical to
the one used for the actual \cmacionize{} program, and provides a single
function that takes an array of positions, smoothing lengths and masses as an
input, and outputs an array of neutral hydrogen fractions. Under the hood, it
converts the positions, smoothing lengths and masses into a density field that
is mapped onto a density grid, and then uses sources from the parameter file to
illuminate this density field and compute self-consistent neutral fractions.
These neutral fractions are then mapped back to the original SPH positions using
the provided smoothing lengths.

It should be fairly straightforward to use the same approach to couple the code
to other types of hydrodynamical codes, like AMR codes.

Some parts of the code are also wrapped into a Python library. This library is
not meant to run full photoionization simulations like the C/C++/Fortran
counterpart, but instead can be used to facilitate the analysis of simulation
snapshots, by e.g. providing access to the line cooling data used by the code.

%%%%%%%%%%%%%%%%%%%%%%%%%%%%%%%%%%%%%%%%%%%%%%%%%%%%%%%%%%%%%%%%%%%%%%%%%%%%%%%%
\subsection{Unstructured grid generation}

An important feature of \cmacionize{} is the option to use an unstructured
Voronoi grid as the main grid structure for both the radiation transfer and the
hydrodynamics. Due to the poor scaling properties of the incremental
construction algorithm used in \citet{2016Vandenbroucke}, we decided to
implement two new algorithms. The first algorithm (which for historical reasons
is called the ``old'' Voronoi algorithm) is our own rewritten version of the no
longer actively supported voro++
library\footnote{\url{http://math.lbl.gov/voro++/}}. This algorithm works in
most cases. However, we found that in some very specific degenerate cases, the
voro++ algorithm can produce the wrong Voronoi grid without crashing (as can be
confirmed graphically, or by computing the total volume of all cells). This is
due to the way the algorithm handles degeneracies.

Since these degenerate cases do in fact happen when starting a moving-mesh
hydrodynamical simulation from a regular initial grid, we also implemented a
new, more scalable version of the incremental construction algorithm used in
\shadowfax{} (called the ``new'' algorithm). The current version of this grid
construction algorithm is about a factor 3 slower than the voro++ algorithm, but
is completely robust to any degeneracies due to the usage of arbitrary precision
arithmetics. We plan to release this algorithm as a standalone library that can
be used as a replacement for voro++ (Vandenbroucke \emph{et al.}, \emph{in
prep.}).

In this work, we will use the old Voronoi algorithm whenever an unstructured
grid is used. This does not affect the results we show in any way, as the
Voronoi grid for a set of points is a unique geometrical structure, and is
independent of the way it is computed.

%%%%%%%%%%%%%%%%%%%%%%%%%%%%%%%%%%%%%%%%%%%%%%%%%%%%%%%%%%%%%%%%%%%%%%%%%%%%%%%%
\section{Benchmarks}
\label{section:benchmarks}
%%%%%%%%%%%%%%%%%%%%%%%%%%%%%%%%%%%%%%%%%%%%%%%%%%%%%%%%%%%%%%%%%%%%%%%%%%%%%%%%

In order to verify that our code produces physically accurate results in an
efficient way, we run a number of benchmark tests. Tests are grouped together
into four categories:
\begin{enumerate}
  \item{} tests that verify the ionization algorithm,
  \item{} tests that verify the combined ionization and temperature computation
  algorithm,
  \item{} tests that verify the radiation hydrodynamics algorithms, and
  \item{} tests that check the parallel efficiency of the code.
\end{enumerate}

For the fourth category, we just reuse tests from the three other categories.
The initial conditions and analysis scripts for all benchmark tests are part of
the public version of the code.

%%%%%%%%%%%%%%%%%%%%%%%%%%%%%%%%%%%%%%%%%%%%%%%%%%%%%%%%%%%%%%%%%%%%%%%%%%%%%%%%
\subsection{Str\"{o}mgren sphere}
\label{subsection:stromgren_sphere}

To test the ionization algorithm, we run a simple test, inspired by the work of
\citet{1939Stromgren}, in which a single ionizing source is at the centre of a
box containing a homogeneous density field consisting of only hydrogen. If we
assume that the source completely ionizes all hydrogen within a radius $R_s$
(the Str\"{o}mgren radius), while the hydrogen outside this region stays
neutral, then the ionization balance equation for the ionized region is
\citep{2006Osterbrock}:
\begin{equation}
Q = \frac{4\pi{}}{3} R_s^3 n_\text{H}^2 \alpha{}_{\text{H}^+ \rightarrow{}
\text{H}^0} (T),
\end{equation}
where we made the assumption that $n_\text{H} = n_{\text{H}^+}$ within the
ionized region. If we also assume a fixed temperature $T_0$, then we get an
analytic expression for the constant Str\"{o}mgren radius:
\begin{equation}
R_s = \left( \frac{3Q}{4\pi{}n_{\text{H}}^2\alpha{}_{\text{H}^+ \rightarrow{}
\text{H}^0}(T_0)} \right)^{\frac{1}{3}}.
\label{equation:stromgren_radius}
\end{equation}

If the ionized region itself emits ionizing radiation (through the diffuse
field), the ionization balance equation changes to
\begin{multline}
Q \left( 1 + P_{r,\text{H}}(T) + P_{r,\text{H}}(T)^2 + ... \right) \\
= \frac{4\pi{}}{3} R_s^3 n_\text{H}^2 \alpha{}_{\text{H}^+ \rightarrow{}
\text{H}^0} (T),
\end{multline}
where $P_{r,\text{H}}(T)$ is the reemission probability for ionizing radiation.

Using $1 + x + x^2 + ... = \frac{1}{1-x}$, \eqref{equation:stromgren_radius} now
becomes:
\begin{equation}
R_s' = \left( \left( \frac{Q}{1 - P_{r,\text{H}}(T_0)} \right)
\frac{3}{4\pi{}n_{\text{H}}^2\alpha{}_{\text{H}^+ \rightarrow{}
\text{H}^0}(T_0)} \right)^{\frac{1}{3}}.
\label{equation:stromgren_radius_diffuse}
\end{equation}

We will hence run two different versions of the test, that test different parts
of the algorithm:
\begin{enumerate}
  \item{} a version that does not include the diffuse reemission field and
  should reproduce \eqref{equation:stromgren_radius}, and
  \item{} a version that includes diffuse radiation and should produce a larger
  ionization region, as given by \eqref{equation:stromgren_radius_diffuse}.
\end{enumerate}

For both tests, we will use a cubic box of $10 \times{} 10 \times{} 10$~pc
containing gas with a hydrogen number density of $n_\text{H} =
100~\text{cm}^{-3}$. At the centre of the box, we put a single source with a
luminosity of $Q = 4.26 \times{} 10^{49}~\text{s}^{-1}$ with a monochromatic
spectrum that emits photons at the ionization threshold energy for hydrogen,
$\nu{} = 13.6~\text{eV}$. We assume a constant photoionization cross section for
neutral hydrogen of $\sigma{}_{\text{H}^0 \rightarrow{} \text{H}^+} = 6.3
\times{} 10^{-18}~\text{cm}^{-2}$, and a constant radiative recombination rate
$\alpha{}_{\text{H}^+ \rightarrow{} \text{H}^0} (T_0) = 4 \times{}
10^{-13}~\text{cm}^3~\text{s}^{-1}$. The abundances, photoionization cross
sections and recombination rates for all other elements and ions are set to
zero.

We use a Cartesian density grid of $64 \times{} 64 \times{} 64$ cells, and use
$10^6$ photon packets for 20 iterations to get a converged result.

\subsubsection{No diffuse field}

\begin{figure}
\centering{}
\includegraphics[width=0.48\textwidth]{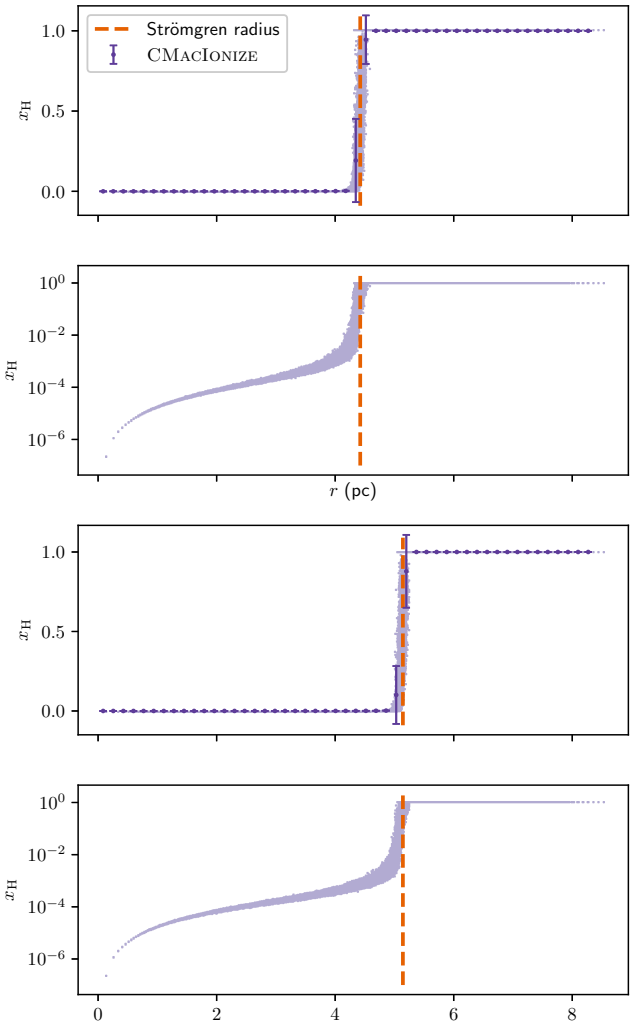}
\caption{Hydrogen neutral fraction as a function of radius for the two versions
of the Str\"{o}mgren benchmark test. \emph{Top panels}: version without diffuse
radiation, \emph{bottom panels}: version with diffuse radiation. The light
purple dots show the simulation results for the individual cells; the dark
purple error bars show the same results in 50 radial bins, with the size of the
error bars showing the scatter within the bin. The orange dashed line is the
corresponding analytic Str\"{o}mgren radius. For clarity, the results are shown
on both a linear and a logarithmic scale.
\label{figure:stromgren_results}}
\end{figure}

\begin{figure}
\centering{}
\includegraphics[width=0.48\textwidth]{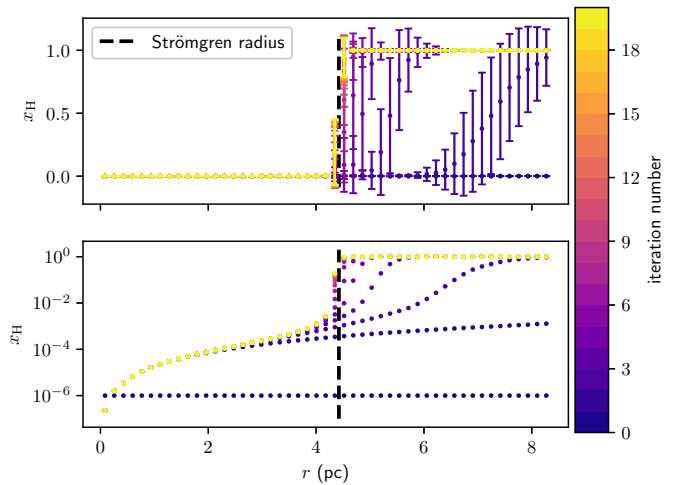}
\caption{Hydrogen neutral fraction as a function of radius for the benchmark
test without diffuse field, for different iteration numbers. For clarity, we
only show the binned results, with the size of the error bar representing the
scatter within the bin. The black dashed line is the analytic Str\"{o}mgren
radius. On the bottom panel we show the same results on a logarithmic scale
without the error bars.
\label{figure:stromgren_convergence}}
\end{figure}

This test corresponds to the benchmark test \code{stromgren}, and the setup is
as described above. The resulting hydrogen neutral fraction as a function of
radius is shown in the top panel of \figureref{figure:stromgren_results}. The
code accurately reproduces the expected Str\"{o}mgren radius given by
\eqref{equation:stromgren_radius}.

\figureref{figure:stromgren_convergence} shows the evolution of the hydrogen
neutral fraction profile with the number of iterations used. Initially, the
neutral fraction is set to a very low value everywhere in the box, so that the
ionizing radiation can illuminate a large region efficiently. After the first
iterations of the ionization state computation, the neutral fraction in the
outer regions quickly goes up until a converged result is reached. The result is
already well converged after 6 iterations.

\subsubsection{Diffuse field}

This test corresponds to the benchmark test \code{stromgren\_diffuse}, and
includes diffuse reemission with a reemission probability $P_{r,\text{H}}(T_0) =
0.364$ (corresponding to $T_0 = 8000~\text{K}$). The resulting hydrogen neutral
fraction profile is shown in the bottom panel of
\figureref{figure:stromgren_results}. As expected, the ionized region is larger
in this case. Our code still accurately reproduces the expected Str\"{o}mgren
radius given by \eqref{equation:stromgren_radius_diffuse}.

%%%%%%%%%%%%%%%%%%%%%%%%%%%%%%%%%%%%%%%%%%%%%%%%%%%%%%%%%%%%%%%%%%%%%%%%%%%%%%%%
\subsection{Lexington benchmarks}
\label{subsection:lexington_benchmarks}

To test the combined temperature and ionization calculation for the full system
including metals, we run two of the benchmark tests that were the result of a
1995 workshop in Lexington and that are known as the Lexington benchmarks
\citep{1995Ferland}. The initial conditions for these tests can be found in
\citet{2001Pequignot}, and correspond to the HII40 and HII20 model in that work.

The tests consist of a uniform density box with a hydrogen number density
$n_\text{H} = 100~\text{cm}^{-3}$, in which a central spherical region with
radius $R_\text{in} = 3 \times{} 10^{16}~\text{m}$ is evacuated. In the centre
of the evacuated region we place a single stellar source with a luminosity $Q =
10^{49}~\text{s}^{-1}$ for the low temperature benchmark, and $Q = 4.26 \times{}
10^{49}~\text{s}^{-1}$ for the high temperature benchmark, with a black body
spectrum. The temperature $T_\text{BB}$ of the black body spectrum is also
different for the two tests, with $T_\text{BB} = 20,000~\text{K}$ for the low
temperature benchmark, and $T_\text{BB} = 40,000~\text{K}$ for the high
temperature version. The abundances (relative to hydrogen) of helium and the
various coolants are set to the values listed in
\tableref{table:lexington_abundances}.

\begin{table}
\centering{}
\caption{Abundances for helium and the different coolants in the Lexington
benchmark tests.
\label{table:lexington_abundances}}

\vspace{12pt}
\begin{tabular}{l l}
\hline
element & abundance \\
\hline
He/H & $0.1$ \\
C/H & $2.2 \times{} 10^{-4}$ \\
N/H & $4 \times{} 10^{-5}$ \\
O/H & $3.3 \times{} 10^{-4}$ \\
Ne/H & $5 \times{} 10^{-5}$ \\
S/H & $9 \times{} 10^{-6}$ \\
\hline
\end{tabular}
\end{table}

To compare the test results, a number of quantities are computed:
\begin{itemize}
  \item{} The total H$\beta{}$ luminosity, which is computed from a power law
  fit to the data of \citet{1995Storey}, using the electron density and
  temperature derived from the photoionization simulation.
  \item{} The height of the Balmer Jump $\Delta{} (BC~3645)$, defined as the
  jump in the Balmer continuum flux in a synthetic spectrum between the flux at
  $3643~\text{\AA}$ and $3681~\text{\AA}$. To obtain synthetic spectra, we use
  the continuum emission coefficients for hydrogen and helium from
  \citet{1970Brown}. Note that \citet{2001Pequignot} and other authors wrongly
  quote this value in units of \AA, while the actual value is in
  $\text{\AA}^{-1}$.
  \item{} The inner temperature $T_\text{inner}$ at the boundary of the
  evacuated region.
  \item{} The average density weighted temperature \citep{1995Ferland}
  \begin{equation}
  \langle{} T [ n_e n_{\text{H}^+} ] \rangle{} = \frac{\int n_e n_{\text{H}^+}
  T \text{d}V}{\int n_e n_{\text{H}^+} \text{d}V}.
  \end{equation}
  \item{} The outer radius of the ionization region, defined as the average
  radius of cells with hydrogen neutral fractions in the range $f_{\text{H}^0}
  \in{} [0.1, 0.2]$.
  \item{} The ratio of the density weighted ionized fractions of hydrogen and
  helium, defined as
  \begin{equation}
  \frac{\langle{} f_{\text{He}^+} \rangle{}}{\langle{} f_{\text{H}^+} \rangle{}}
  = \frac{\int n_e f_{\text{He}^+} \text{d}V}{\int n_e n_{\text{H}^+}
  \text{d}V}.
  \end{equation}
\end{itemize}

Apart from those, we also compute the line strengths of a number of emission
lines, relative to the total H$\beta{}$ luminosity. These lines are a subset of
the emission lines that are used for the metal line cooling (see
\ref{subsection:line_cooling}), and their strength is computed in the same way
(summed over all cells).

For both tests, we set up a box of size $6 \times{} 6 \times{} 6~\text{pc}$,
using a Cartesian density grid of $64 \times{} 64 \times{} 64$ cells. We use
$10^8$ photons and 20 iterations to get a converged result for all coolants.

To set up the initial condition with a vacuum region in the centre, we use a
special implementation of the \code{DensityFunction} used to set up the density
field, called \code{BlockSyntaxDensityFunction}. This implementation uses a very
simple geometrical block description of the initial condition, which is the same
as used by the initial condition generator of \shadowfax{}
\citep{2016Vandenbroucke}.

\subsubsection{Low temperature benchmark}

\begin{figure*}
\centering{}
\includegraphics[width=0.98\textwidth]{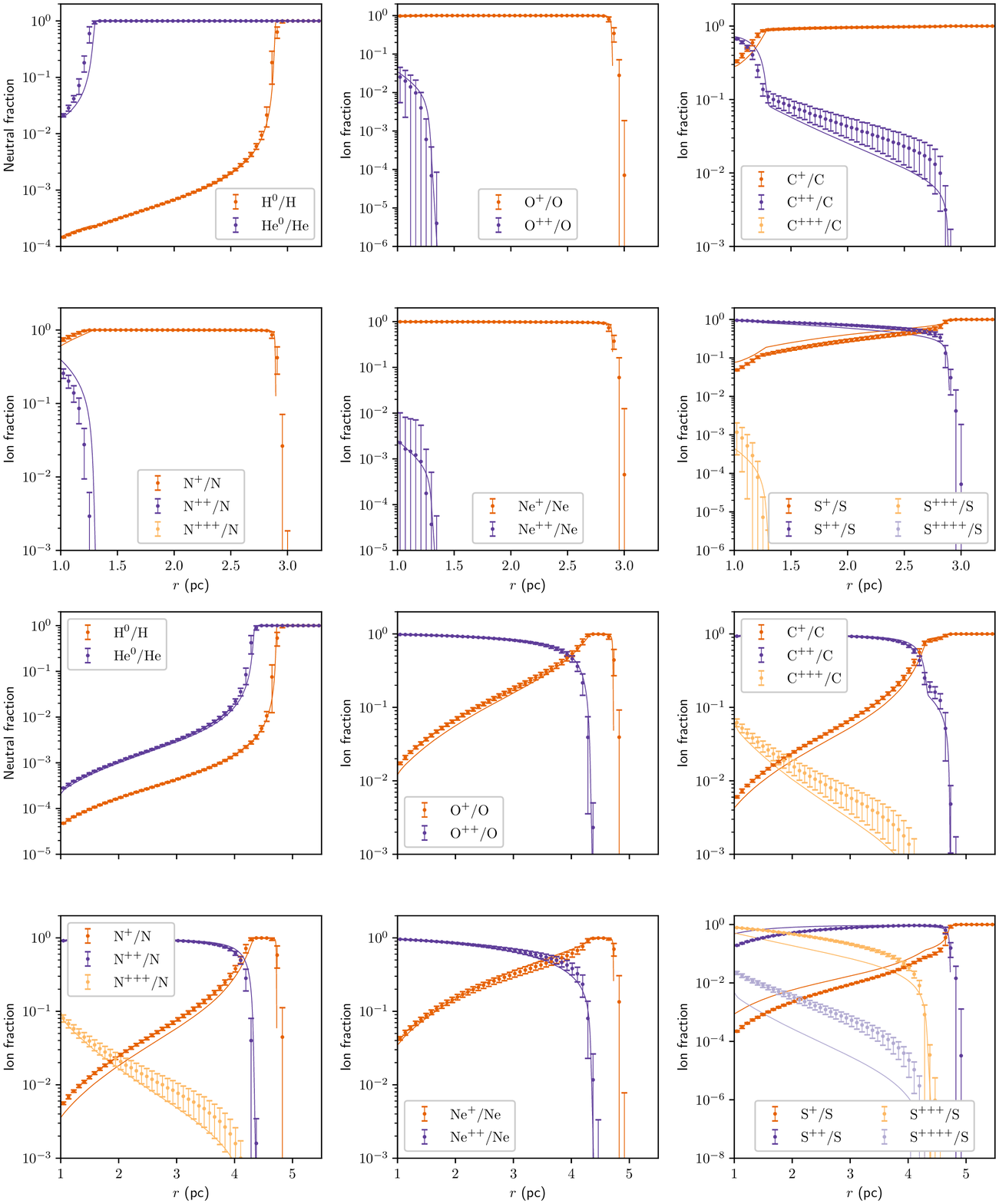}
\caption{Ionic fraction of hydrogen, helium, and various coolants as a function
of radius, as indicated on the figures. \emph{Top rows}: low temperature
benchmark, \emph{bottom rows}: high temperature benchmark. The error bars show
the simulation results, while the full lines are the equivalent \textsc{Cloudy}
results.
\label{figure:lexington_profiles}}
\end{figure*}

\begin{figure}
\centering{}
\includegraphics[width=0.48\textwidth]{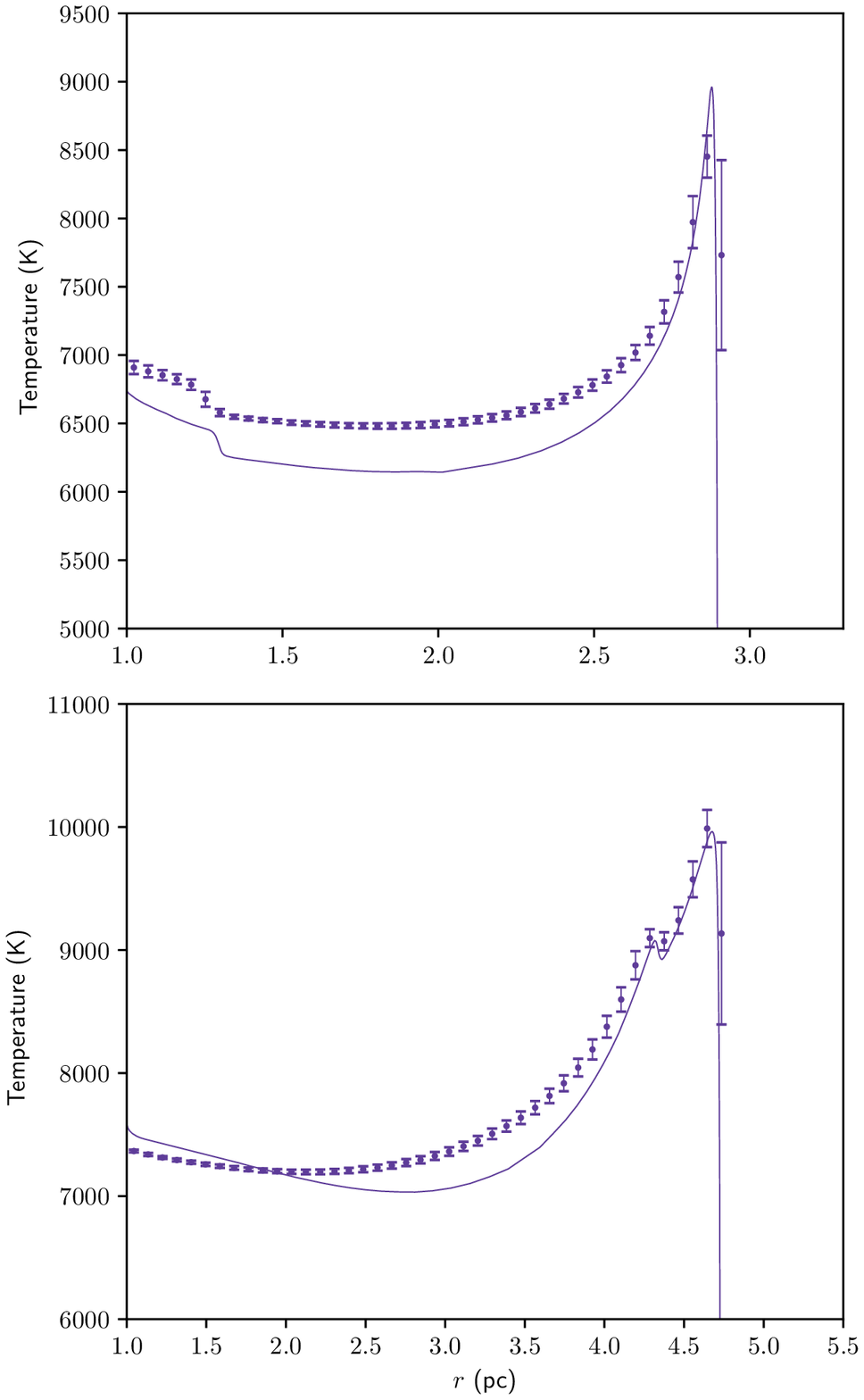}
\caption{Temperature as a function of radius for the Lexington benchmarks.
\emph{Top rows}: low temperature benchmark, \emph{bottom rows}: high temperature
benchmark. The error bars show the simulation results, while the full line
represents the equivalent \textsc{Cloudy} result.
\label{figure:lexington_temperatures}}
\end{figure}

This corresponds to the \code{lexingtonHII20} benchmark test, and uses a black
body spectrum with $T_\text{BB} = 20,000~\text{K}$. The resulting ionic
fraction profiles for hydrogen, helium and several coolants are shown in the top
panel of \figureref{figure:lexington_profiles}, together with reference values
from the 1D code \textsc{Cloudy}. They follow the same trends as observed in
e.g. \citet{2004Wood}, and overall agreement is pretty good. The resulting
temperature profile is shown in the top panel of
\figureref{figure:lexington_temperatures}. Our result follows the same overall
trend as the reference curve, but we systematically overestimate the temperature
in the ionized region. We do reproduce the correct peak temperature at the
ionization radius. This difference can be attributed to the different cooling
rates we use compared to \textsc{Cloudy}.

\tableref{table:lexingthonHII20} lists the line strengths and comparison
quantities used by \citet{2001Pequignot}, and compares them with the median
values given in that paper. All values have the expected order of magnitude,
although some values deviate significantly. This can be attributed to a
combination of our overall higher temperature, and different line emission
rates.

\begin{table*}
\centering{}
\caption{Line strengths and comparison quantities for the low temperature
Lexington benchmark test. P2001 denotes the median value as given in
\citet{2001Pequignot}.
\label{table:lexingthonHII20}}

\vspace{12pt}
\begin{tabular}{l c c}
\hline
Quantity & \cmacionize{} value & P2001 value \\
\hline
H$\beta{}$ luminosity &
% Hbeta
$4.81 \times{} 10^{36}~\text{erg}~\text{s}^{-1}$ &
$4.89 \times{} 10^{36}~\text{erg}~\text{s}^{-1}$ \\
$\Delta{}(\text{BC}~3645)$ &
% Delta BC3645
$5.44 \times{} 10^{-3}~\text{\AA}^{-1}$ &
$5.56 \times{} 10^{-3}~\text{\AA}^{-1}$ \\
$T_\text{inner}$ &
% Tinner
$7052~\text{K}$ &
$6789~\text{K}$ \\
$\langle{} T [ n_e n_{\text{H}^+} ] \rangle{}$ &
% T[NpNe]
$6843~\text{K}$ &
$6663~\text{K}$ \\
$R_\text{out}$ &
% Rout
$8.83 \times{} 10^{16}~\text{m}$ &
$8.89 \times{} 10^{16}~\text{m}$ \\
$\langle{} f_{\text{He}^+} \rangle{} / \langle{} f_{\text{H}^+} \rangle{}$ &
% <He+>/<H+>
0.047 &
0.049 \\
\hline
\\
\hline
Line & \cmacionize{} line strength & P2001 line strength \\
\hline
$[\text{C~II}]$ $2325~\text{\AA}$ multiplet &
% CII 2325+
0.066 &
0.047 \\
$[\text{N~II}]$ $122~\mu{}\text{m}$ &
% NII 122 um
0.068 &
0.071 \\
$[\text{N~II}]$ $6584~\text{\AA}$ and $6548~\text{\AA}$ &
% NII 6584 + 6548
0.845 &
0.803 \\
$[\text{N~II}]$ $5755~\text{\AA}$ &
% NII 5755
0.0029 &
0.0029 \\
$[\text{N~III}]$ $57.3~\mu{}\text{m}$ &
% NIII 57.3 um
0.0030 &
0.0031 \\
$[\text{O~I}]$ $6300~\text{\AA}$ and $6363~\text{\AA}$ &
% OI 6300 + 6363
0.0052 &
0.0060 \\
$[\text{O~II}]$ $7320~\text{\AA}$ and $7330~\text{\AA}$ &
% OII 7320 + 7330
0.0103 &
0.0087 \\
$[\text{O~II}]$ $3726~\text{\AA}$ and $3729~\text{\AA}$ &
% OII 3726 + 3729
1.33 &
1.10 \\
$[\text{O~III}]$ $51.8~\mu{}\text{m}$ &
% OIII 52 um
0.0013 &
0.0012 \\
$[\text{O~III}]$ $88.3~\mu{}\text{m}$ &
% OIII 88 um
0.0016 &
0.0014 \\
$[\text{O~III}]$ $5007~\text{\AA}$ and $4959~\text{\AA}$ &
% OIII 5007 + 4959
0.0018 &
0.0015 \\
$[\text{Ne~II}]$ $12.8~\mu{}\text{m}$ &
% NeII 12.8 um
0.297 &
0.271 \\
$[\text{S~II}]$ $6716~\text{\AA}$ and $6731~\text{\AA}$ &
% SII 6716 + 6731
0.459 &
0.492 \\
$[\text{S~II}]$ $4068~\text{\AA}$ and $4076~\text{\AA}$ &
% SII 4068 + 4076
0.014 &
0.017 \\
$[\text{S~III}]$ $18.7~\mu{}\text{m}$ &
% SIII 18.7 um
0.333 &
0.420 \\
$[\text{S~III}]$ $33.6~\mu{}\text{m}$ &
% SIII 33.6 um
0.558 &
0.750 \\
$[\text{S~III}]$ $9532~\text{\AA}$ and $9069~\text{\AA}$ &
% SIII 9532 + 9069
0.479 &
0.525 \\
\hline
\end{tabular}
\end{table*}

\subsubsection{High temperature benchmark}

This corresponds to the \code{lexingtonHII40} benchmark test, and uses a black
body spectrum with $T_\text{BB} = 40,000~\text{K}$. The resulting ionic
fraction profiles for hydrogen, helium and several coolants are shown in the
bottom panel of \figureref{figure:lexington_profiles}, and they again follow the
same trends as observed in \citet{2004Wood}. The resulting temperature profile
is shown in the bottom panel of \figureref{figure:lexington_temperatures}. We
again notice an overall higher temperature in most of the ionized region
(although we actually underestimate the central temperature), which again is due
to the different data values used by our code.

\tableref{table:lexingthonHII40} lists the line strengths and comparison
quantities. Note that we cannot compute the $[\text{C~II}]$ $1335~\text{\AA}$
line strength given in \citet{2001Pequignot}, as this corresponds to a
transition outside the five lowest lying levels of C$^+$. Neither can we compute
the $[\text{O~IV}]$ $25.9~\mu{}\text{m}$ line strength, since O$^{++}$ has an
ionization potential that is (just) higher than the $54.4~\text{eV}$ upper limit
of our energy interval, and we hence cannot track O$^{+++}$. The values
generally agree with the \citet{2001Pequignot} values, although there are again
significant differences, especially for the sulphur lines.

\begin{table*}
\centering{}
\caption{Line strengths and comparison quantities for the high temperature
Lexington benchmark test. P2001 denotes the median value as given in
\citet{2001Pequignot}.
\label{table:lexingthonHII40}}

\vspace{12pt}
\begin{tabular}{l c c}
\hline
Quantity & \cmacionize{} value & P2001 value \\
\hline
H$\beta{}$ luminosity &
% H beta
$2.01 \times{} 10^{37}~\text{erg}~\text{s}^{-1}$ &
$2.05 \times{} 10^{37}~\text{erg}~\text{s}^{-1}$ \\
$\Delta{}(\text{BC}~3645)$ &
% Delta BC3645
$4.99 \times{} 10^{-3}~\text{\AA}^{-1}$ &
$4.97 \times{} 10^{-3}~\text{\AA}^{-1}$ \\
$T_\text{inner}$ &
% Tinner
$7410~\text{K}$ &
$7663~\text{K}$ \\
$\langle{} T [ n_e n_{\text{H}^+} ] \rangle{}$ &
% T[NpNe]
$8127~\text{K}$ &
$8030~\text{K}$ \\
$R_\text{out}$ &
% Rout
$1.44 \times{} 10^{17}~\text{m}$ &
$1.46 \times{} 10^{17}~\text{m}$ \\
$\langle{} f_{\text{He}^+} \rangle{} / \langle{} f_{\text{H}^+} \rangle{}$ &
% <He+>/<H+>
0.784 &
0.770 \\
\hline
\\
\hline
Line & \cmacionize{} line strength & P2001 line strength \\
\hline
$[\text{He~I}]$ $5876~\text{\AA}$ &
% HeI 5876
0.116 &
0.116 \\
$[\text{C~II}]$ $2325~\text{\AA}$ multiplet &
% CII 2325
0.184 &
0.140 \\
$[\text{C~III}]$ $1907~\text{\AA}$ and $1909~\text{\AA}$ &
% CIII 1907 + 1909
0.073 &
0.071 \\
$[\text{N~II}]$ $122~\mu{}\text{m}$ &
% NII 122 um
0.029 &
0.033 \\
$[\text{N~II}]$ $6584~\text{\AA}$ and $6548~\text{\AA}$ &
% NII 6584 + 6548
0.712 &
0.725 \\
$[\text{N~II}]$ $5755~\text{\AA}$ &
% NII 5755
0.0056 &
0.0052 \\
$[\text{N~III}]$ $57.3~\mu{}\text{m}$ &
% NIII 57.3
0.304 &
0.297 \\
$[\text{O~I}]$ $6300~\text{\AA}$ and $6363~\text{\AA}$ &
% OI 6300 + 6363
0.0091 &
0.0087 \\
$[\text{O~II}]$ $7320~\text{\AA}$ and $7330~\text{\AA}$ &
% OII 7320 + 7330
0.031 &
0.030 \\
$[\text{O~II}]$ $3726~\text{\AA}$ and $3729~\text{\AA}$ &
% OII 3726 + 3729
2.19 &
2.12 \\
$[\text{O~III}]$ $51.8~\mu{}\text{m}$ &
% OIII 52 um
1.21 &
1.06 \\
$[\text{O~III}]$ $88.3~\mu{}\text{m}$ &
% OIII 88 um
1.43 &
1.23 \\
$[\text{O~III}]$ $5007~\text{\AA}$ and $4959~\text{\AA}$ &
% OIII 5007 + 4959
2.46 &
2.20 \\
$[\text{O~III}]$ $4363~\text{\AA}$ &
% OIII 4363
0.0043 &
0.0040 \\
$[\text{Ne~II}]$ $12.8~\mu{}\text{m}$ &
% NeII 12.8 um
0.180 &
0.194 \\
$[\text{Ne~III}]$ $15.5~\mu{}\text{m}$ &
% NeIII 15.5 um
0.326 &
0.350 \\
$[\text{Ne~III}]$ $3869~\text{\AA}$ and $3968~\text{\AA}$ &
% NeIII 3869 + 3968
0.088 &
0.086 \\
$[\text{S~II}]$ $6716~\text{\AA}$ and $6731~\text{\AA}$ &
% SII 6716 + 6731
0.129 &
0.153 \\
$[\text{S~II}]$ $4068~\text{\AA}$ and $4076~\text{\AA}$ &
% SII 4068 + 4076
0.0060 &
0.0090 \\
$[\text{S~III}]$ $18.7~\mu{}\text{m}$ &
% SIII 18.7 um
0.480 &
0.580 \\
$[\text{S~III}]$ $33.6~\mu{}\text{m}$ &
% SIII 33.6 um
0.772 &
0.936 \\
$[\text{S~III}]$ $9532~\text{\AA}$ and $9069~\text{\AA}$ &
% SIII 9532 + 9069
0.989 &
1.23 \\
$[\text{S~IV}]$ $10.5~\mu{}\text{m}$ &
% SIV 10.5 um
0.589 &
0.330 \\
\hline
\end{tabular}
\end{table*}

%%%%%%%%%%%%%%%%%%%%%%%%%%%%%%%%%%%%%%%%%%%%%%%%%%%%%%%%%%%%%%%%%%%%%%%%%%%%%%%%
\subsection{STARBENCH benchmark}
\label{subsection:STARBENCH}

To test the coupling between the radiation transfer algorithm and our
hydrodynamical integration scheme, we use the benchmark D-type expansion of an
HII region which is part of the STARBENCH project \citep{2015Bisbas}. The setup
is similar to the Str\"{o}mgren test introduced above, but instead of assuming a
static solution, we allow the gas to react to the increased pressure due to the
higher temperature of the ionised region, and study the expansion of the
resulting ionized region over time.

We put a central source with a luminosity of $10^{49}~\text{s}^{-1}$ and a
monochromatic spectrum that emits at $13.6~\text{eV}$ in a box of $2.512
\times{} 2.512 \times{} 2.512~\text{pc}$ containing only hydrogen with a density
of $3113~\text{cm}^{-3}$. The hydrogen photoionization cross section is set to
$6.3\times{}10^{-18}~\text{cm}^2$, while the hydrogen recombination rate is set
to $2.7\times{}10^{-13}~\text{cm}^3~\text{s}^{-1}$. We assume there is no
diffuse radiation field.

We assume a very simple isothermal equation of state (in practice this is
realized by setting the adiabatic index to $\gamma{} = 1.0001$), and assume that
the ionized region has a constant temperature $T_i = 10,000~\text{K}$, while the
neutral region has a constant temperature $T_n = 100~\text{K}$.

The system is evolved in time until $t = 0.141~\text{Myr}$, and we keep track of
the evolution of the ionization front.

There is no strict analytic solution for this problem, but there are two
reference solutions for the evolution of the ionization front as a function of
time. The first is the so called Spitzer solution \citep{1978Spitzer}
\begin{equation}
R_\text{Sp}(t) = R_s \left( 1 + \frac{7}{4} \frac{c_{s,i} t}{R_s}
\right)^\frac{4}{7} \label{equation:spitzer},
\end{equation}
where $c_{s,i}$ is the (constant) sound speed in the ionized region, and $R_s$
is the Str\"{o}mgren radius, as defined in \eqref{equation:stromgren_radius}.

The second solution is due to \citet{2006Hosokawa}:
\begin{equation}
R_\text{HI}(t) = R_s \left( 1 + \frac{7}{4} \sqrt{\frac{4}{3}} \frac{c_{s,i}
t}{R_s} \right)^\frac{4}{7} \label{equation:hosokawa_inutsuka},
\end{equation}
and evolves at a somewhat faster rate. It is worth pointing out that we do not
require our simulation to reproduce any one of these solutions, but we do
require it to be close to them.

As a measure of the ionization front radius, we will use the average radius of
cells with neutral fractions in the range $[0.8, 0.9]$. Due to the sharp
transition from ionized to neutral (as can be seen from
\figureref{figure:starbench_profile}), using different boundaries for this
interval does not change the ionization radius much, as long as we make sure we
exclude noisy cells with $x_\text{H} < 0.1$ or $x_\text{H} > 0.9$.

We will run two versions of this test: a version that uses a static Cartesian
grid of $64 \times{} 64 \times{} 64$ cells (the Eulerian solution), and a
version that uses a co-moving Voronoi mesh with 10,000 grid generator positions
sampled from a uniform distribution and regularized using Lloyd's algorithm for
10 iterations (the Lagrangian solution). For both, we apply the photoionisation
algorithm after every hydrodynamics step, using 10 iterations. The Eulerian
version uses $10^6$ photon packets, while the Lagrangian version (with a lower
effective grid resolution) uses $10^5$. These values were found to give a good
trade-off between accuracy and computational efficiency.

\subsubsection{Eulerian solution}

\begin{figure}
\centering{}
\includegraphics[width=0.48\textwidth]{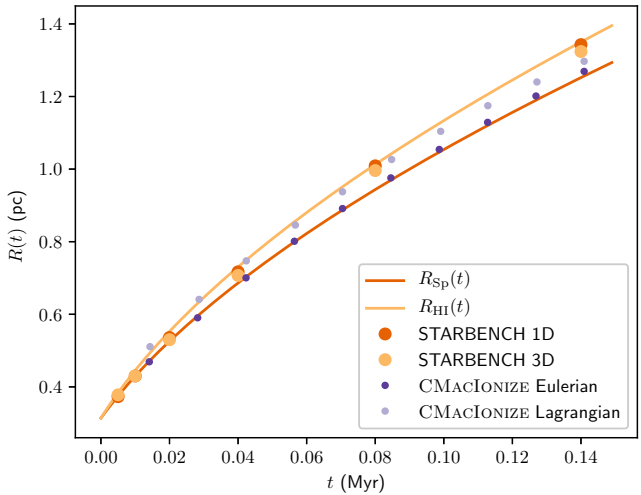}
\caption{Position of the ionization front as a function of time for the
STARBENCH benchmark. The full orange lines show the reference evolution curves,
the purple dots are the simulation results. The orange circles show the average
1D and 3D results from \citet{2015Bisbas}.
\label{figure:starbench_radius}}
\end{figure}

\begin{figure*}
\centering{}
\includegraphics[width=0.98\textwidth]{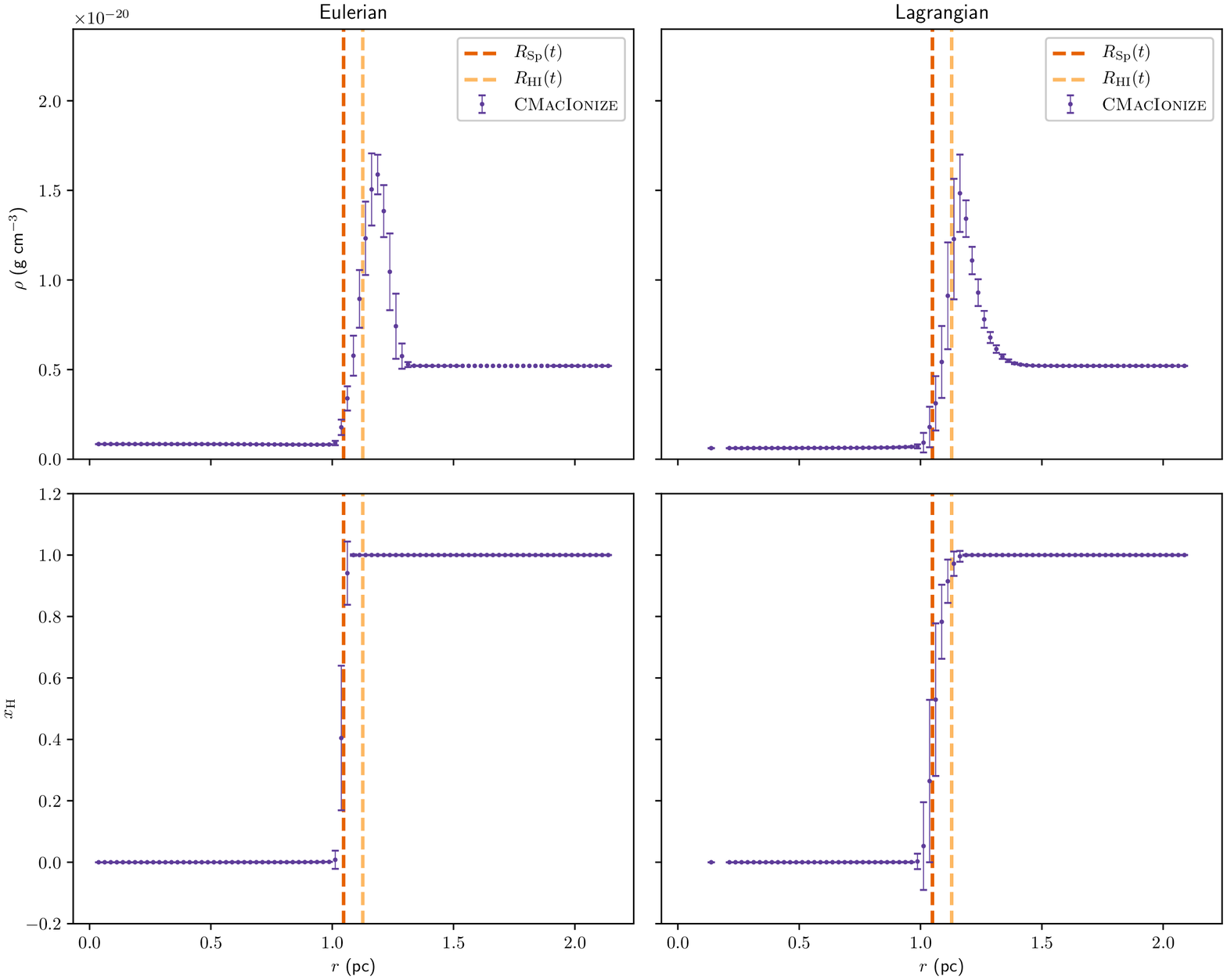}
\caption{Density (\emph{top}) and neutral fraction (\emph{bottom}) as a function
of radius for the STARBENCH benchmark test at $t=0.0987~\text{Myr}$.
\emph{Left}: Eulerian result, \emph{right}: Lagrangian result. The orange dashed
lines are the reference radii, the purple error bars are the simulation
results.
\label{figure:starbench_profile}}
\end{figure*}

This version corresponds to benchmark test \code{starbench}. It evolves the
system forward in time using 2,048 fixed size time steps. The evolution of the
ionization front is shown in \figureref{figure:starbench_radius}, and very
closely follows the Spitzer solution \eqref{equation:spitzer}. The left panels
in \figureref{figure:starbench_profile} show the density and neutral fraction as
a function of radius for $t=0.0987~\text{Myr}$.

\subsubsection{Lagrangian solution}

This version corresponds to benchmark test \code{star\-bench\_\-vo\-ro\-noi}. It
evolves the system forward in time using only 256 fixed size time steps (since
the average cell size is larger in this case). The evolution of the ionization
front is again shown in \figureref{figure:starbench_radius}. This time, the
ionization front first follows the Hosokawa-Inutsuka solution
\eqref{equation:hosokawa_inutsuka}, and then slows down to the Spitzer solution
\eqref{equation:spitzer}. The density and neutral fraction profiles are shown in
the right panels of \figureref{figure:starbench_profile}.

%%%%%%%%%%%%%%%%%%%%%%%%%%%%%%%%%%%%%%%%%%%%%%%%%%%%%%%%%%%%%%%%%%%%%%%%%%%%%%%%
\subsection{Parallel efficiency}

As mentioned in \ref{subsubsection:task_based_design}, we have made most of our
algorithm inherently parallel by designing it in terms of small tasks. In this
subsection, we will illustrate how this affects the parallel scaling of the
code. Since different modes of the code use different parts of the algorithm, we
will rerun all of the benchmarks tests introduced above with different numbers
of shared memory \emph{threads} and distributed memory \emph{processes}.

For all runs, we use a single node of our local high performance computing
cluster Kennedy. This node has a 2.10 GHz Intel Xeon E5-2683 v4 processor with
16 physical cores, hyper-threaded to run 32 threads in parallel. Note that
because of the hyper-threading, we do not expect perfect scaling if more than 16
cores are used, since then different threads will be competing for resources.

We will quantify the scaling through the \emph{speed up} $S(n)$ of the algorithm
as a function of the number of computing units $n$. It is given by
\begin{equation}
S(n) = \frac{t(1)}{t(n)},
\end{equation}
where $t(n)$ is the total runtime of the algorithm when using $n$ computing
units. For a hypothetical code that scales perfectly, the speed up is simply
given by $S_p(n)=n$.

In practice, there are a number of important factors that can affect the
parallel scaling, and cause $S(n)$ to lie below $S_p(n)$:
\begin{itemize}
  \item{} The existence of serial parts of the code that are only executed by
  one computing unit while the other computing units are idle, or that are
  executed by all computing units (and hence duplicate work). These are located
  in parts of the algorithm that are not parallel. If $t_s(n)$ and $t_p(n)$ are
  respectively the runtime of the serial and the parallel part of the code when
  using $n$ computing units, then the theoretical maximum speed up is always
  lower than $S_p(n)$ (assuming $t_p(n) = \frac{t_p(1)}{n}$):
  \begin{equation}
  S_{p,s}(n) = \left( \frac{t_s(1) + t_p(1)}{nt_s(1) + t_p(1)} \right) n.
  \end{equation}
  \item{} The \emph{overhead} caused by running the algorithm in parallel. This
  overhead can be caused by extra code that needs to be executed as part of the
  parallelization strategy, or by delays caused by different computing units
  fighting over hardware access. Overhead will cause $t_{p,o}(n) >
  \frac{t_p(1)}{n}$, i.e. $t_{p,o}(n) = \frac{t_p(1)}{n} + t_o(n)$ (with $t_o(1)
  = 0$).
  If the overhead is \emph{parallel}, it is shared among the various computing
  units ($t_o(n) = \frac{t_{o,c}}{n}$). We can determine the constant from the 2
  computing units measurement, and the speed up including overhead is
  \begin{equation}
  S_{p,s+op}(n) = \left( \frac{t_s(1) + t_p(1)}{n t_s(1) + 2 t_{p,o}(2)} \right)
  n.
  \end{equation}
  If on the other hand the overhead is serial, it is constant per computing unit
  ($t_o(n) = t_{o,u}$), and the speed up is given by
  \begin{multline}
  S_{p,s+os}(n) = \\
  \left( \frac{t_s(1) + t_p(1)}{n t_s(1) + n t_{p,o}(2) + \left( 1 - \frac{n}{2}
  \right) t_p(1)} \right) n.
  \end{multline}
  \item{} The occurrence of load imbalances between different computing units
  that cause some computing units to sit idle while they are waiting for other
  computing units to finish a task. While load imbalances affect the scaling in
  the same way as serial parts of the code, we cannot directly measure them, and
  treat them in the same way as the overhead.
\end{itemize}

For this version of \cmacionize{}, we made the decision to only parallelize
those parts of the algorithm that would lead to the most significant speed up,
i.e. the photon traversal algorithm and the cell based computations. This means
that there is still a significant fraction of serial code left in the algorithm,
which we will address in future versions of the code. This will limit the
expected scaling behaviour of the code, especially in simulations with small
grid sizes and small photon packet numbers.

Since the serial version of the code uses the same task based strategy as the
shared memory parallel version, the latter has no code based overhead (unless
the code is configured without OpenMP, in which case the OpenMP calls will be
absent). The only overhead will be caused by hardware concurrency issues: since
different threads might attempt to write to the same grid cell during the photon
traversal step, we need to lock each cell upon access. If another thread already
obtained access to the cell, the thread trying to acquire the lock will sit idle
until that cell becomes available, causing a small overhead.

The distributed memory parallel version of the code has a much larger code based
overhead, as it requires communication between the different parallel processes
which is completely absent from the serial version. The overhead is particularly
large in our code, as we made the decision to store variables per cell rather
than per variable type, which means we need to repack variables into separate
communication buffers before we can send them to another process. We plan to
address this issue as part of a future distributed memory parallel version that
includes a domain decomposition to distribute the grid across multiple
processes. We hence do not expect very good distributed memory scaling for our
current version.

The task based shared memory parallelization strategy we use automatically takes
care of the load balancing on a single node, as different threads get tasks from
a shared task pool. The time a single thread will need to wait cannot be longer
than the longest time it takes to finish a single task, so we can control the
former by adapting the size of the latter. When a \code{JobMarket} (see
\ref{subsubsection:task_based_design}) starts spawning tasks of a specific type,
it usually starts with reasonably large task sizes, and then gradually makes the
tasks smaller, until some lower limit is reached. This is illustrated in
\figureref{figure:jobtimes}. Using large initial task sizes limits the overhead
caused by calls to the \code{JobMarket}, while the lower size limit sets the
maximum load imbalance between different threads.

We do not perform any specific load balancing between distributed memory
parallel processes, and simply try to assign equal photon numbers and cell
numbers to each process when executing a step in parallel. A better load
balancing scheme will be part of the future domain decomposed distributed memory
algorithm.

Below, we will give the scaling results for the various benchmark tests. We will
focus on shared memory parallelization for all tests, and only show distributed
memory parallelization scaling for the Str\"{o}mgren benchmark, as we still plan
to change our distributed memory parallelization strategy in future versions of
the code.

\subsubsection{Photoionization only}

\paragraph{Shared memory scaling}

\begin{table}
\centering{}
\caption{Timing information for the shared memory scaling run of the
Str\"{o}mgren benchmark test (without diffuse field) on a system with 32
available cores.
\label{table:stromgren_scaling_openmp}}

\vspace{12pt}
\begin{tabular}{c c c c}
\hline
$n$ & $t(n)$ (s) & $t_s(n)$ (s) & $t_o(n)$ (s) \\
\hline
1 & 158.457 & 1.764 & 0.000\\
2 & 107.071 & 1.674 & 26.961\\
3 & 70.670 & 1.577 & 16.675\\
4 & 56.178 & 1.644 & 15.241\\
5 & 44.539 & 1.726 & 11.437\\
6 & 37.249 & 1.683 & 9.370\\
7 & 33.128 & 1.623 & 8.980\\
8 & 29.450 & 2.032 & 8.100\\
9 & 26.531 & 1.628 & 7.357\\
10 & 25.540 & 1.743 & 8.107\\
11 & 22.625 & 1.817 & 6.616\\
12 & 20.642 & 1.676 & 5.821\\
13 & 19.453 & 1.679 & 5.636\\
14 & 18.350 & 1.732 & 5.394\\
15 & 17.511 & 1.718 & 5.301\\
16 & 16.526 & 1.674 & 4.969\\
17 & 15.924 & 1.768 & 4.943\\
18 & 15.074 & 1.740 & 4.605\\
19 & 14.758 & 1.783 & 4.747\\
20 & 14.081 & 1.758 & 4.483\\
21 & 13.663 & 1.696 & 4.438\\
22 & 13.197 & 1.606 & 4.311\\
23 & 12.743 & 1.555 & 4.167\\
24 & 12.644 & 1.684 & 4.351\\
25 & 12.328 & 1.639 & 4.297\\
26 & 12.274 & 1.662 & 4.484\\
27 & 12.241 & 1.691 & 4.674\\
28 & 12.066 & 1.701 & 4.706\\
29 & 12.042 & 1.702 & 4.875\\
30 & 12.336 & 1.814 & 5.349\\
31 & 12.319 & 1.697 & 5.501\\
32 & 12.592 & 2.006 & 5.932\\
\hline
\end{tabular}
\end{table}

\begin{figure}
\centering{}
\includegraphics[width=0.48\textwidth]{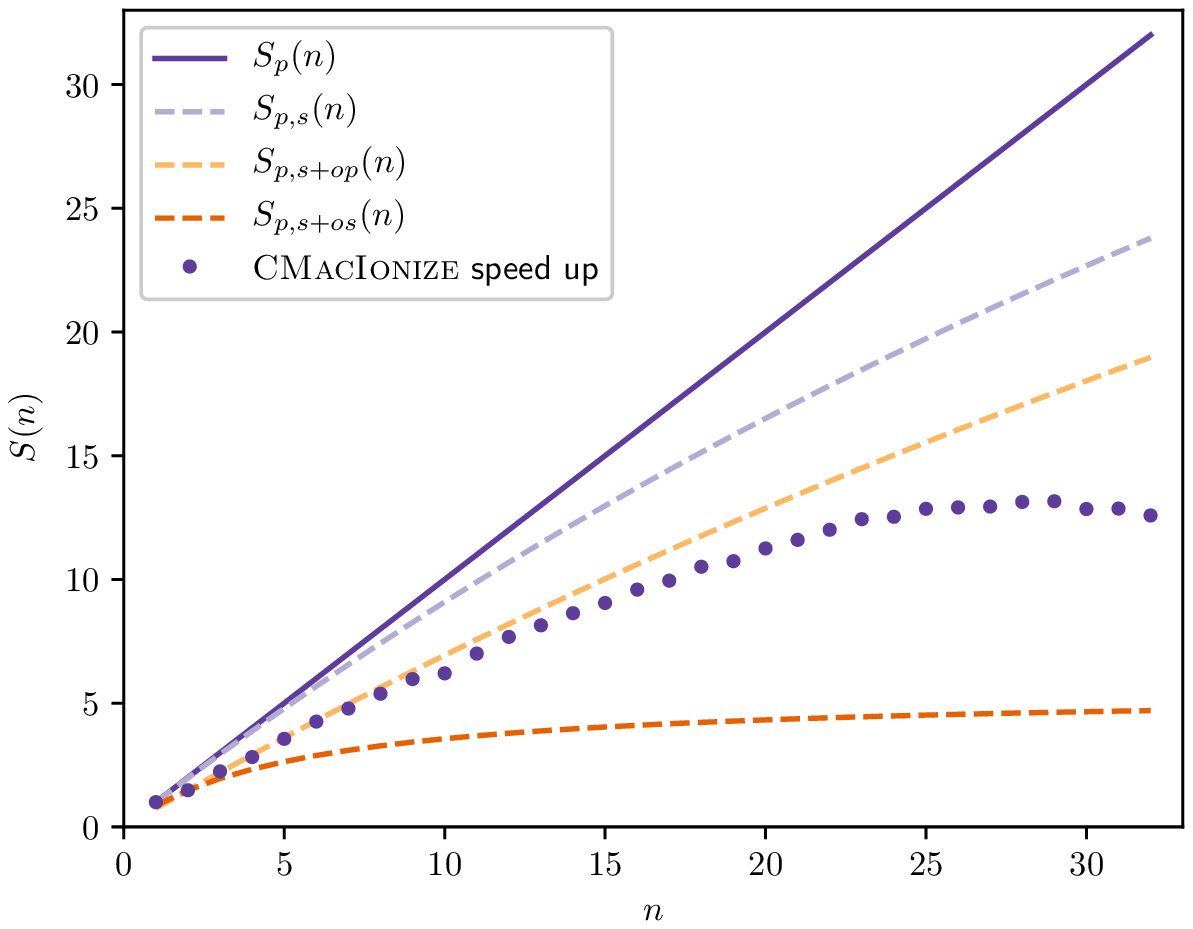}
\caption{Speed up as a function of number of shared memory parallel threads for
the Str\"{o}mgren benchmark test (without diffuse field). The full purple line
shows the theoretical perfect speed up, the dashed purple line is the perfect
speed up taking into account the serial part of the code, as measured from the 1
thread run. The purple dots are the actual code results. In orange we show the
speed up curves for parallel and serial overhead as estimated from the 2 thread
run.
\label{figure:stromgren_scaling_openmp}}
\end{figure}

\tableref{table:stromgren_scaling_openmp} and
\figureref{figure:stromgren_scaling_openmp} show the shared memory scaling
measurements for the default version of the Str\"{o}mgren benchmark test
(without diffuse field). It is immediately obvious that there is a large
overhead in the parallel runs. Comparing the speed up with the two speed up
curves including overhead, we conclude that this overhead is parallel, and hence
shared among the threads.

The most likely cause of the constant overhead is the locking mechanism, since
the number of times a lock is used depends on the number of visited cells and is
hence constant for a given simulation. In the serial run, no locks need to be
set, and there is almost no overhead when locking and unlocking a cell. As soon
as more than 1 thread is used, locking is necessary and the constant overhead
will be added to the simulation time.

If we compare the scaling to $S_{p,s+op}(n)$, the parallel overhead scaling
curve, we see good scaling up to 20 threads. After that, scaling decreases
significantly. This is expected, as the 32 threads have to compete for the 16
available physical cores.

\paragraph{Distributed memory scaling}

\begin{table}
\centering{}
\caption{Timing information for the distributed memory scaling run of the
Str\"{o}mgren benchmark test (without diffuse field) on a system with 32
available cores.
\label{table:stromgren_scaling_mpi}}

\vspace{12pt}
\begin{tabular}{c c c c}
\hline
$n$ & $t(n)$ (s) & $t_s(n)$ (s) & $t_o(n)$ (s) \\
\hline
1 & 160.613 & 1.708 & 0.000\\
2 & 108.900 & 2.192 & 27.740\\
3 & 86.124 & 1.744 & 31.448\\
4 & 73.623 & 1.657 & 32.189\\
5 & 66.784 & 1.732 & 33.295\\
6 & 62.263 & 1.841 & 34.071\\
7 & 60.972 & 2.580 & 36.563\\
8 & 56.100 & 2.048 & 34.529\\
9 & 55.613 & 2.133 & 36.249\\
10 & 54.367 & 2.665 & 36.769\\
11 & 56.128 & 2.949 & 39.974\\
12 & 52.518 & 3.297 & 37.568\\
13 & 54.639 & 1.970 & 40.708\\
14 & 53.979 & 4.033 & 40.921\\
15 & 51.954 & 3.197 & 39.653\\
16 & 50.357 & 2.745 & 38.718\\
17 & 50.973 & 2.532 & 39.918\\
18 & 49.690 & 2.353 & 39.154\\
19 & 49.313 & 2.177 & 39.242\\
20 & 49.338 & 2.310 & 39.685\\
21 & 48.730 & 3.318 & 39.455\\
22 & 48.250 & 3.046 & 39.319\\
23 & 46.816 & 2.259 & 38.199\\
24 & 47.624 & 2.747 & 39.295\\
25 & 47.438 & 2.419 & 39.374\\
26 & 47.331 & 3.552 & 39.511\\
27 & 46.753 & 2.846 & 39.160\\
28 & 45.965 & 2.721 & 38.582\\
29 & 45.232 & 2.563 & 38.045\\
30 & 46.483 & 2.605 & 39.478\\
31 & 44.717 & 2.662 & 37.883\\
32 & 44.706 & 3.822 & 38.032\\
\hline
\end{tabular}
\end{table}

\begin{figure}
\centering{}
\includegraphics[width=0.48\textwidth]{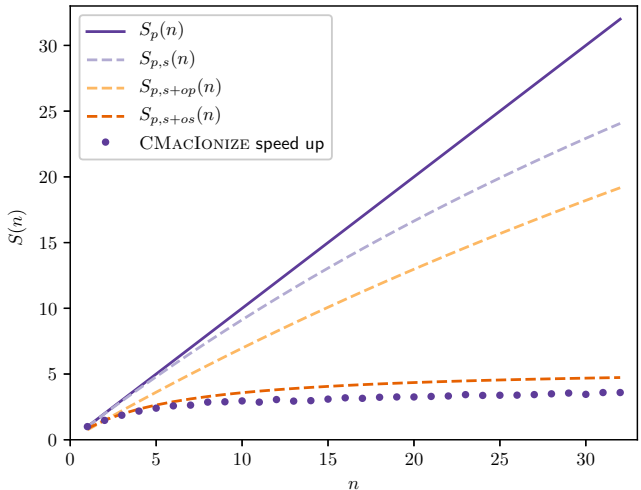}
\caption{Speed up as a function of number of distributed memory parallel
processes for the Str\"{o}mgren benchmark test (without diffuse field). The full
purple line shows the theoretical perfect speed up, the dashed purple line is
the perfect speed up taking into account the serial part of the code, as
measured from the 1 process run. The purple dots are the actual code results. In
orange we show the speed up curves for parallel and serial overhead as estimated
from the 2 processes run.
\label{figure:stromgren_scaling_mpi}}
\end{figure}

\tableref{table:stromgren_scaling_mpi} and
\figureref{figure:stromgren_scaling_mpi} show the distributed memory scaling
measurements for the default version of the Str\"{o}mgren benchmark test
(without diffuse field). Again there is a large overhead, but this time it is a
serial overhead that is almost constant per process. This overhead is due to the
repacking and communication, and due to load imbalances. Although there is a
speed up and hence some gain from using multiple processes, the parallel scaling
is poor.

\paragraph{Time line}

\begin{figure*}
\centering{}
\includegraphics[width=0.98\textwidth]{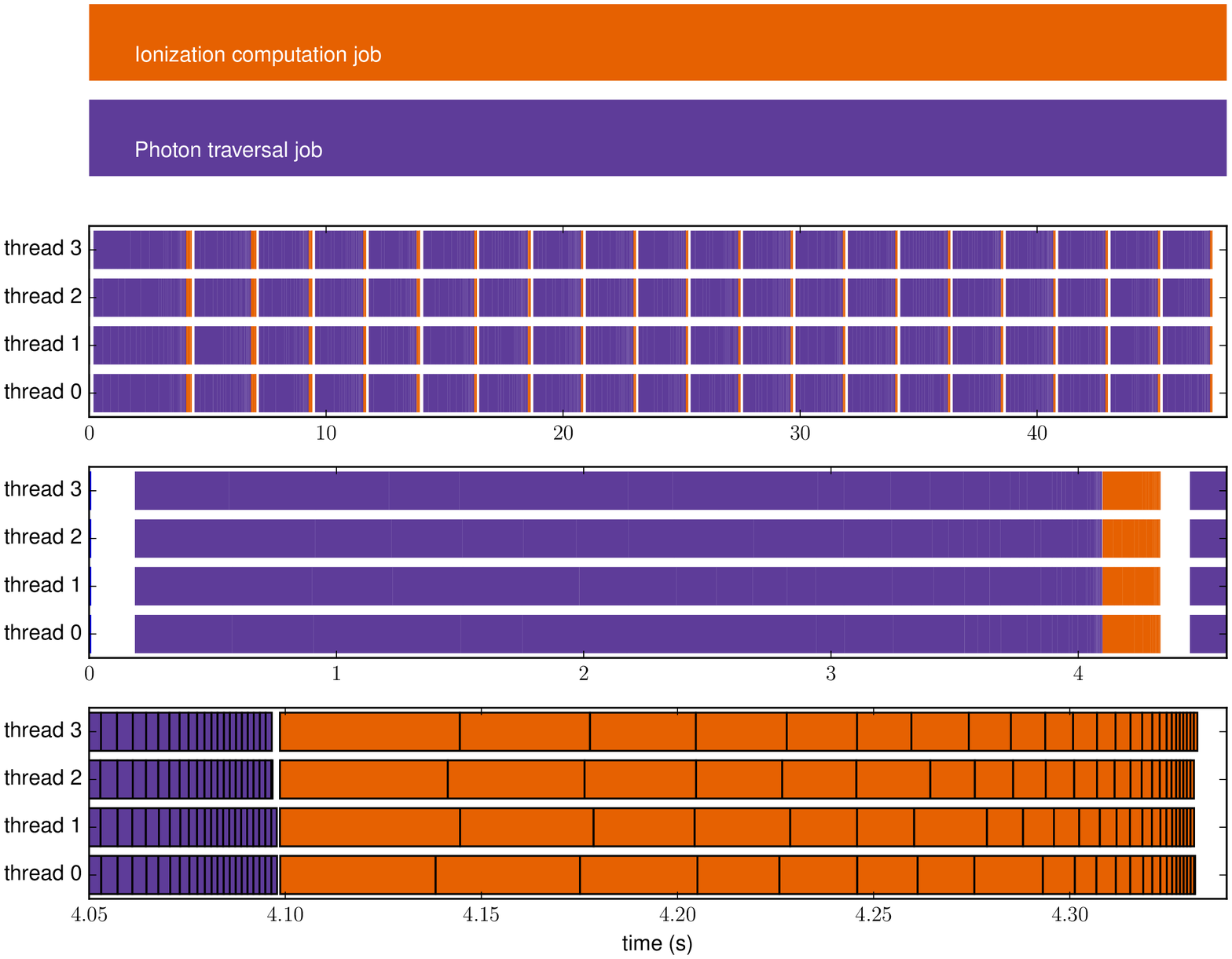}
\caption{Time line of a shared memory parallel run, showing the various tasks
being executed by different threads. \emph{Top}: total time line, \emph{middle}:
zoom on the first photoionization iteration, showing the large serial part at
the start of the simulation and in between subsequent iterations, \emph{bottom}:
zoom on the end of the first iteration, showing the small load imbalance between
the various threads at the end of the different task categories.
\label{figure:jobtimes}}
\end{figure*}

\figureref{figure:jobtimes} shows a time line of the Str\"{o}mgren benchmark
test, run on 4 shared memory threads on the same node. The different coloured
bars represent different tasks being executed, while the whitespace represents
serial parts of the code, or parts where the threads are effectively waiting
until all threads finished a specific type of jobs. Overall, the task based
parallelism works well to reduce load imbalances between different threads.
However, there are still some serial parts of the code that limit the
scalability. This graph also does not show the time threads spend waiting to
acquire locked resources, which can affect the summed total runtime of the
photon traversal task.

\subsubsection{Cooling and heating}

\begin{table}
\centering{}
\caption{Timing information for the shared memory scaling run of the high
temperature Lextington benchmark test on a system with 32 available cores.
\label{table:lexington_scaling}}

\vspace{12pt}
\begin{tabular}{c c c c}
\hline
$n$ & $t(n)$ (s) & $t_s(n)$ (s) & $t_o(n)$ (s) \\
\hline
1 & 383.991 & 2.050 & 0.000\\
2 & 206.586 & 1.694 & 13.566\\
3 & 137.661 & 1.552 & 8.297\\
4 & 105.118 & 1.635 & 7.583\\
5 & 83.918 & 1.710 & 5.480\\
6 & 70.396 & 1.666 & 4.689\\
7 & 60.710 & 1.674 & 4.097\\
8 & 53.373 & 1.716 & 3.580\\
9 & 47.803 & 1.756 & 3.315\\
10 & 43.367 & 1.830 & 3.123\\
11 & 39.631 & 1.819 & 2.859\\
12 & 36.558 & 1.775 & 2.680\\
13 & 33.864 & 1.753 & 2.434\\
14 & 31.836 & 1.678 & 2.505\\
15 & 29.818 & 1.722 & 2.305\\
16 & 28.278 & 1.772 & 2.357\\
17 & 26.602 & 1.813 & 2.085\\
18 & 25.422 & 1.976 & 2.153\\
19 & 24.243 & 1.863 & 2.091\\
20 & 23.183 & 1.900 & 2.036\\
21 & 22.158 & 1.745 & 1.920\\
22 & 21.354 & 1.766 & 1.943\\
23 & 20.804 & 1.693 & 2.148\\
24 & 20.312 & 1.645 & 2.348\\
25 & 19.954 & 1.737 & 2.626\\
26 & 19.630 & 1.726 & 2.890\\
27 & 19.125 & 1.598 & 2.929\\
28 & 18.883 & 1.711 & 3.192\\
29 & 18.620 & 1.752 & 3.400\\
30 & 18.380 & 1.778 & 3.599\\
31 & 18.363 & 2.031 & 3.992\\
32 & 18.503 & 1.792 & 4.517\\
\hline
\end{tabular}
\end{table}

\begin{figure}
\centering{}
\includegraphics[width=0.48\textwidth]{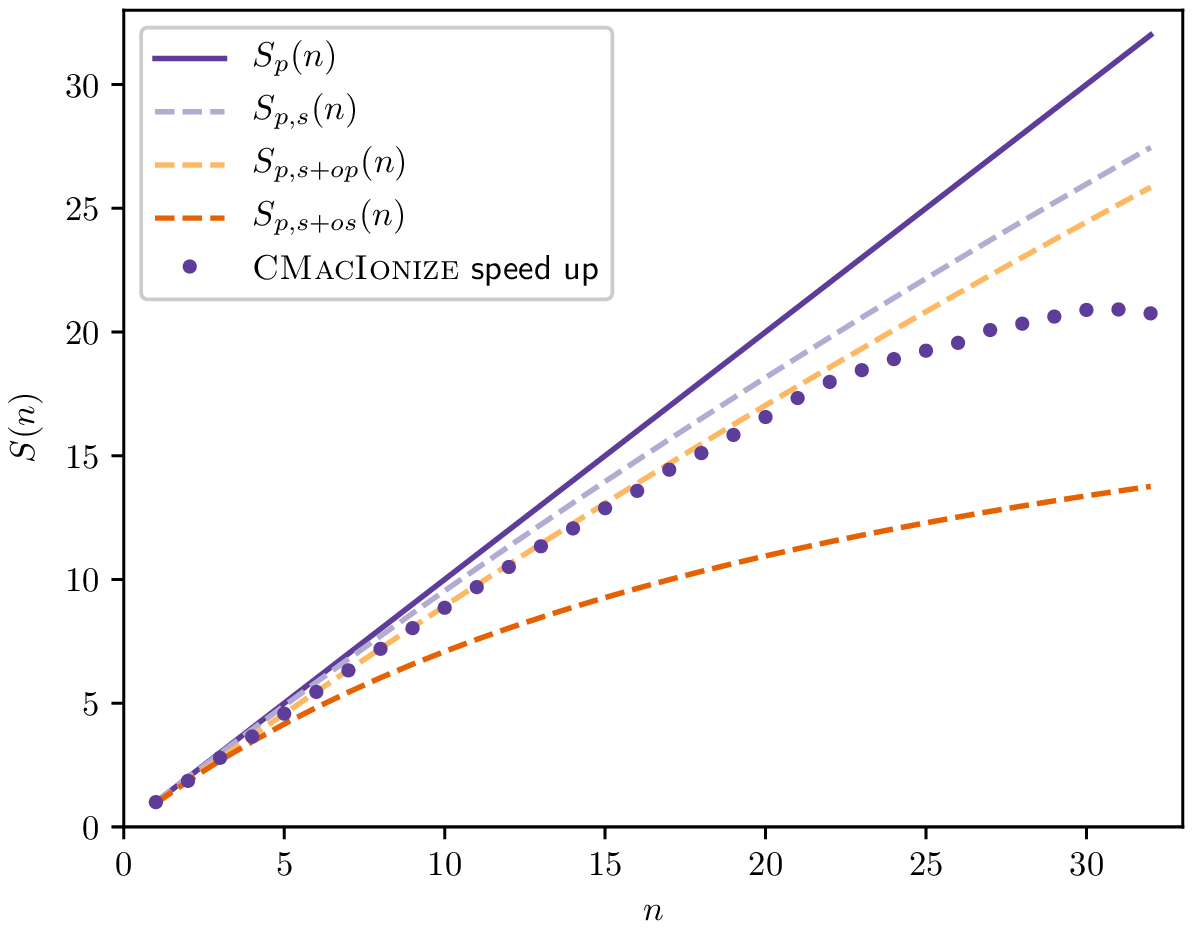}
\caption{Speed up as a function of number of shared memory parallel threads for
the high temperature Lexington benchmark. The full purple line shows the
theoretical perfect speed up, the dashed purple line is the perfect speed up
taking into account the serial part of the code, as measured from the 1 thread
run. The purple dots are the actual code results. In orange we show the speed up
curves for parallel and serial overhead as estimated from the 2 thread run.
\label{figure:lexington_scaling}}
\end{figure}

\tableref{table:lexington_scaling} and \figureref{figure:lexington_scaling} show
the shared memory scaling measurements for the default high temperature
Lexington benchmark test. As in the Str\"{o}mgren run, there is a considerable
overhead due to the locking mechanism, but since we do a lot more work, this
overhead is less noticeable. Overall, the code scales well up to 20 threads.

\subsubsection{Radiation hydrodynamics}

\begin{table}
\centering{}
\caption{Timing information for the shared memory scaling run of the STARBENCH
benchmark test on a system with 32 available cores.
\label{table:starbench_scaling}}

\vspace{12pt}
\begin{tabular}{c c c c}
\hline
$n$ & $t(n)$ (s) & $t_s(n)$ (s) & $t_o(n)$ (s) \\
\hline
1 & 266.779 & 7.688 & 0.000\\
2 & 180.832 & 6.798 & 43.598\\
3 & 127.426 & 7.062 & 33.374\\
4 & 103.211 & 6.752 & 30.750\\
5 & 86.684 & 6.761 & 27.178\\
6 & 77.835 & 6.742 & 26.965\\
7 & 70.042 & 7.042 & 25.341\\
8 & 63.489 & 6.756 & 23.415\\
9 & 61.477 & 7.031 & 25.001\\
10 & 61.021 & 6.711 & 27.424\\
11 & 59.448 & 6.948 & 28.206\\
12 & 58.676 & 6.677 & 29.397\\
13 & 58.356 & 7.043 & 30.738\\
14 & 58.028 & 6.808 & 31.833\\
15 & 58.679 & 6.795 & 33.719\\
16 & 57.233 & 6.716 & 33.352\\
17 & 57.437 & 6.774 & 34.508\\
18 & 57.136 & 7.279 & 35.054\\
19 & 59.106 & 6.672 & 37.782\\
20 & 57.044 & 6.697 & 36.402\\
21 & 56.868 & 6.768 & 36.843\\
22 & 57.060 & 6.920 & 37.595\\
23 & 58.660 & 7.086 & 39.707\\
24 & 59.691 & 6.924 & 41.207\\
25 & 56.175 & 6.936 & 38.124\\
26 & 58.387 & 6.700 & 40.734\\
27 & 57.452 & 6.989 & 40.168\\
28 & 58.521 & 6.748 & 41.580\\
29 & 59.500 & 6.834 & 42.878\\
30 & 59.405 & 7.067 & 43.080\\
31 & 58.506 & 7.201 & 42.460\\
32 & 59.636 & 7.227 & 43.851\\
\hline
\end{tabular}
\end{table}

\begin{figure}
\centering{}
\includegraphics[width=0.48\textwidth]{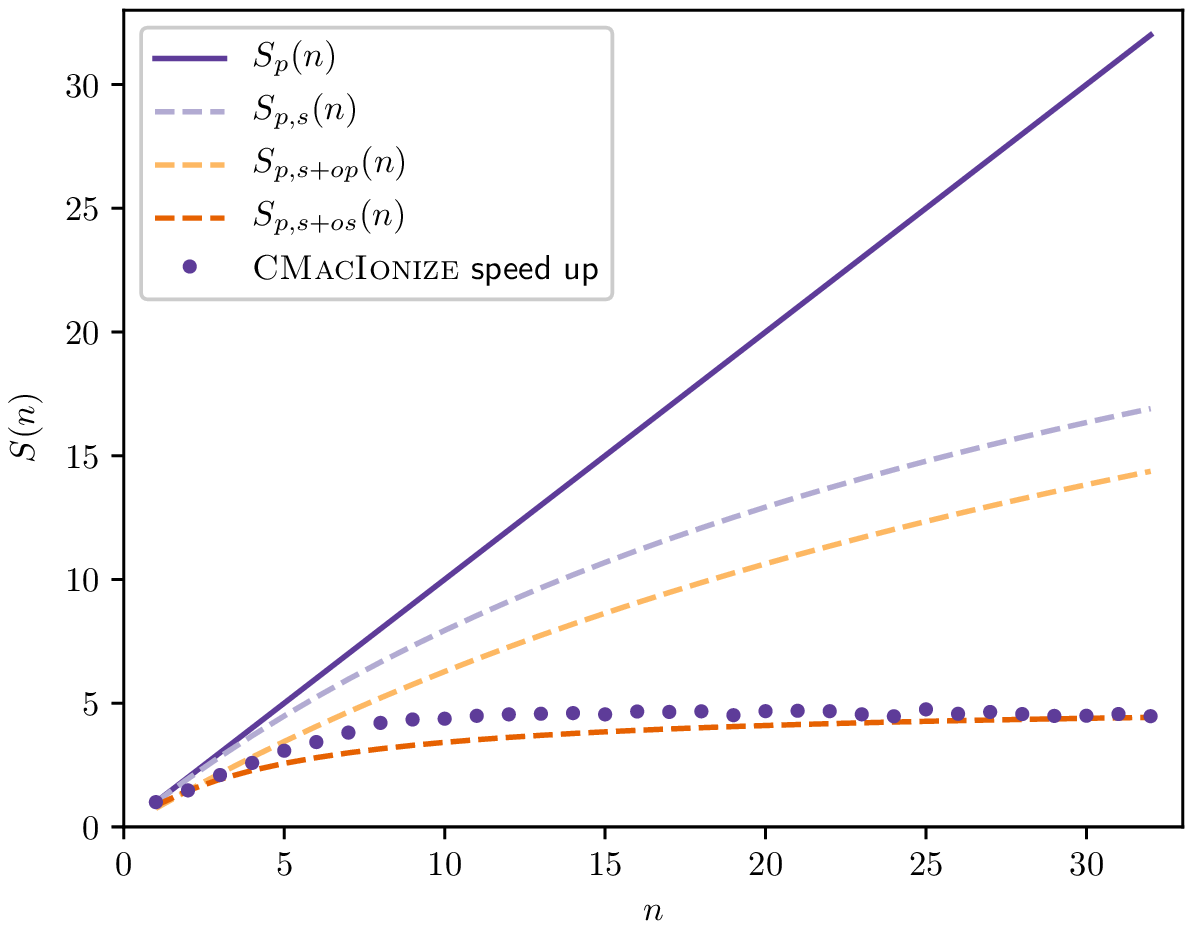}
\caption{Speed up as a function of number of shared memory parallel threads for
the STARBENCH benchmark. The full purple line shows the theoretical perfect
speed up, the dashed purple line is the perfect speed up taking into account the
serial part of the code, as measured from the 1 thread run. The purple dots are
the actual code results. In orange we show the speed up curves for parallel and
serial overhead as estimated from the 2 thread run.
\label{figure:starbench_scaling}}
\end{figure}

\tableref{table:starbench_scaling} and \figureref{figure:starbench_scaling} show
the shared memory scaling measurements for a short (10 time step) version of the
default STARBENCH benchmark test. In this case, there is a significant serial
fraction, and a considerable amount of overhead as well. The overhead is
parallel at first, but switches to serial for high thread number. This is likely
due to hardware issues, as the hydrodynamical integration scheme is more
computation bound, and hence does not benefit from core hyper threading.
Overall, there is still a lot of room for improvement of the scalability.

%%%%%%%%%%%%%%%%%%%%%%%%%%%%%%%%%%%%%%%%%%%%%%%%%%%%%%%%%%%%%%%%%%%%%%%%%%%%%%%%
\section{Conclusion}
%%%%%%%%%%%%%%%%%%%%%%%%%%%%%%%%%%%%%%%%%%%%%%%%%%%%%%%%%%%%%%%%%%%%%%%%%%%%%%%%

We presented the public Monte Carlo photoionization and moving-mesh RHD code
\cmacionize{}, highlighted the implemented physics, and the most important
design considerations during code development. We illustrated the usage and
performance of the code with a number of relevant benchmark tests, and showed
that it produces accurate and reproducable scientific results.

This work accompanies the first official release of the code (\cmacionize{}
1.0), and describes the code as it is for this version. A number of important
improvements to the code are already on the way:
\begin{itemize}
  \item{} implementation of a second order hydrodynamics solver,
  \item{} implementation of a distributed memory domain decomposition, using a
  scheme similar to \citet{2015Harries}, which would enable us to overcome
  current memory limitations that make it impossible to run large grids,
  \item{} optimization and full implementation of an alternative Voronoi grid
  construction algorithm (Vandenbroucke \emph{et al.}, \emph{in prep.}), and
  \item{} implementation of more extensive atomic data using the \textsc{Stout}
  database \citep{2015Lykins}.
\end{itemize}
These will be part of future code releases.

%%%%%%%%%%%%%%%%%%%%%%%%%%%%%%%%%%%%%%%%%%%%%%%%%%%%%%%%%%%%%%%%%%%%%%%%%%%%%%%%
\section*{Acknowledgements}
%%%%%%%%%%%%%%%%%%%%%%%%%%%%%%%%%%%%%%%%%%%%%%%%%%%%%%%%%%%%%%%%%%%%%%%%%%%%%%%%

We want to thank the anonymous referee for a positive and constructive report
that helped clarify some key points in the paper.
We acknowledge support from STFC grant ST/M001296/1. We would like to thank Maya
Petkova for trying out the code and coupling it to \textsc{PHANTOM}, Daniel
Hanaway for suffering through the complications of trying to get the code to
work under Windows, and Diego Gon\c{c}alves and Nina Sartorio for useful
discussions about the coupling between radiation and hydrodynamics. Many thanks
as well to Pedro Gonnet and Matthieu Schaller for their constructive input about
task based parallelism; the latter also for advocating the use of a linear scale
in scaling plots.

%%%%%%%%%%%%%%%%%%%%%%%%%%%%%%%%%%%%%%%%%%%%%%%%%%%%%%%%%%%%%%%%%%%%%%%%%%%%%%%%
\section*{References}
%%%%%%%%%%%%%%%%%%%%%%%%%%%%%%%%%%%%%%%%%%%%%%%%%%%%%%%%%%%%%%%%%%%%%%%%%%%%%%%%

\bibliography{main}{}
\bibliographystyle{elsarticle-harv}

%%%%%%%%%%%%%%%%%%%%%%%%%%%%%%%%%%%%%%%%%%%%%%%%%%%%%%%%%%%%%%%%%%%%%%%%%%%%%%%%
\appendix{}
%%%%%%%%%%%%%%%%%%%%%%%%%%%%%%%%%%%%%%%%%%%%%%%%%%%%%%%%%%%%%%%%%%%%%%%%%%%%%%%%

\section{Fitting parameters for metal line cooling velocity-averaged collision
strengths}

Below we list the fit parameters to the collision strength data for all elements
with low lying excited levels that are used in the code, grouped together per
element and per ion of that element. The figures show the data points that were
used in the fit, the actual fitting curve (with the same parameter precision as
displayed in the tables), and the relative difference between the data values
and the fitting curve, given by
\begin{equation}
\xi{}_{i \rightarrow{} j} (T) = \frac{\left| d(i \rightarrow{} j, T) - f(i
\rightarrow{} j, T)\right|}{d(i \rightarrow{} j, T) + f(i \rightarrow{} j, T)},
\end{equation}
where $d(i \rightarrow{} j, T)$ and $f(i \rightarrow{} j, T)$ are the data value
and fit at temperature $T$ for the transition from level $i$ to level $j$
respectively.

\begin{table*}
\centering{}
\caption{Fit parameters for the velocity-averaged collision strength data for
carbon used in the code.}
\label{table:collision_strength_fits_carbon}

\vspace{12pt}
\begin{tabular}{c c c c c c c}
\hline
ion & & $0 \rightarrow{} 1$ & $0 \rightarrow{} 2$ & $0 \rightarrow{} 3$ & $0
\rightarrow{} 4$ & $1 \rightarrow{} 2$ \\
& & $1 \rightarrow{} 3$ & $1 \rightarrow{} 4$ & $2 \rightarrow{} 3$ & $2
\rightarrow{} 4$ & $3 \rightarrow{} 4$ \\
\hline
C$^+$ &
$\begin{matrix}
a \\
b \\
c \\
d \\
e \\
f \\
g \\
\end{matrix}$
&
\begin{tabular}{l}-9.519e-01 \\-5.097e+00 \\1.436e+03 \\7.424e-01 \\
-5.391e+01 \\3.673e-06 \\2.806e-07 \\\end{tabular}
&
\begin{tabular}{l}1.247e-01 \\-2.710e-05 \\1.110e-01 \\2.784e-06 \\
1.410e-02 \\-3.198e-08 \\-2.437e-09 \\\end{tabular}
&
\begin{tabular}{l}3.496e-07 \\-2.076e-05 \\3.990e-01 \\2.426e-06 \\
-2.832e+00 \\2.765e-10 \\2.159e-11 \\\end{tabular}
&
\begin{tabular}{l}-2.814e-06 \\-2.407e-05 \\2.474e-01 \\2.906e-06 \\
1.393e+00 \\-6.962e-10 \\-5.438e-11 \\\end{tabular}
&
\begin{tabular}{l}1.949e-01 \\-9.811e-06 \\4.074e-02 \\1.003e-06 \\
7.165e-03 \\-2.276e-08 \\-1.738e-09 \\\end{tabular}
\vspace{12pt}
\\
&
$\begin{matrix}
a \\
b \\
c \\
d \\
e \\
f \\
g \\
\end{matrix}$
&
\begin{tabular}{l}1.449e-01 \\-3.673e-05 \\1.713e-01 \\3.763e-06 \\
-9.812e-03 \\6.304e-08 \\4.812e-09 \\\end{tabular}
&
\begin{tabular}{l}-2.743e-06 \\-1.291e-04 \\1.183e+00 \\1.437e-05 \\
1.598e+00 \\-2.263e-09 \\-1.753e-10 \\\end{tabular}
&
\begin{tabular}{l}-1.195e-03 \\-2.027e-04 \\6.479e-01 \\2.660e-05 \\
-7.636e-01 \\1.257e-08 \\9.795e-10 \\\end{tabular}
&
\begin{tabular}{l}-4.787e-01 \\-1.490e-03 \\1.501e+01 \\9.750e-04 \\
3.103e+00 \\-2.437e-07 \\-1.908e-08 \\\end{tabular}
&
\begin{tabular}{l}-2.430e-03 \\-3.839e-04 \\1.605e+00 \\5.173e-05 \\
-2.446e-01 \\8.372e-08 \\6.553e-09 \\\end{tabular}
\vspace{12pt}
\\
C$^{++}$ &
$\begin{matrix}
a \\
b \\
c \\
d \\
e \\
f \\
g \\
\end{matrix}$
&
\begin{tabular}{l}1.265e-07 \\5.762e-06 \\1.080e-01 \\-5.626e-07 \\
1.412e+00 \\1.950e-11 \\1.323e-12 \\\end{tabular}
&
\begin{tabular}{l}3.793e-07 \\1.727e-05 \\3.239e-01 \\-1.686e-06 \\
1.095e+00 \\7.533e-11 \\5.111e-12 \\\end{tabular}
&
\begin{tabular}{l}3.942e-07 \\2.877e-05 \\5.399e-01 \\-2.809e-06 \\
6.095e-01 \\2.255e-10 \\1.530e-11 \\\end{tabular}
&
\begin{tabular}{l}8.424e-06 \\7.004e-05 \\3.767e+00 \\-5.480e-06 \\
-4.018e-01 \\-4.228e-10 \\-2.906e-11 \\\end{tabular}
&
\begin{tabular}{l}5.033e-06 \\1.611e-04 \\6.600e-01 \\-1.477e-05 \\
3.546e+00 \\1.577e-10 \\1.063e-11 \\\end{tabular}
\vspace{12pt}
\\
&
$\begin{matrix}
a \\
b \\
c \\
d \\
e \\
f \\
g \\
\end{matrix}$
&
\begin{tabular}{l}-6.251e-05 \\3.034e-04 \\1.039e-01 \\-2.744e-05 \\
2.856e-01 \\3.444e-09 \\2.317e-10 \\\end{tabular}
&
\begin{tabular}{l}1.741e-06 \\-1.148e-05 \\4.633e-01 \\7.853e-07 \\
-4.244e+00 \\-5.205e-12 \\-3.781e-13 \\\end{tabular}
&
\begin{tabular}{l}1.932e-05 \\8.772e-04 \\1.032e+00 \\-7.960e-05 \\
3.735e+00 \\7.737e-10 \\5.206e-11 \\\end{tabular}
&
\begin{tabular}{l}3.777e-04 \\-3.470e-05 \\1.386e+00 \\2.382e-06 \\
4.838e-02 \\1.336e-09 \\9.718e-11 \\\end{tabular}
&
\begin{tabular}{l}7.721e-06 \\-5.752e-05 \\2.317e+00 \\3.938e-06 \\
5.449e-01 \\2.013e-10 \\1.463e-11 \\\end{tabular}
\\
\hline
\end{tabular}
\end{table*}

\begin{figure*}
\centering{}
\includegraphics[width=0.98\textwidth]{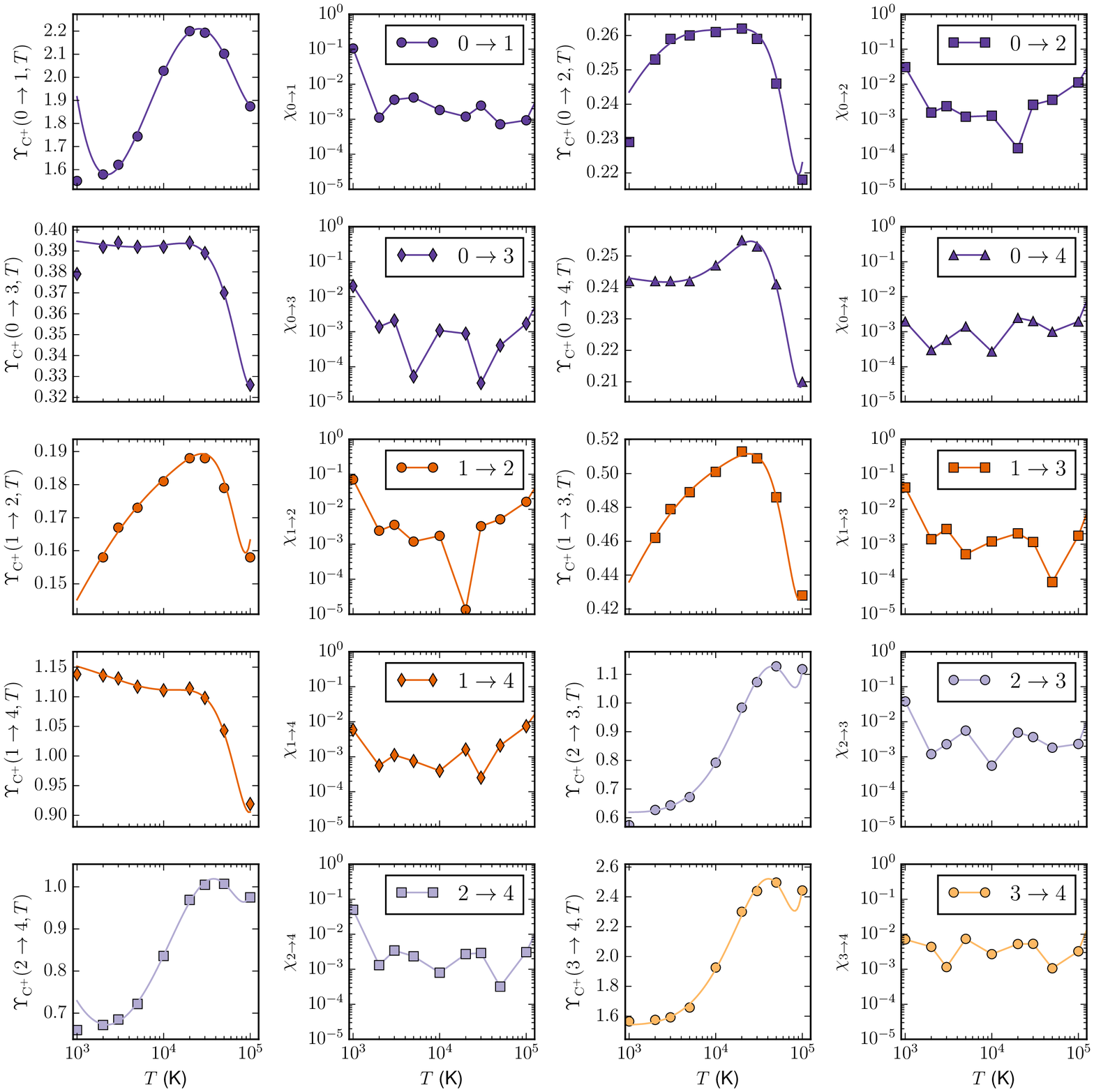}
\caption{Fits for C$^{+}$.
\label{figure:CII_fit}}
\end{figure*}

\begin{figure*}
\centering{}
\includegraphics[width=0.98\textwidth]{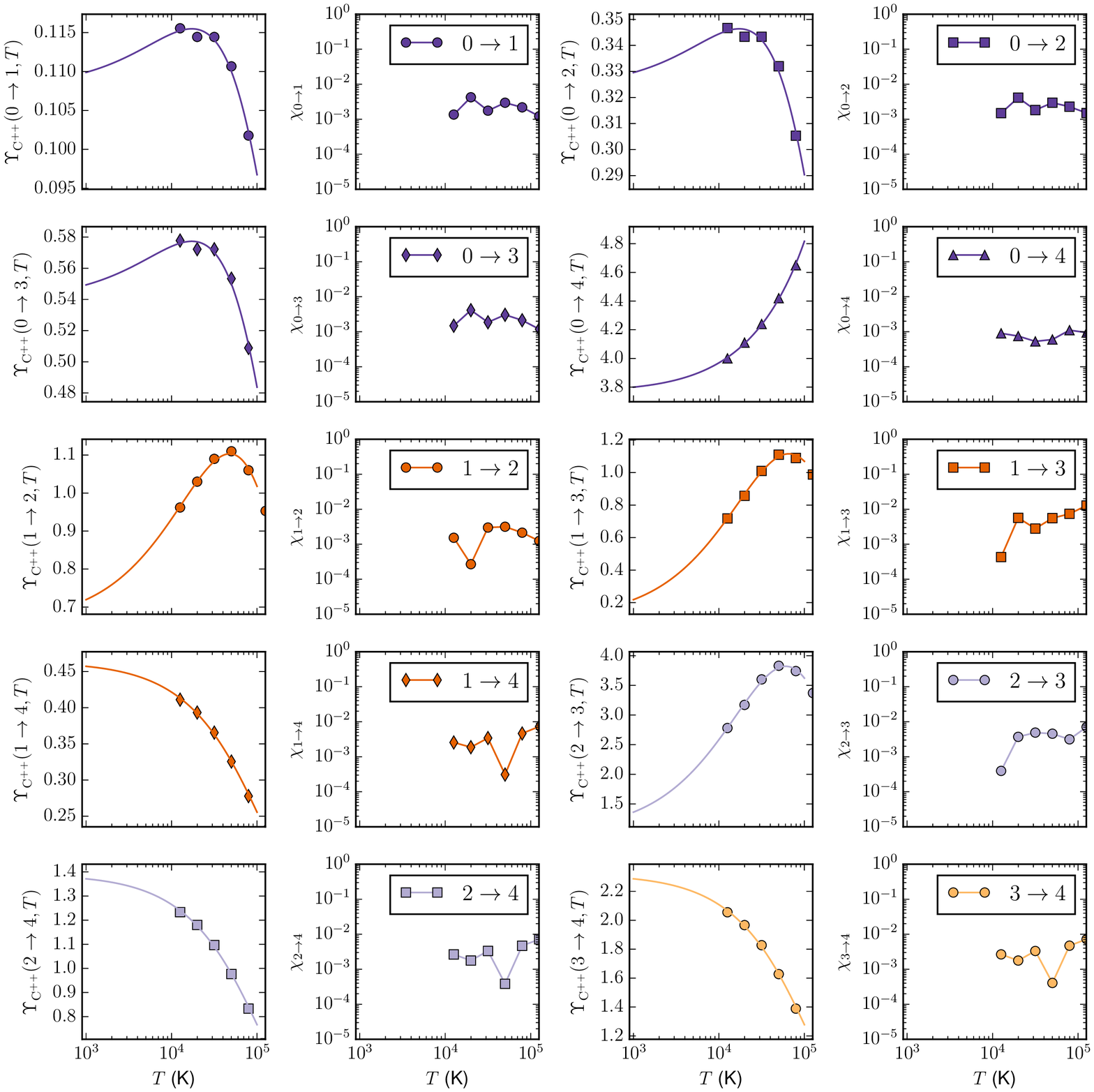}
\caption{Fits for C$^{++}$.
\label{figure:CIII_fit}}
\end{figure*}

\begin{table*}
\centering{}
\caption{Fit parameters for the velocity-averaged collision strength data for
nitrogen used in the code.}
\label{table:collision_strength_fits_nitrogen}

\vspace{12pt}
\begin{tabular}{c c c c c c c}
\hline
ion & & $0 \rightarrow{} 1$ & $0 \rightarrow{} 2$ & $0 \rightarrow{} 3$ & $0
\rightarrow{} 4$ & $1 \rightarrow{} 2$ \\
& & $1 \rightarrow{} 3$ & $1 \rightarrow{} 4$ & $2 \rightarrow{} 3$ & $2
\rightarrow{} 4$ & $3 \rightarrow{} 4$ \\
\hline
N$^{0}$ &
$\begin{matrix}
a \\
b \\
c \\
d \\
e \\
f \\
g \\
\end{matrix}$
&
\begin{tabular}{l}4.350e-04 \\-4.500e-04 \\2.206e-01 \\4.958e-05 \\
7.786e-02 \\-1.099e-07 \\-8.271e-09 \\\end{tabular}
&
\begin{tabular}{l}2.751e-06 \\-3.007e-04 \\1.474e-01 \\3.312e-05 \\
3.882e-01 \\-1.470e-08 \\-1.106e-09 \\\end{tabular}
&
\begin{tabular}{l}2.254e-07 \\-3.336e-05 \\2.277e-02 \\3.636e-06 \\
8.211e-02 \\-5.171e-09 \\-3.695e-10 \\\end{tabular}
&
\begin{tabular}{l}6.567e-06 \\-6.814e-05 \\4.632e-02 \\7.429e-06 \\
5.470e-02 \\-1.612e-08 \\-1.155e-09 \\\end{tabular}
&
\begin{tabular}{l}7.907e-02 \\-8.603e-03 \\6.355e+00 \\9.366e-04 \\
1.259e-01 \\-1.315e-06 \\-9.950e-08 \\\end{tabular}
\vspace{12pt}
\\
&
$\begin{matrix}
a \\
b \\
c \\
d \\
e \\
f \\
g \\
\end{matrix}$
&
\begin{tabular}{l}-1.420e-02 \\-8.743e-03 \\3.581e+00 \\9.893e-04 \\
8.734e-01 \\-2.580e-07 \\-1.983e-08 \\\end{tabular}
&
\begin{tabular}{l}-3.258e-01 \\-1.290e+00 \\5.455e+02 \\1.451e-01 \\
-1.122e+01 \\2.792e-06 \\2.138e-07 \\\end{tabular}
&
\begin{tabular}{l}-3.592e-01 \\-9.058e-01 \\3.817e+02 \\1.019e-01 \\
-1.055e+01 \\2.096e-06 \\1.606e-07 \\\end{tabular}
&
\begin{tabular}{l}-1.910e-01 \\-9.570e-02 \\4.070e+01 \\1.077e-02 \\
-5.010e+00 \\4.665e-07 \\3.574e-08 \\\end{tabular}
&
\begin{tabular}{l}-2.313e-03 \\-2.099e-02 \\1.122e+01 \\2.337e-03 \\
2.240e+00 \\-2.197e-07 \\-1.682e-08 \\\end{tabular}
\vspace{12pt}
\\
N$^+$ &
$\begin{matrix}
a \\
b \\
c \\
d \\
e \\
f \\
g \\
\end{matrix}$
&
\begin{tabular}{l}-3.483e-01 \\4.384e-03 \\2.059e+00 \\-4.032e-04 \\
-3.391e-05 \\-1.192e-03 \\-8.966e-05 \\\end{tabular}
&
\begin{tabular}{l}-2.779e-03 \\-2.640e-06 \\2.127e-01 \\1.368e-06 \\
-7.504e-04 \\1.570e-06 \\1.229e-07 \\\end{tabular}
&
\begin{tabular}{l}-1.472e-02 \\2.665e-05 \\2.965e-01 \\-2.560e-06 \\
1.796e-07 \\1.600e-03 \\1.209e-04 \\\end{tabular}
&
\begin{tabular}{l}2.426e-03 \\-1.170e-07 \\3.066e-02 \\2.803e-08 \\
2.259e-09 \\-2.924e-03 \\-2.031e-04 \\\end{tabular}
&
\begin{tabular}{l}-1.015e-01 \\6.264e-04 \\1.559e+00 \\-5.445e-05 \\
1.428e-06 \\5.929e-04 \\3.183e-05 \\\end{tabular}
\vspace{12pt}
\\
&
$\begin{matrix}
a \\
b \\
c \\
d \\
e \\
f \\
g \\
\end{matrix}$
&
\begin{tabular}{l}-1.472e-02 \\7.995e-05 \\8.894e-01 \\-7.680e-06 \\
5.191e-07 \\1.661e-03 \\1.255e-04 \\\end{tabular}
&
\begin{tabular}{l}2.426e-03 \\-3.510e-07 \\9.197e-02 \\8.408e-08 \\
7.197e-09 \\-2.753e-03 \\-1.912e-04 \\\end{tabular}
&
\begin{tabular}{l}-1.473e-02 \\1.333e-04 \\1.482e+00 \\-1.280e-05 \\
-1.036e-06 \\-1.390e-03 \\-1.048e-04 \\\end{tabular}
&
\begin{tabular}{l}2.425e-03 \\-5.846e-07 \\1.533e-01 \\1.401e-07 \\
-1.952e-07 \\1.690e-04 \\1.175e-05 \\\end{tabular}
&
\begin{tabular}{l}2.226e-02 \\-3.918e-04 \\1.121e+00 \\3.934e-05 \\
6.245e-06 \\-7.473e-04 \\-5.553e-05 \\\end{tabular}
\vspace{12pt}
\\
N$^{++}$ &
$\begin{matrix}
a \\
b \\
c \\
d \\
e \\
f \\
g \\
\end{matrix}$
&
\begin{tabular}{l}-3.257e-02 \\-3.573e-04 \\1.653e+00 \\4.822e-05 \\
-7.529e-06 \\3.018e-03 \\2.453e-04 \\\end{tabular}
& & & \\
\hline
\end{tabular}
\end{table*}

\begin{figure*}
\centering{}
\includegraphics[width=0.98\textwidth]{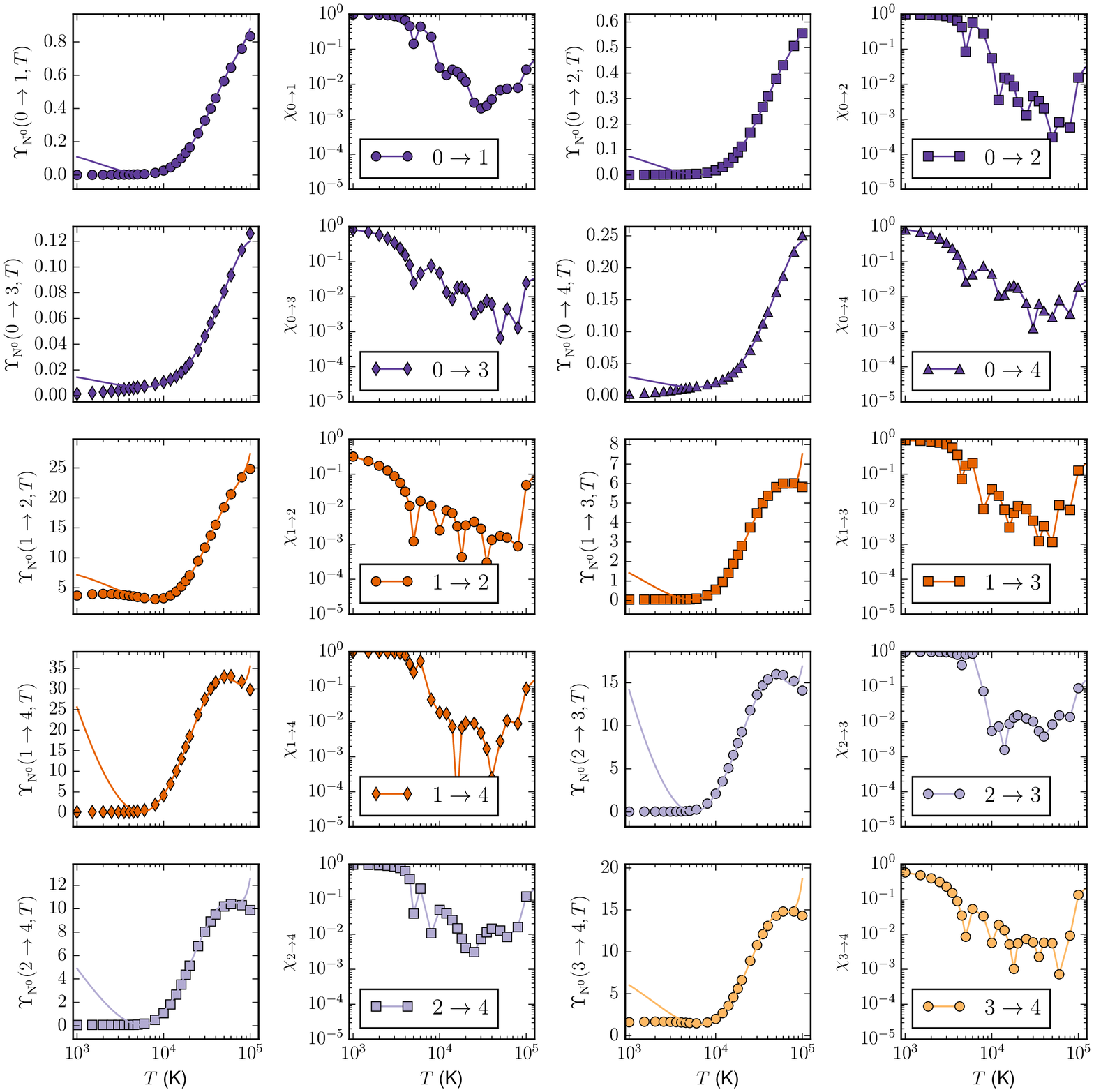}
\caption{Fits for N$^{0}$.
\label{figure:NI_fit}}
\end{figure*}

\begin{figure*}
\centering{}
\includegraphics[width=0.98\textwidth]{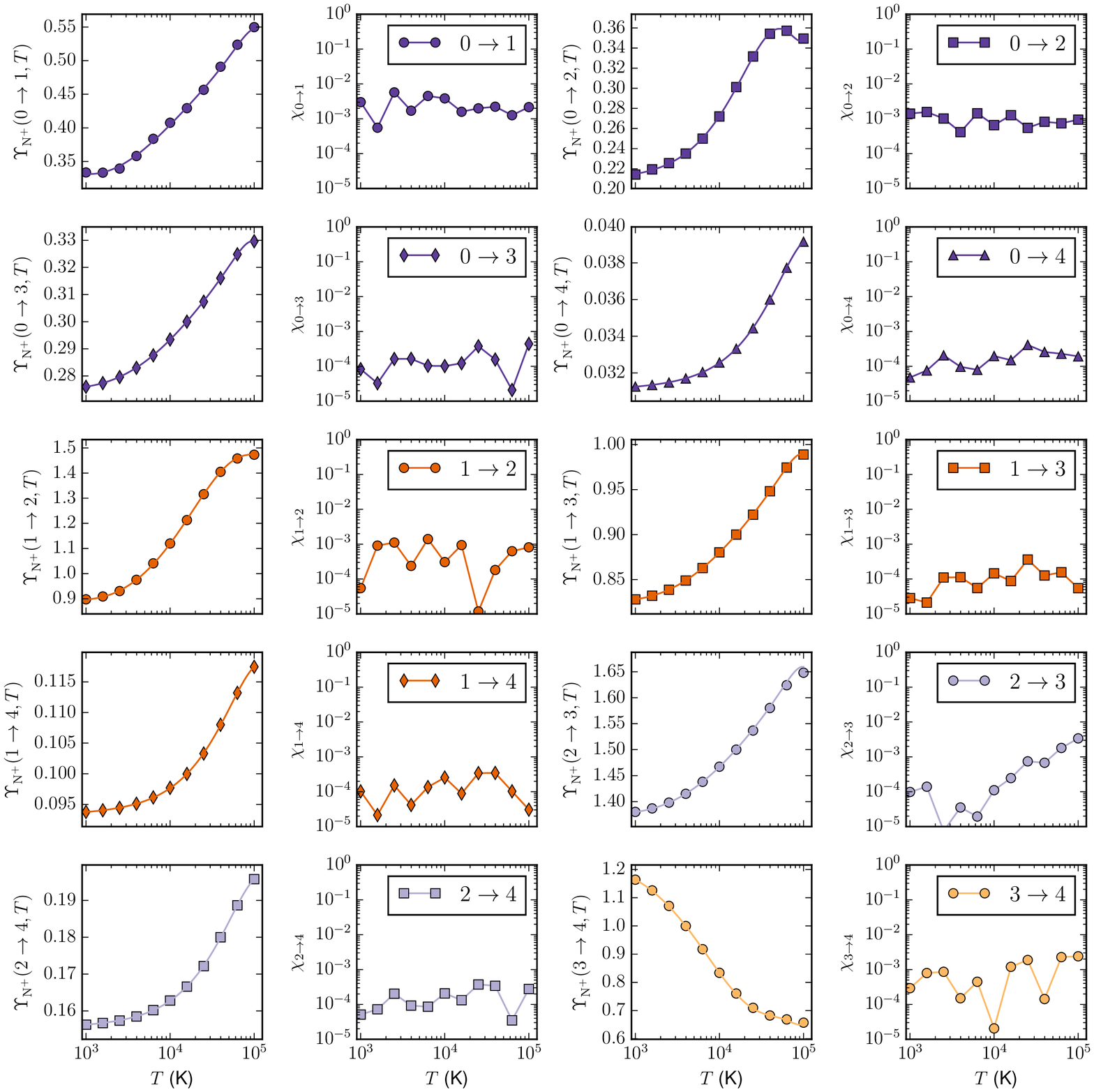}
\caption{Fits for N$^{+}$.
\label{figure:NII_fit}}
\end{figure*}

\begin{figure}
\centering{}
\includegraphics[width=0.48\textwidth]{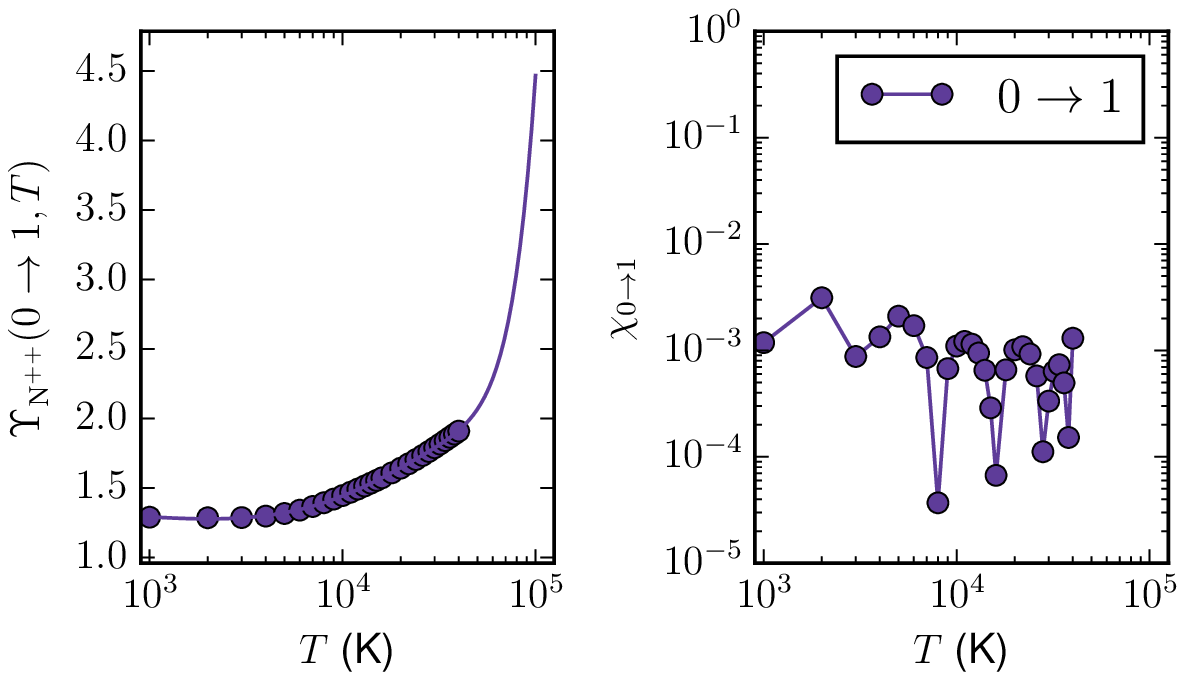}
\caption{Fits for N$^{++}$.
\label{figure:NIII_fit}}
\end{figure}

\begin{table*}
\centering{}
\caption{Fit parameters for the velocity-averaged collision strength data for
oxygen used in the code.}
\label{table:collision_strength_fits_oxygen}

\vspace{12pt}
\begin{tabular}{c c c c c c c}
\hline
ion & & $0 \rightarrow{} 1$ & $0 \rightarrow{} 2$ & $0 \rightarrow{} 3$ & $0
\rightarrow{} 4$ & $1 \rightarrow{} 2$ \\
& & $1 \rightarrow{} 3$ & $1 \rightarrow{} 4$ & $2 \rightarrow{} 3$ & $2
\rightarrow{} 4$ & $3 \rightarrow{} 4$ \\
\hline
O$^{0}$ &
$\begin{matrix}
a \\
b \\
c \\
d \\
e \\
f \\
g \\
\end{matrix}$
&
\begin{tabular}{l}-8.839e-01 \\-6.783e-03 \\7.448e-02 \\1.578e-03 \\
3.239e-01 \\-2.549e-05 \\-3.727e-06 \\\end{tabular}
&
\begin{tabular}{l}-7.697e-01 \\-9.788e-04 \\1.200e-02 \\2.477e-04 \\
9.490e-01 \\-4.462e-07 \\-7.755e-08 \\\end{tabular}
&
\begin{tabular}{l}-1.093e+00 \\1.413e+00 \\-1.373e+02 \\-1.998e-01 \\
9.365e+00 \\3.197e-05 \\2.514e-06 \\\end{tabular}
&
\begin{tabular}{l}-1.219e+00 \\6.529e-01 \\-5.311e+01 \\-9.393e-02 \\
4.506e+00 \\2.706e-05 \\2.088e-06 \\\end{tabular}
&
\begin{tabular}{l}-1.016e+00 \\-6.509e-03 \\6.884e-02 \\1.457e-03 \\
1.015e+00 \\-1.947e-05 \\-2.390e-06 \\\end{tabular}
\vspace{12pt}
\\
&
$\begin{matrix}
a \\
b \\
c \\
d \\
e \\
f \\
g \\
\end{matrix}$
&
\begin{tabular}{l}-1.093e+00 \\8.480e-01 \\-8.239e+01 \\-1.199e-01 \\
3.496e+00 \\5.140e-05 \\4.042e-06 \\\end{tabular}
&
\begin{tabular}{l}-1.219e+00 \\3.917e-01 \\-3.187e+01 \\-5.636e-02 \\
3.302e+00 \\2.216e-05 \\1.709e-06 \\\end{tabular}
&
\begin{tabular}{l}-1.093e+00 \\2.827e-01 \\-2.746e+01 \\-3.996e-02 \\
2.681e+00 \\2.234e-05 \\1.757e-06 \\\end{tabular}
&
\begin{tabular}{l}-1.219e+00 \\1.306e-01 \\-1.062e+01 \\-1.879e-02 \\
1.964e+00 \\1.242e-05 \\9.578e-07 \\\end{tabular}
&
\begin{tabular}{l}-1.056e+00 \\-3.587e-01 \\3.534e+01 \\4.778e-02 \\
1.295e+00 \\1.089e-05 \\6.545e-07 \\\end{tabular}
\vspace{12pt}
\\
O$^+$ &
$\begin{matrix}
a \\
b \\
c \\
d \\
e \\
f \\
g \\
\end{matrix}$
&
\begin{tabular}{l}-2.523e-03 \\-6.241e-06 \\8.502e-01 \\7.233e-07 \\
1.444e-03 \\9.709e-09 \\1.522e-09 \\\end{tabular}
&
\begin{tabular}{l}-1.125e-02 \\-8.446e-07 \\6.036e-01 \\2.181e-07 \\
-1.635e-04 \\4.311e-08 \\-2.238e-10 \\\end{tabular}
&
\begin{tabular}{l}1.161e-02 \\-4.213e-06 \\2.279e-01 \\4.878e-07 \\
-1.033e-11 \\1.000e+00 \\2.265e+00 \\\end{tabular}
&
\begin{tabular}{l}7.789e-02 \\-1.026e-05 \\7.464e-02 \\1.046e-06 \\
1.431e-07 \\-8.913e-04 \\-6.598e-05 \\\end{tabular}
&
\begin{tabular}{l}-8.588e-01 \\-4.648e+00 \\-3.736e+02 \\-1.190e+00 \\
-1.466e-06 \\-5.766e+06 \\-9.308e-01 \\\end{tabular}
\vspace{12pt}
\\
&
$\begin{matrix}
a \\
b \\
c \\
d \\
e \\
f \\
g \\
\end{matrix}$
&
\begin{tabular}{l}-1.028e-02 \\-7.342e-05 \\8.844e-01 \\9.292e-06 \\
-3.079e-04 \\7.387e-06 \\5.594e-07 \\\end{tabular}
&
\begin{tabular}{l}-5.239e-02 \\-6.736e-06 \\4.666e-01 \\1.976e-06 \\
-8.040e-04 \\1.289e-06 \\9.867e-08 \\\end{tabular}
&
\begin{tabular}{l}-3.494e-02 \\-2.313e-05 \\5.770e-01 \\3.760e-06 \\
-1.210e-03 \\1.098e-06 \\8.364e-08 \\\end{tabular}
&
\begin{tabular}{l}1.038e-02 \\-3.425e-05 \\2.996e-01 \\4.003e-06 \\
-2.143e-04 \\3.819e-06 \\2.881e-07 \\\end{tabular}
&
\begin{tabular}{l}6.465e-02 \\-3.643e-05 \\1.838e-01 \\3.856e-06 \\
-1.932e-07 \\2.875e-03 \\2.147e-04 \\\end{tabular}
\vspace{12pt}
\\
O$^{++}$ &
$\begin{matrix}
a \\
b \\
c \\
d \\
e \\
f \\
g \\
\end{matrix}$
&
\begin{tabular}{l}-2.517e-01 \\2.326e-03 \\2.174e+00 \\-2.246e-04 \\
-4.349e-03 \\-5.803e-06 \\-4.351e-07 \\\end{tabular}
&
\begin{tabular}{l}-2.886e-01 \\1.462e-03 \\1.373e+00 \\-1.364e-04 \\
6.062e-03 \\2.338e-06 \\1.764e-07 \\\end{tabular}
&
\begin{tabular}{l}-7.878e-01 \\-3.075e-02 \\5.015e+01 \\7.448e-03 \\
-3.189e-03 \\6.917e-04 \\5.198e-05 \\\end{tabular}
&
\begin{tabular}{l}-5.259e-01 \\1.618e-04 \\1.045e+00 \\2.345e-05 \\
-7.038e-04 \\3.326e-05 \\2.552e-06 \\\end{tabular}
&
\begin{tabular}{l}-2.714e-01 \\6.306e-03 \\5.834e+00 \\-5.990e-04 \\
-5.646e-03 \\-1.140e-05 \\-8.562e-07 \\\end{tabular}
\vspace{12pt}
\\
&
$\begin{matrix}
a \\
b \\
c \\
d \\
e \\
f \\
g \\
\end{matrix}$
&
\begin{tabular}{l}-7.879e-01 \\-9.239e-02 \\1.505e+02 \\2.237e-02 \\
1.919e-02 \\-3.453e-04 \\-2.593e-05 \\\end{tabular}
&
\begin{tabular}{l}-5.259e-01 \\4.853e-04 \\3.136e+00 \\7.039e-05 \\
-1.210e-03 \\5.806e-05 \\4.455e-06 \\\end{tabular}
&
\begin{tabular}{l}-7.879e-01 \\-1.541e-01 \\2.510e+02 \\3.730e-02 \\
6.860e-02 \\-1.610e-04 \\-1.209e-05 \\\end{tabular}
&
\begin{tabular}{l}-5.260e-01 \\8.087e-04 \\5.229e+00 \\1.175e-04 \\
-1.729e-03 \\6.777e-05 \\5.200e-06 \\\end{tabular}
&
\begin{tabular}{l}6.242e-02 \\1.702e-04 \\1.766e-01 \\-1.776e-05 \\
-2.314e-03 \\-1.121e-06 \\-8.415e-08 \\\end{tabular}
\\
\hline
\end{tabular}
\end{table*}

\begin{figure*}
\centering{}
\includegraphics[width=0.98\textwidth]{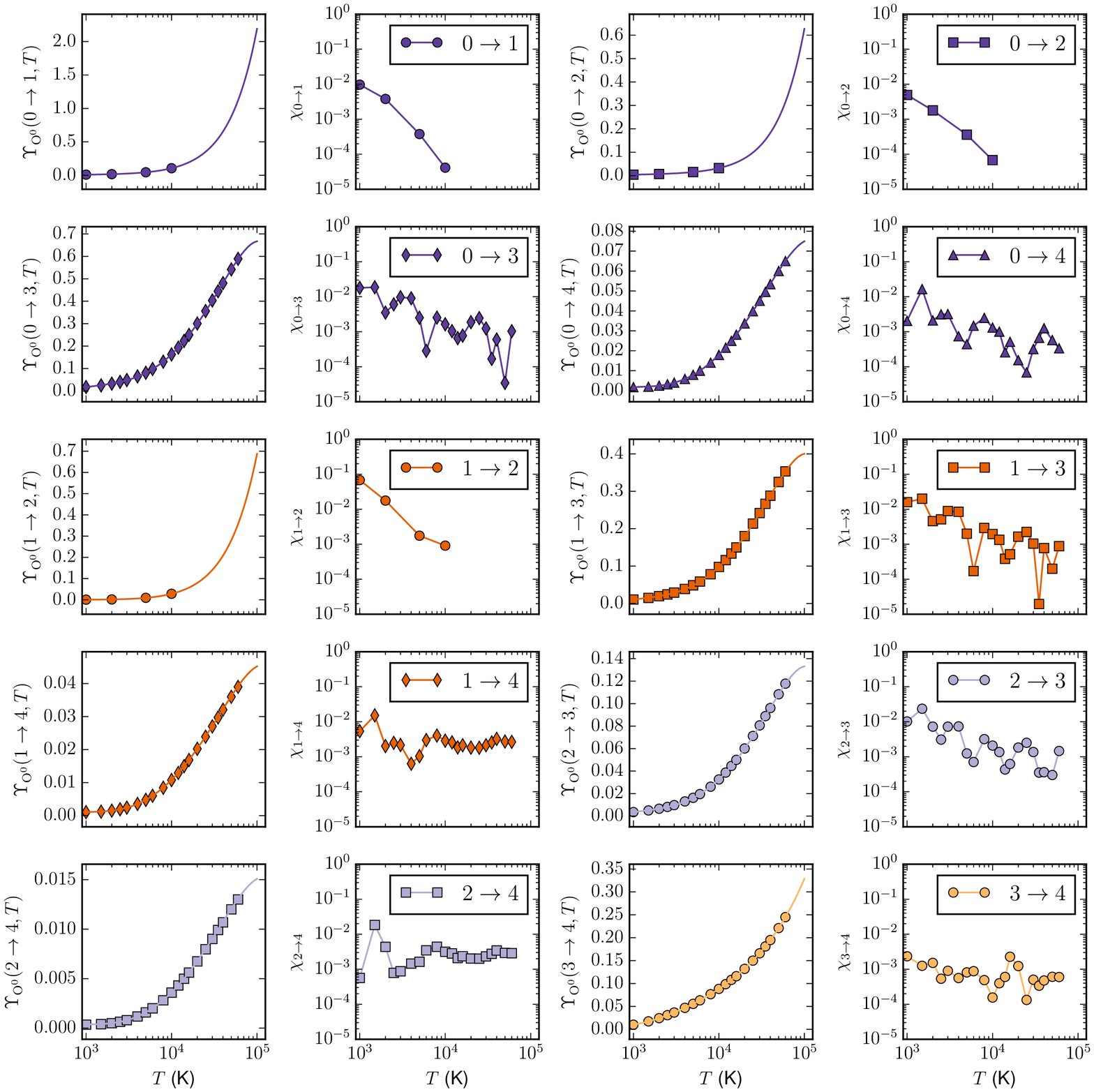}
\caption{Fits for O$^{0}$.
\label{figure:OI_fit}}
\end{figure*}

\begin{figure*}
\centering{}
\includegraphics[width=0.98\textwidth]{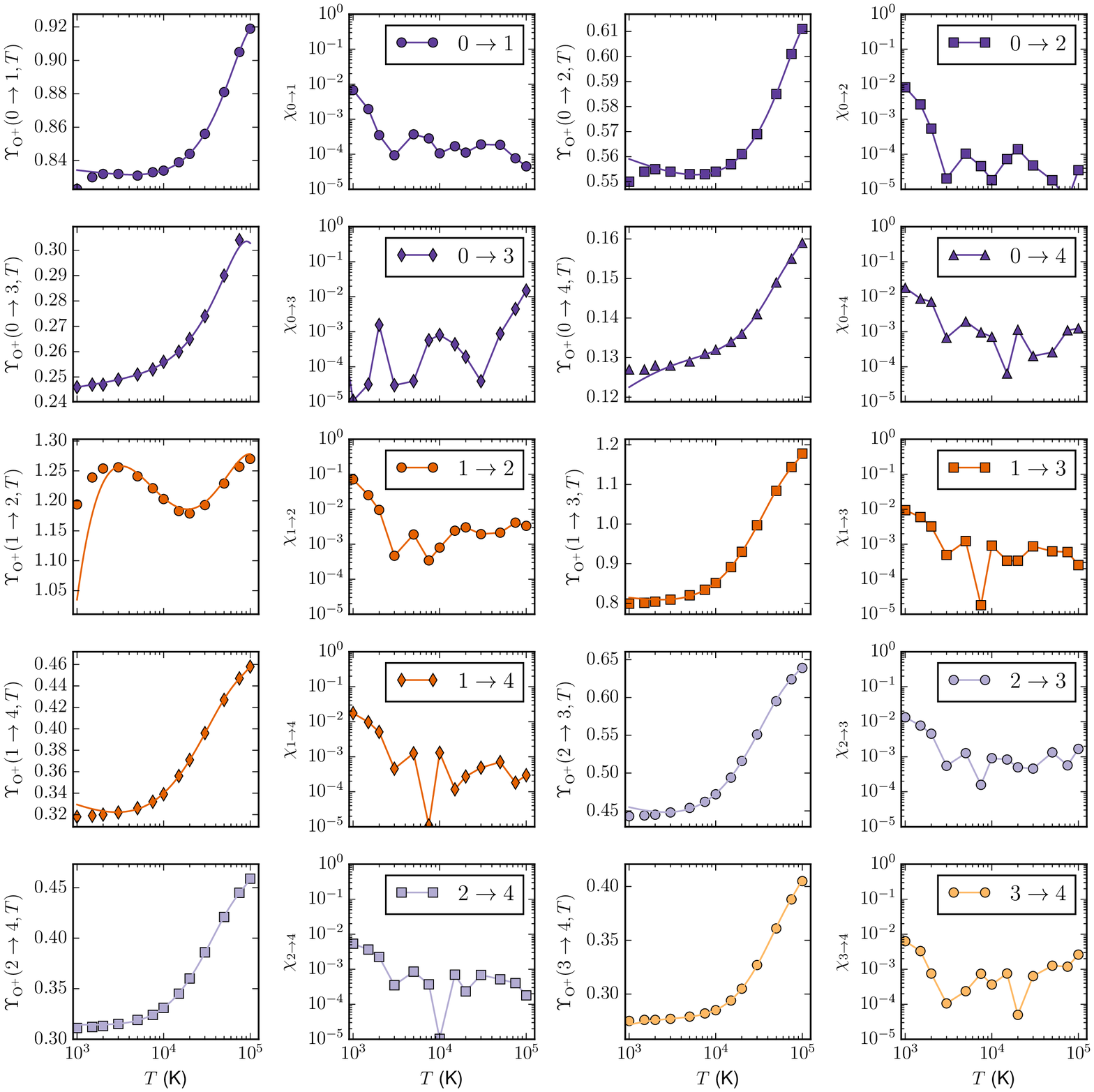}
\caption{Fits for O$^{+}$.
\label{figure:OII_fit}}
\end{figure*}

\begin{figure*}
\centering{}
\includegraphics[width=0.98\textwidth]{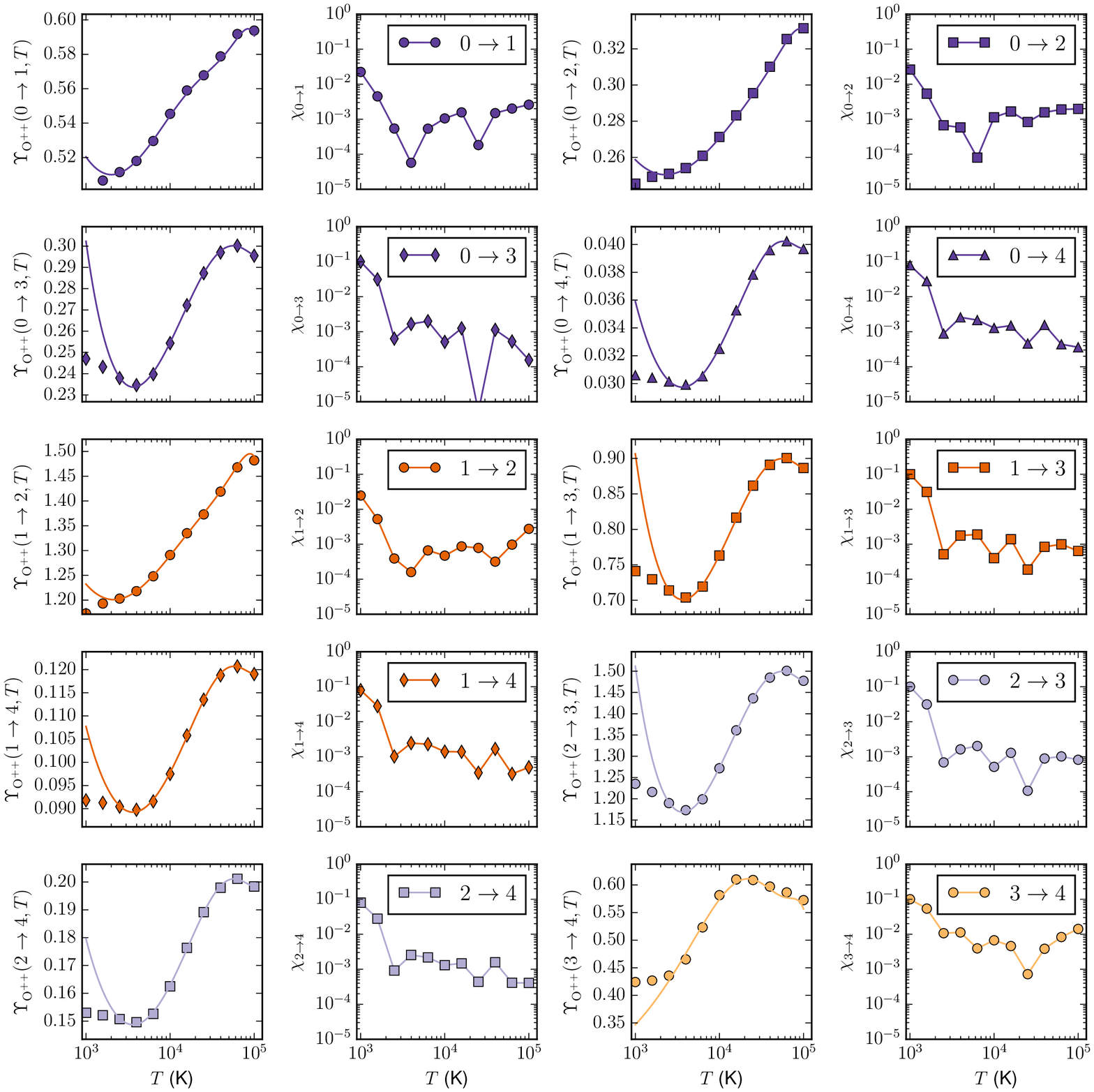}
\caption{Fits for O$^{++}$.
\label{figure:OIII_fit}}
\end{figure*}

\begin{table*}
\centering{}
\caption{Fit parameters for the velocity-averaged collision strength data for
neon used in the code.}
\label{table:collision_strength_fits_neon}

\vspace{12pt}
\begin{tabular}{c c c c c c c}
\hline
ion & & $0 \rightarrow{} 1$ & $0 \rightarrow{} 2$ & $0 \rightarrow{} 3$ & $0
\rightarrow{} 4$ & $1 \rightarrow{} 2$ \\
& & $1 \rightarrow{} 3$ & $1 \rightarrow{} 4$ & $2 \rightarrow{} 3$ & $2
\rightarrow{} 4$ & $3 \rightarrow{} 4$ \\
\hline
Ne$^+$ &
$\begin{matrix}
a \\
b \\
c \\
d \\
e \\
f \\
g \\
\end{matrix}$
&
\begin{tabular}{l}-8.548e-01 \\1.938e-01 \\-1.038e+01 \\-1.254e-02 \\
2.715e-02 \\5.755e-05 \\4.171e-06 \\\end{tabular}
& & & &
\vspace{12pt}
\\
Ne$^{++}$ &
$\begin{matrix}
a \\
b \\
c \\
d \\
e \\
f \\
g \\
\end{matrix}$
&
\begin{tabular}{l}-6.515e-01 \\1.958e-01 \\-3.645e+01 \\-1.812e-02 \\
-1.815e-03 \\-1.043e-03 \\-7.735e-05 \\\end{tabular}
&
\begin{tabular}{l}-7.577e-01 \\1.184e-01 \\-2.862e+01 \\-1.047e-02 \\
-9.100e-04 \\-1.094e-03 \\-8.029e-05 \\\end{tabular}
&
\begin{tabular}{l}-5.448e-03 \\-9.752e-06 \\8.134e-01 \\7.937e-07 \\
4.823e-06 \\3.627e-05 \\3.005e-06 \\\end{tabular}
&
\begin{tabular}{l}1.019e-01 \\-8.068e-06 \\4.190e-02 \\8.074e-07 \\
1.758e-08 \\-5.000e-03 \\-3.672e-04 \\\end{tabular}
&
\begin{tabular}{l}-6.173e-01 \\4.546e-02 \\-8.088e+00 \\-4.215e-03 \\
-2.781e-04 \\-1.537e-03 \\-1.139e-04 \\\end{tabular}
\vspace{12pt}
\\
&
$\begin{matrix}
a \\
b \\
c \\
d \\
e \\
f \\
g \\
\end{matrix}$
&
\begin{tabular}{l}-1.932e-03 \\-8.154e-06 \\4.758e-01 \\7.006e-07 \\
-9.413e-05 \\-8.679e-07 \\-7.422e-08 \\\end{tabular}
&
\begin{tabular}{l}1.161e-01 \\-5.770e-06 \\2.333e-02 \\5.826e-07 \\
-6.898e-04 \\1.004e-07 \\7.445e-09 \\\end{tabular}
&
\begin{tabular}{l}-2.118e-02 \\7.071e-06 \\1.788e-01 \\-7.768e-07 \\
1.263e-07 \\1.534e-03 \\1.191e-04 \\\end{tabular}
&
\begin{tabular}{l}9.843e-02 \\-1.648e-06 \\8.827e-03 \\1.628e-07 \\
1.792e-08 \\-8.454e-04 \\-6.126e-05 \\\end{tabular}
&
\begin{tabular}{l}5.812e-02 \\-4.021e-05 \\1.876e-01 \\4.268e-06 \\
3.941e-03 \\-1.725e-07 \\-1.298e-08 \\\end{tabular}
\\
\hline
\end{tabular}
\end{table*}

\begin{figure}
\centering{}
\includegraphics[width=0.48\textwidth]{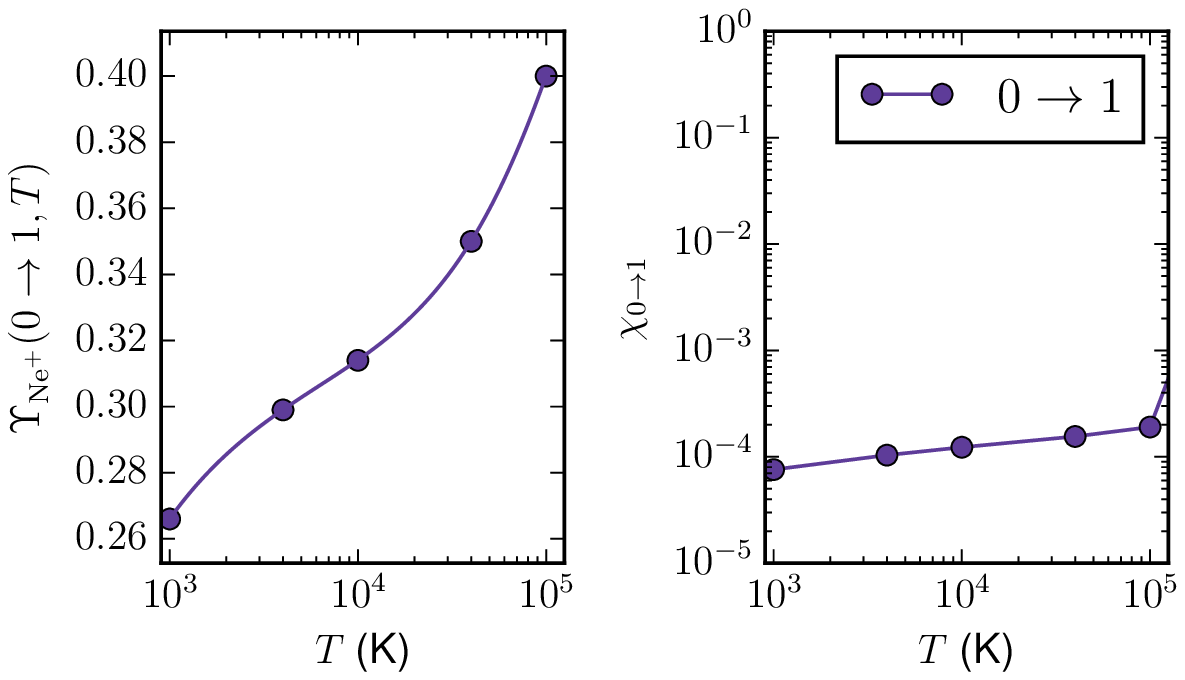}
\caption{Fits for Ne$^{+}$.
\label{figure:NeII_fit}}
\end{figure}

\begin{figure*}
\centering{}
\includegraphics[width=0.98\textwidth]{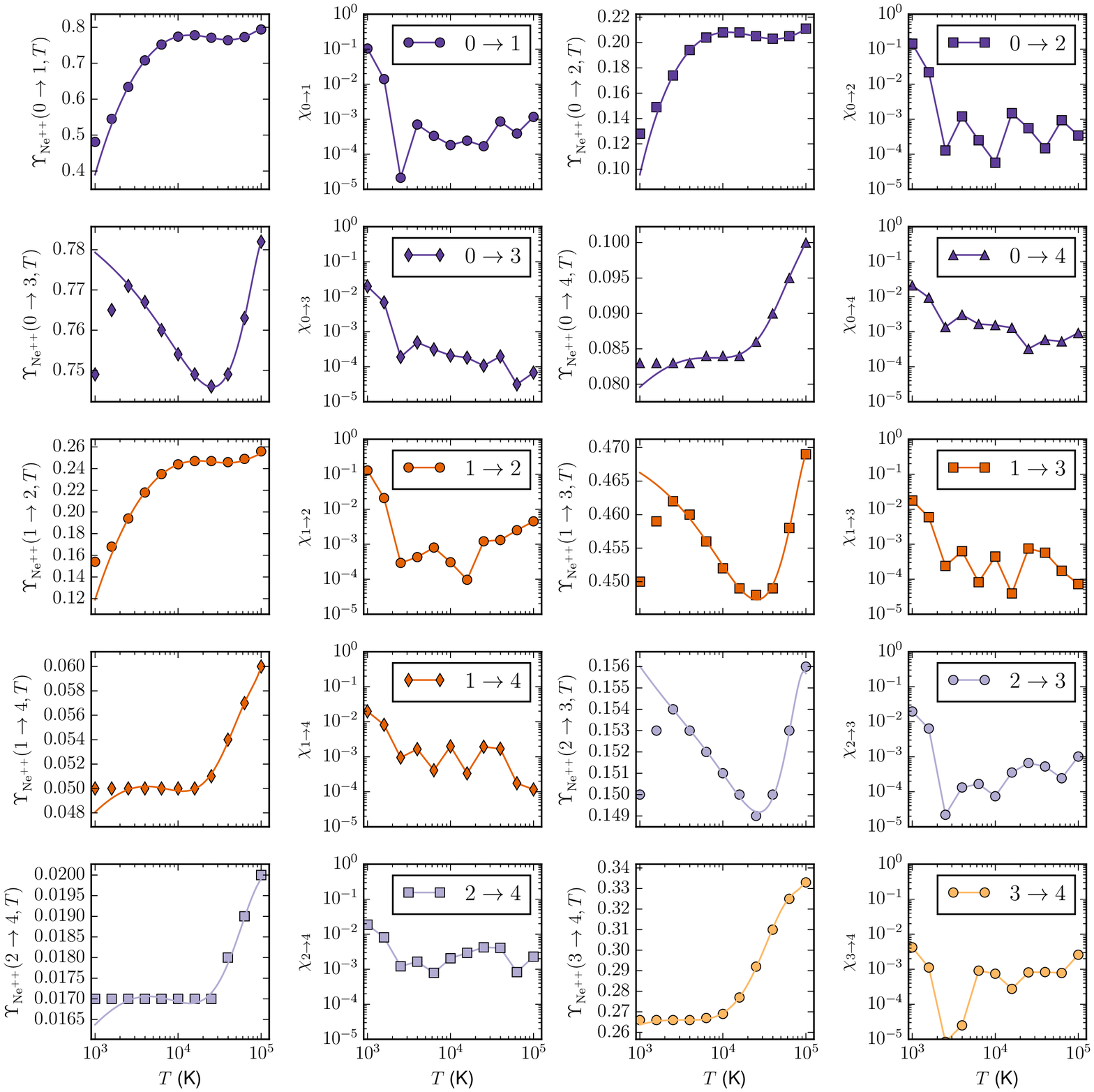}
\caption{Fits for Ne$^{++}$.
\label{figure:NeIII_fit}}
\end{figure*}

\begin{table*}
\centering{}
\caption{Fit parameters for the velocity-averaged collision strength data for
sulphur used in the code.}
\label{table:collision_strength_fits_sulphur}

\vspace{12pt}
\begin{tabular}{c c c c c c c}
\hline
ion & & $0 \rightarrow{} 1$ & $0 \rightarrow{} 2$ & $0 \rightarrow{} 3$ & $0
\rightarrow{} 4$ & $1 \rightarrow{} 2$ \\
& & $1 \rightarrow{} 3$ & $1 \rightarrow{} 4$ & $2 \rightarrow{} 3$ & $2
\rightarrow{} 4$ & $3 \rightarrow{} 4$ \\
\hline
S$^+$ &
$\begin{matrix}
a \\
b \\
c \\
d \\
e \\
f \\
g \\
\end{matrix}$
&
\begin{tabular}{l}-2.378e-02 \\-3.780e-05 \\3.350e+00 \\2.326e-06 \\
0.000e+00 \\0.000e+00 \\0.000e+00 \\\end{tabular}
&
\begin{tabular}{l}-2.591e-02 \\-5.072e-05 \\5.088e+00 \\3.003e-06 \\
0.000e+00 \\0.000e+00 \\0.000e+00 \\\end{tabular}
&
\begin{tabular}{l}-4.744e-01 \\1.594e-02 \\1.361e+01 \\-1.274e-03 \\
0.000e+00 \\0.000e+00 \\0.000e+00 \\\end{tabular}
&
\begin{tabular}{l}-4.730e-01 \\3.177e-02 \\2.699e+01 \\-2.544e-03 \\
0.000e+00 \\0.000e+00 \\0.000e+00 \\\end{tabular}
&
\begin{tabular}{l}-8.272e-02 \\3.140e-05 \\1.492e+01 \\-5.602e-06 \\
0.000e+00 \\0.000e+00 \\0.000e+00 \\\end{tabular}
\vspace{12pt}
\\
&
$\begin{matrix}
a \\
b \\
c \\
d \\
e \\
f \\
g \\
\end{matrix}$
&
\begin{tabular}{l}-5.368e-02 \\8.155e-05 \\2.216e+00 \\-6.673e-06 \\
0.000e+00 \\0.000e+00 \\0.000e+00 \\\end{tabular}
&
\begin{tabular}{l}-2.183e-02 \\5.571e-05 \\2.822e+00 \\-4.915e-06 \\
0.000e+00 \\0.000e+00 \\0.000e+00 \\\end{tabular}
&
\begin{tabular}{l}-1.569e-02 \\2.911e-05 \\2.012e+00 \\-2.687e-06 \\
0.000e+00 \\0.000e+00 \\0.000e+00 \\\end{tabular}
&
\begin{tabular}{l}-3.268e-02 \\1.339e-04 \\5.190e+00 \\-1.130e-05 \\
0.000e+00 \\0.000e+00 \\0.000e+00 \\\end{tabular}
&
\begin{tabular}{l}-5.249e-01 \\6.595e-02 \\4.606e+01 \\-5.206e-03 \\
0.000e+00 \\0.000e+00 \\0.000e+00 \\\end{tabular}
\vspace{12pt}
\\
S$^{++}$ &
$\begin{matrix}
a \\
b \\
c \\
d \\
e \\
f \\
g \\
\end{matrix}$
&
\begin{tabular}{l}8.018e-02 \\-4.930e-05 \\1.206e+00 \\3.800e-06 \\
0.000e+00 \\0.000e+00 \\0.000e+00 \\\end{tabular}
&
\begin{tabular}{l}-2.250e-01 \\1.746e-03 \\3.814e+00 \\-1.408e-04 \\
0.000e+00 \\0.000e+00 \\0.000e+00 \\\end{tabular}
&
\begin{tabular}{l}-3.470e-02 \\6.561e-05 \\8.851e-01 \\-5.617e-06 \\
0.000e+00 \\0.000e+00 \\0.000e+00 \\\end{tabular}
&
\begin{tabular}{l}1.134e-01 \\3.383e-06 \\3.649e-02 \\-2.851e-07 \\
0.000e+00 \\0.000e+00 \\0.000e+00 \\\end{tabular}
&
\begin{tabular}{l}-4.587e-03 \\2.257e-04 \\4.846e+00 \\-1.949e-05 \\
0.000e+00 \\0.000e+00 \\0.000e+00 \\\end{tabular}
\vspace{12pt}
\\
&
$\begin{matrix}
a \\
b \\
c \\
d \\
e \\
f \\
g \\
\end{matrix}$
&
\begin{tabular}{l}-4.454e-02 \\2.365e-04 \\2.823e+00 \\-2.014e-05 \\
0.000e+00 \\0.000e+00 \\0.000e+00 \\\end{tabular}
&
\begin{tabular}{l}1.665e-01 \\2.041e-06 \\6.825e-02 \\-1.900e-07 \\
0.000e+00 \\0.000e+00 \\0.000e+00 \\\end{tabular}
&
\begin{tabular}{l}-3.303e-02 \\3.185e-04 \\4.890e+00 \\-2.742e-05 \\
0.000e+00 \\0.000e+00 \\0.000e+00 \\\end{tabular}
&
\begin{tabular}{l}1.765e-01 \\2.565e-06 \\1.041e-01 \\-2.451e-07 \\
0.000e+00 \\0.000e+00 \\0.000e+00 \\\end{tabular}
&
\begin{tabular}{l}-2.273e-01 \\3.338e-03 \\2.390e+00 \\-2.696e-04 \\
0.000e+00 \\0.000e+00 \\0.000e+00 \\\end{tabular}
\vspace{12pt}
\\
S$^{+++}$ &
$\begin{matrix}
a \\
b \\
c \\
d \\
e \\
f \\
g \\
\end{matrix}$
&
\begin{tabular}{l}-9.667e-01 \\-1.721e+01 \\2.979e+03 \\2.954e+00 \\
-1.593e+01 \\9.973e-05 \\8.092e-06 \\\end{tabular}
& & & & \\
\hline
\end{tabular}
\end{table*}

\begin{figure*}
\centering{}
\includegraphics[width=0.98\textwidth]{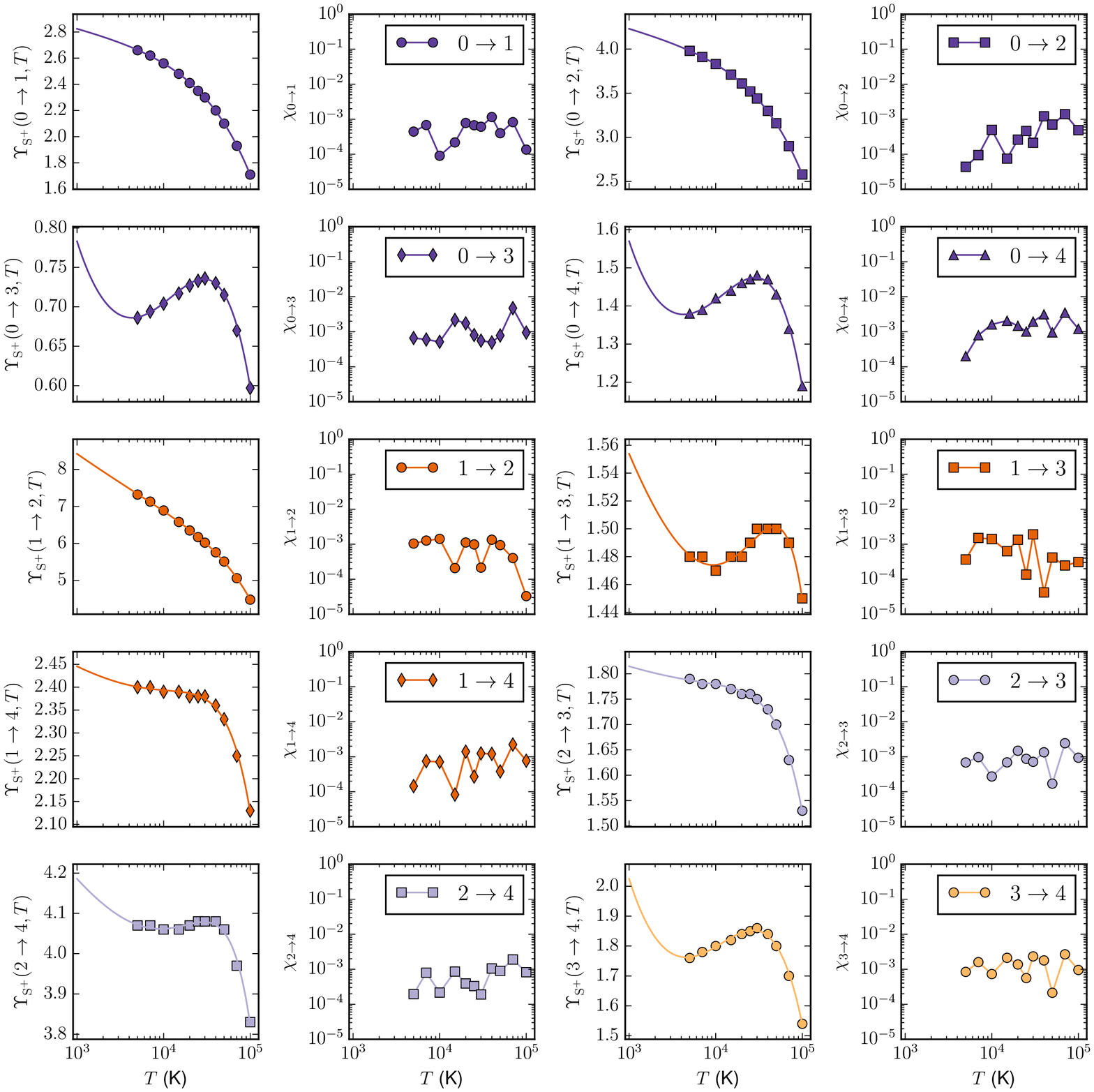}
\caption{Fits for S$^{+}$.
\label{figure:SII_fit}}
\end{figure*}

\begin{figure*}
\centering{}
\includegraphics[width=0.98\textwidth]{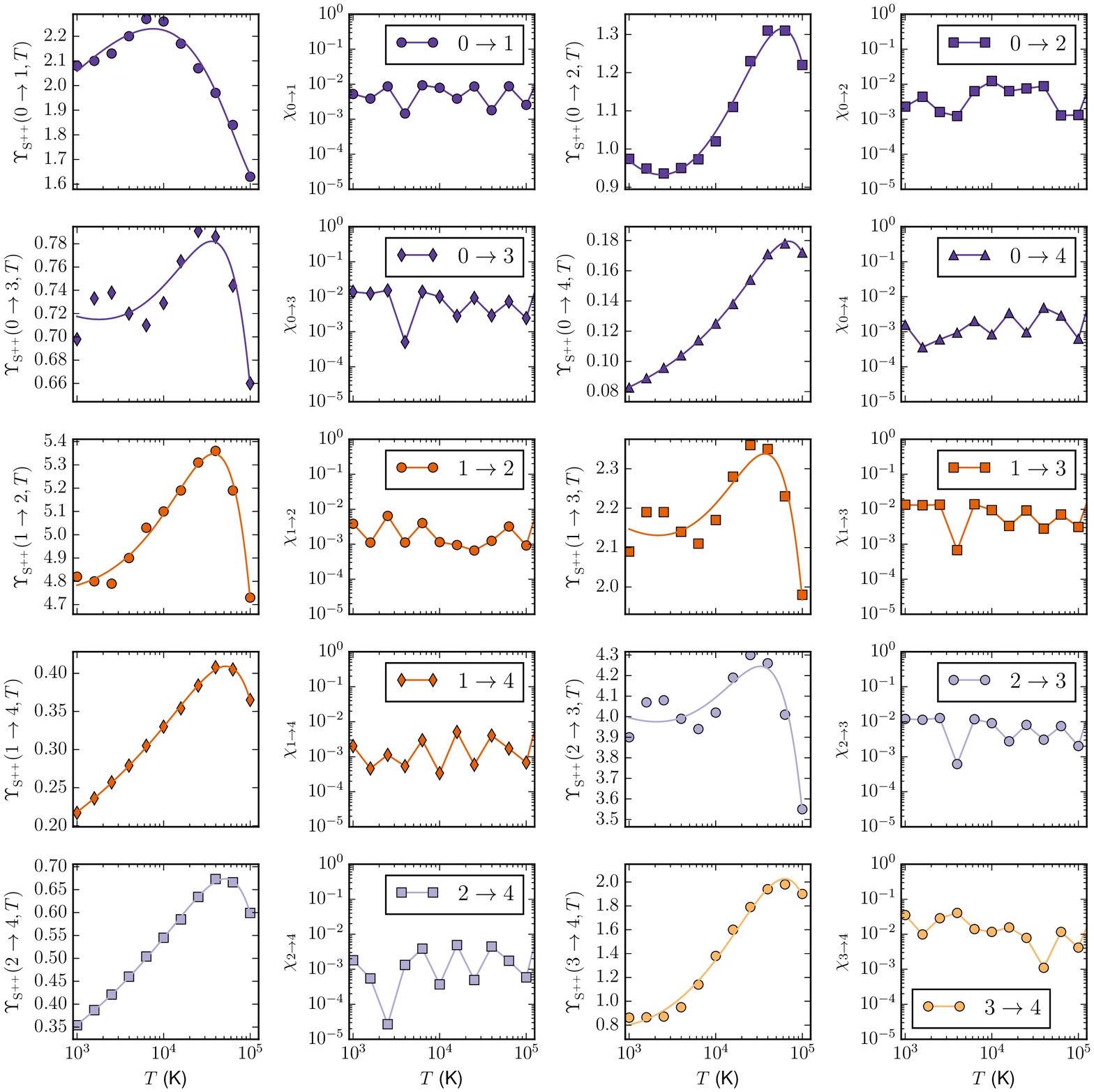}
\caption{Fits for S$^{++}$.
\label{figure:SIII_fit}}
\end{figure*}

\begin{figure}
\centering{}
\includegraphics[width=0.48\textwidth]{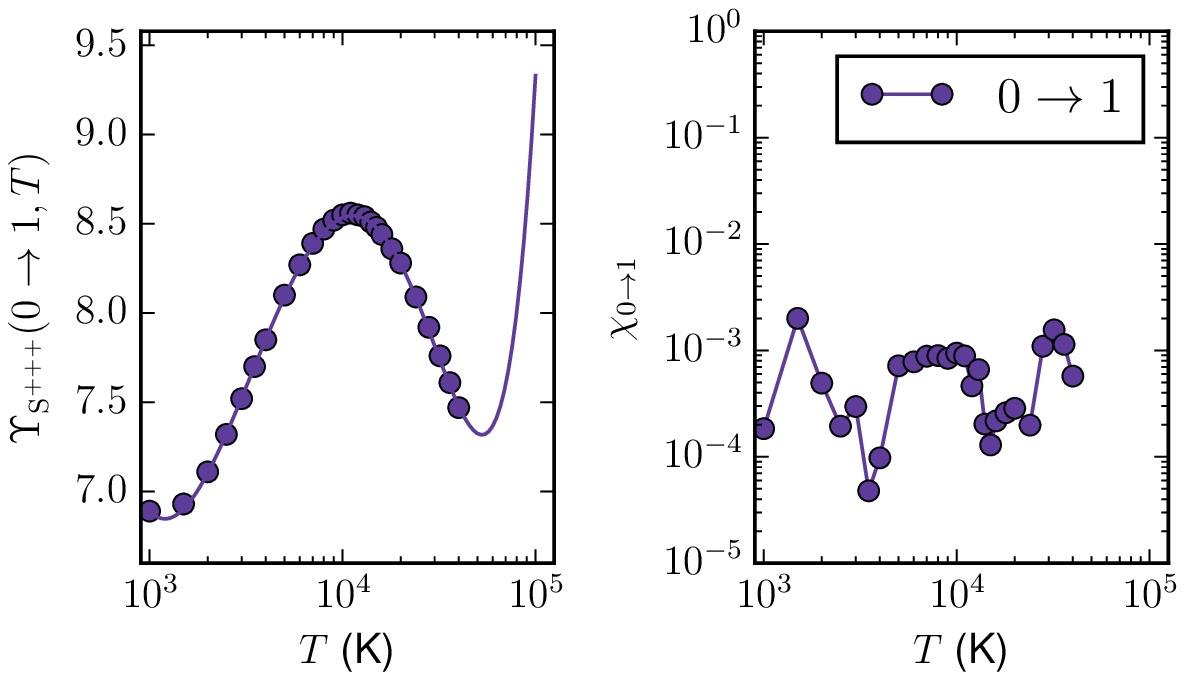}
\caption{Fits for S$^{+++}$.
\label{figure:SIV_fit}}
\end{figure}

\end{document}